\documentclass{elsart}
\usepackage{amsmath,amssymb}
\usepackage{graphicx}

\journal{Annals of Physics}

\renewcommand{\vec}[1]{\mathbf{#1}}
\newcommand{\unitvec}[1]{\mathbf{\hat{#1}}}
\newcommand{\matr}[1]{\hat{#1}}
\newcommand{\transpose}[1]{\mathop{{#1}^t}\nolimits}

\newcommand{\Id}{\matr{\mathbb{I}}}
\newcommand{\matS}{\matr{\mathcal{S}}}
\newcommand{\matC}{\matr{\mathcal{C}}}
\newcommand{\matE}{\matr{\mathcal{E}}}
\newcommand{\matU}{\matr{\mathcal{U}}}

\newcommand{\rmi}{\mathrm{i}}
\newcommand{\rme}{\mathrm{e}}
\newcommand{\rmd}{\mathrm{d}}

\newcommand{\ArcXY}{\mathrm{ArcXY}}
\newcommand{\atand}{\mathrm{atan2}}

\newlength\figwidth
\newcommand{\OneWithLabel}[2]
{(#2)
\begin{minipage}{0.9\figwidth}
\centering
\includegraphics*[angle=270,width=\textwidth]{#1.eps}
\end{minipage}
\\}

\newcommand{\TwoUnRelated}[4]
{(#3)
\begin{minipage}{0.4\figwidth}
\centering
\includegraphics*[angle=270,width=\textwidth]{#1.eps}
\end{minipage}
\hspace{0.075\figwidth}
(#4)
\begin{minipage}{0.4\figwidth}
\centering
\includegraphics*[angle=270,width=\textwidth]{#2.eps}
\end{minipage}
\\}

\newcommand{\ThreeUnRelated}[6]
{(#4)
\begin{minipage}{0.28\figwidth}
\centering
\includegraphics*[angle=270,width=\textwidth]{#1.eps}
\end{minipage}
\hfill
(#5)
\begin{minipage}{0.28\figwidth}
\centering
\includegraphics*[angle=270,width=\textwidth]{#2.eps}
\end{minipage}
\hfill
(#6)
\begin{minipage}{0.28\figwidth}
\centering
\includegraphics*[angle=270,width=\textwidth]{#3.eps}
\end{minipage}
\\}

\newcommand{\FiveNoLabel}[5]
{
\begin{minipage}{0.17\figwidth}
\centering
\includegraphics*[angle=270,width=\textwidth]{#1.eps}
\end{minipage}
&
\begin{minipage}{0.17\figwidth}
\centering
\includegraphics*[angle=270,width=\textwidth]{#2.eps}
\end{minipage}
&
\begin{minipage}{0.17\figwidth}
\centering
\includegraphics*[angle=270,width=\textwidth]{#3.eps}
\end{minipage}
&
\begin{minipage}{0.17\figwidth}
\centering
\includegraphics*[angle=270,width=\textwidth]{#4.eps}
\end{minipage}
&
\begin{minipage}{0.17\figwidth}
\centering
\includegraphics*[angle=270,width=\textwidth]{#5.eps}
\end{minipage}
\\}

\newcommand{\TwoByTwo}[6]
{\begin{minipage}{0.2\figwidth}
\centering
\includegraphics*[angle=270,width=\textwidth]{#1.eps}
\end{minipage}
(#5)
\begin{minipage}{0.2\figwidth}
\centering
\includegraphics*[angle=270,width=\textwidth]{#2.eps}
\end{minipage}
\hfill
\begin{minipage}{0.2\figwidth}
\centering
\includegraphics*[angle=270,width=\textwidth]{#3.eps}
\end{minipage}
(#6)
\begin{minipage}{0.2\figwidth}
\centering
\includegraphics*[angle=270,width=\textwidth]{#4.eps}
\end{minipage}
\\}

\begin{document}

\begin{frontmatter}

\title{Unified theory of bound and scattering molecular Rydberg states as
quantum maps}

\author[Darmstadt,CIC]{Barbara Dietz}
\ead{dietz@linix6.ikp.physik.tu-darmstadt.de}
\author[Spectro,CIC]{Maurice Lombardi\corauthref{Lombardi}}
\ead{Maurice.Lombardi@ujf-grenoble.fr}
\author[fisica,CIC]{Thomas H. Seligman}
\ead{seligman@ce.fis.unam.mx}

\address[Darmstadt]{Institute for Nuclear Physics. Technische Universitat
Darmstadt. Schlossgartenstrasse 9. Darmstadt. Germany}
\address[Spectro]{Laboratoire de Spectrom\'etrie Physique(CNRS UMR 5588),
Universit\'e Joseph-Fourier de Grenoble, BP87, F-38402
Saint~Martin~d'H\`eres~C\'edex, France}
\address[fisica]{Centro de Ciencias F\'\i sicas, UNAM, Av. Universidad
s/n, Col. Chamilpa, Morelos, Mexico}
\address[CIC]{Centro Internacional de Ciencias, Cuernavaca, Mexico}
\corauth[Lombardi]{Corresponding author. Fax: +33~476~635~495}

\begin{abstract}
Using a representation of multichannel quantum defect theory in terms of a
quantum Poincar\'e map for bound Rydberg molecules, we apply Jung's scattering
map to derive a generalized quantum map, that includes the continuum. We show,
that this representation not only simplifies the understanding of the method,
but moreover produces considerable numerical advantages. Finally we show under
what circumstances the usual semi-classical approximations yield satisfactory
results. In particular we see that singularities that cause problems in
semi-classics are irrelevant to the quantum map.
\end{abstract}

\begin{keyword}
Rydberg molecule \sep Semi classical physics \sep Quantum chaos
\PACS 5.45.Mt \sep 03.65.Sq \sep 31.15.Gy \sep 33.80.Rv \sep 34.80.Gs
\end{keyword}

\end{frontmatter}

\section{Introduction}

Electronic states of molecules are called Rydberg states, as opposed to valence
states (whether covalent or ionically bound), when an outer
electron moves far away from the remaining ionic core. These states form
electronic series which converge towards the ionization limit of the molecule.

The starting point of the quantum analysis of such states was the Quantum
Defect Theory (see e.g. the review article by Seaton \cite{Seaton:RPG83-167}), 
established first for atoms.
It was shown that, due to the non zero spatial extension of the ionic core, the
levels near the ionization limit follow the hydrogenic Rydberg law
$E_n=-\mathrm{Ry}/(n+d)^2$, with only a constant (or nearly so) shift $d$ of
the principal quantum number $n$, entitled Quantum Defect. Quantum Defect Theory
was extended to
a Multichannel Quantum Defect Theory (MQDT), for the case that there are 
several series
which converge to nearby states of the ion, and interact strongly. It was shown
that MQDT gives a unified theory of bound states, autoionizing states and
electron-ion scattering cross sections. This theory depends only on a small
number of parameters, basically one quantum defect per interacting series.
Practically, all is solved with matrices whose size is the number of series,
while ``brute force'' methods would in principle try to diagonalize a matrix
which contains an infinite number of levels for each series.

This theory was extended to molecules by Fano
\cite{Fano:PRA70-253,Fano:JOSA75-979}. There are always many interacting series
corresponding to the rotational states of the ionic core. Indeed the slow
velocity of the core rotation leads to a splitting of the rotational states of 
the
core which is of the same order of magnitude as the splitting between high lying
electronic Rydberg states. The novelty were the implications
of the anisotropy of the core. The effect of this anisotropy on the ionic
potential decays faster with distance $r$ than the point charge $1/r$ Coulomb
potential, at least as $1/r^2$ or $1/r^3$. Fano showed that the key point
of the analysis is the existence of a cut off distance $r_0$. Below this
distance the motion of the outer electron is tightly bound to the
\emph{direction} of the ionic core, above it the two become independent. Many
detailed studies have followed on moderately excited Rydberg states of
molecules, see e.g. reviews in refs.~\cite{Greene:AAMP85-51,Jungen96}.

A novelty resulting from the studies of very high lying states of
molecular Rydberg series was the experimental observation of 
``clear zones'' in such interacting series. It was soon understood that this so
called ``stroboscopic effect'' corresponds to resonances between the period of
rotation of the core and the period of the outer electron orbit
\cite{Labastie:PRL84-1681,Bordas:JP85-27}. In order to 
study the relationship of this
phenomenon with classical and quantum chaos, the classical limit of the MQDT
theory was established \cite{Lombardi:JCP88-3479}. A Poincar\'e surface
of section was introduced, whose coordinates correspond to parameters of the
molecular system when the electron leaves the sphere of radius $r_0$.
The studies showed that the ``stroboscopic effect'' corresponds to a periodic
reestablishment of a nearly integrable phase space structure at each resonance,
while the classical phase space is completely chaotic in between.

The impetus for the present study was given by papers of Bogomolny where he
introduces a
semi-classical method for the quantization of a classical Poincar\'e surface 
of section Map
(PM) \cite{Bogomolny:CAMP90-67,Bogomolny:N92-805,Bogomolny:C92-5}, and some
other follow up papers, e.g. \cite{Prosen:PD96-22}. The original semi-classical
method is not unique and is prone to singularity problems, 
thereby raising the question about its relationship with a true quantization. We
followed exactly the opposite route, starting from a
purely quantum method (MQDT), establishing its classical limit, and at last
studying Poincar\'e surfaces of sections. Our system is ideal for the 
study of
this problem because we know in advance the correct result. Furthermore our 
system has an unusual and non trivial geometry, the \emph{phase space} being a
sphere. For bound systems, a preliminary report of the interpretation of MQDT 
as
a Quantum Poincar\'e Map (QPM) was provided in ref.~\cite{Leyvraz:PLA00-309}.
However, the relationship with the theory of Bogomolny was only a formal 
analogy of
the final quantum and semi-classical formulae used to compute the levels. In
this paper we study and compare step by step classical, semi-classical and
quantum descriptions of the same system.

Another important property of the MQDT is, that it describes  
bound and ionized states in a unified way. Hence, we may use it 
to study the quantization of the so
called Jung Scattering Map (JSM) for ionized states \cite{Jung:JPA86-1345}.
Indeed Jung proposed to send back ionizing trajectories onto the molecule in a
well prescribed way, in order to obtain a of compactification of the phase space
for above threshold states. The role of the JSM in this context, has been
touched upon previously to elucidate certain properties of the $S$-matrix
\cite{Dietz:JPA96-95}. In the light of the reinterpretation of MQDT as a QPM
for bound systems it seems very attractive to attempt a comprehensive 
description,
which extends the advantages outlined in ref.~\cite{Leyvraz:PLA00-309} to
include both scattering, bound states and resonances.

The purpose of this paper is to present a classical map which combines
the PM for the part of the phase space corresponding to negative energies of
the excited electron with the JSM for the positive energy region, and then to
interpret MQDT as the quantization of the resulting symplectic map. This
will provide basically two advantages. On one hand a conceptual advantage, as
the method and its approximation can be formulated in the context of a
classical map and its subsequent quantization. In particular, we will see that
phase space representations of wavefunctions for open channels quantize on
trajectories of the JSM in the same way as bound states and ionization
resonances quantize on trajectories of the PM. On the other hand on a more
practical level, the dynamics is shown to be semi-separable
\cite{Prosen:PD96-22,Prosen:EL01-12} which permits a decomposition into
separable steps, and suggests to choose a surface of section such that it
separates these steps. Consequently, the QPM for half an evolution step of 
the system
contains the relevant information, and its diagonalization will permit to find
eigenvalues, eigenvectors, the level dynamics, \textit{etc} in a simplified way.
This will be particularly useful when dealing with near degenerate levels or
resonances which occur frequently and cause considerable difficulties in the
conventional approach \cite{Leyvraz:PLA00-309,Lombardi:PRA93-3571}.

In the next section we shall develop the generalized classical PM, which
includes the JSM for the particular conditions of a Rydberg molecule. In the
third section we shall present the main result of this paper. MQDT for bound
states and open systems will be formulated as a generalized QPM. In both these
sections attention will be given to relaxing the common approximation to keep
the absolute value of the electron angular momentum fixed. In section four we
shall discuss applications and show numerical examples that demonstrate the
practical advantages of this method and may also give some new insight in
situations of quantum chaotic scattering. These examples will exclusively be
given within the approximation of fixed absolute value of the electron angular
momentum, mainly to show nice two dimensional plots of surfaces of section. As
the QPM was originally proposed in a semi-classical approximation
\cite{Bogomolny:CAMP90-67,Bogomolny:N92-805,Bogomolny:C92-5}, we shall in
section~\ref{sec:SemiClassics} compare the approximate results of semi-classics
to the ones of the "exact" QPM. We shall see that singularities causing
problems in semi-classics are absent in the QPM. Therefore the QPM can be used
to test regularization procedures for semi-classics.

\section{\label{sec:classical}Generalized classical Poincar\'e map for Rydberg
molecules}

\subsection{Principle}

\begin{figure}[!htb]
  \centering
  \setlength{\figwidth}{\textwidth}
  \TwoUnRelated{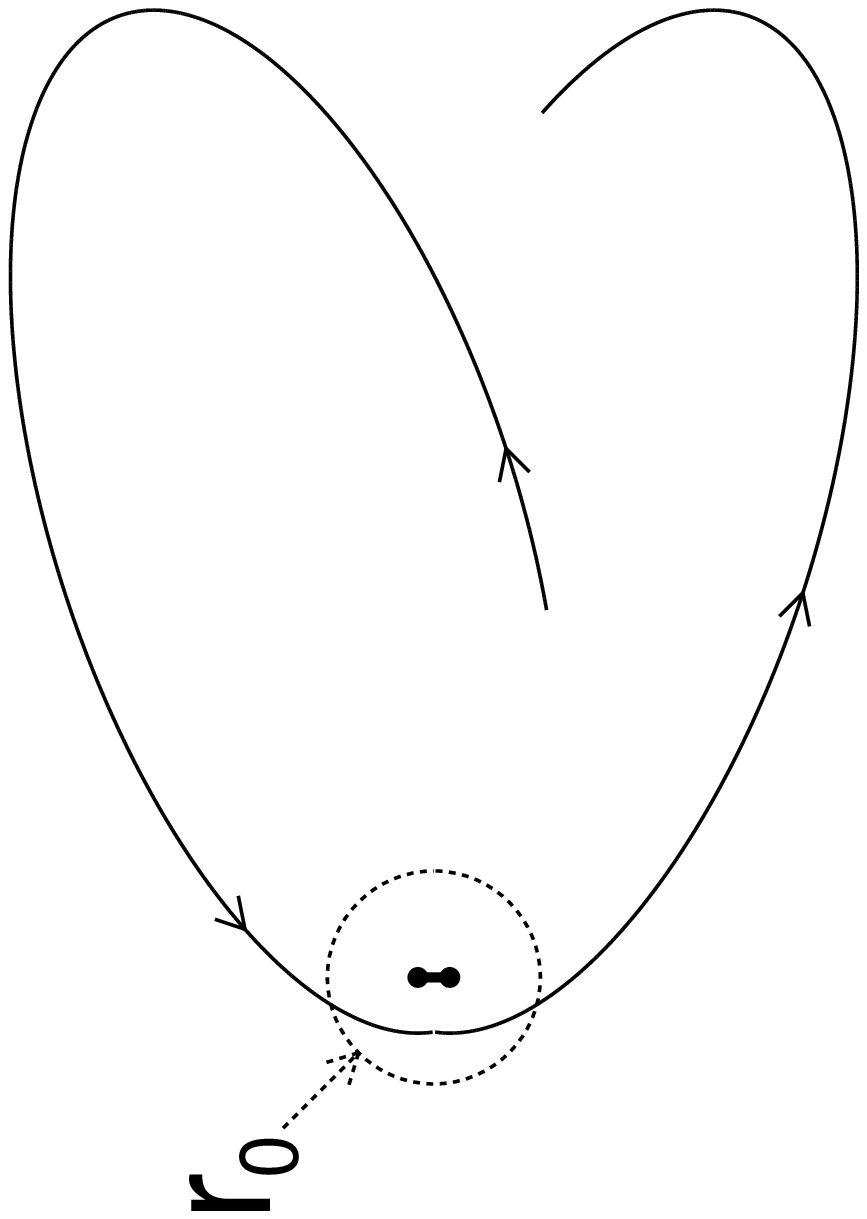}{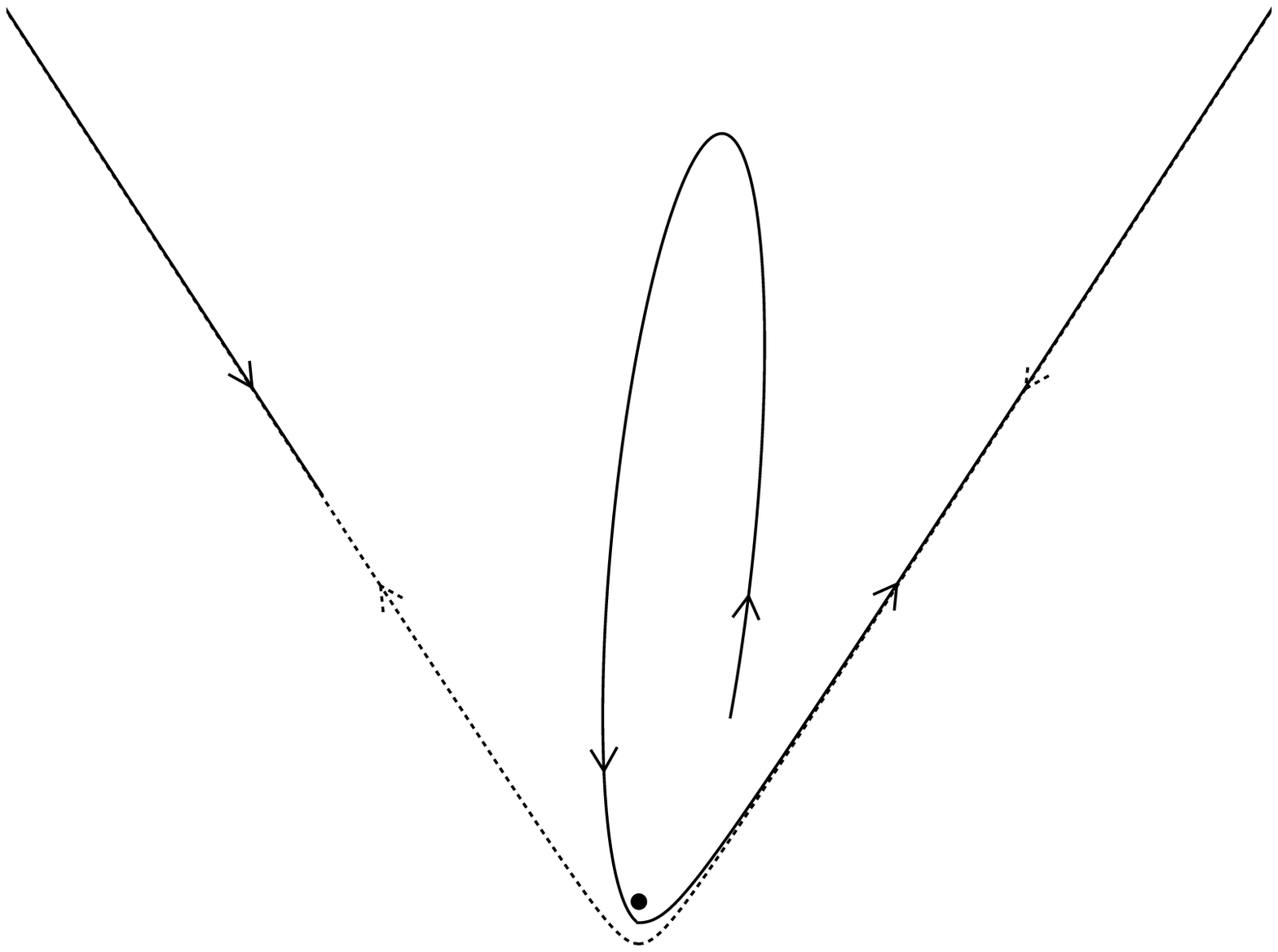}{a}{b}
  \caption{\label{fig:Principle} (a) The outer electron follows a Kepler orbit,
fixed in space, for large distances $r>r_0$ and changes the Kepler orbit for
$r<r_0$ by interaction with the short range part of the ionic core potential.
(b) When the electron ionizes the Jung's recipe is to feed back the electron
which escapes to infinity by doing a step backwards in time without collision
with the core (i.e. a pure Kepler hyperbola with reverted velocity), and then
re launch the electron towards the molecule along the resulting asymptotic
trajectory for the next collision.}
\end{figure}

A Rydberg molecule is composed of an ionic positively charged molecular core
and an outer electron. It is specifically given the name ``Rydberg'' when one
considers states just below or above the ionization threshold, where the
outer electron performs large excursions far away from the core. This situation
allows to define a distance $r_0$ which separates the potential felt by the
outer electron into a long range Coulomb part, and a short range or
collision part. In the long range part, the outer electron only feels the
spherically symmetric $-1/r$ Coulomb potential of the core (we use throughout
the ``atomic units'' (a.u.) $e=\hbar=m=1$). Its motion is comparatively slow
because one considers states near the ionization limit, but it can be either
bound or unbound. The electron moves along a Kepler orbit whose plane and 
axis are fixed in
the laboratory frame. Meanwhile, the ionic core rotates freely around its
angular momentum $\vec{N}$. In the collision area the electron feels the lower
symmetry, faster decreasing, part of the potential of the core (dipolar
($\propto 1/r^2$) or quadrupolar ($\propto 1/r^3$) for a diatomic molecule) and
it is strongly accelerated. This strong acceleration leads to a short range
collision motion, which is approximately the same for a total energy slightly
below or above the ionization threshold, since the difference between the
corresponding velocities is negligible compared to 
the velocity produced by the acceleration. This is basically the reason 
why in quantum mechanics MQDT treats bound and unbound states in a unified 
way. During the
collision there is exchange both of angular momentum and energy between the
electron and the core. The exchange of angular momentum implies a change of
the plane of the
Kepler orbit, since the angular momentum of the electron $\vec{L}$ is
perpendicular to its plane (Fig.~\ref{fig:Principle}(a)). The exchange of
energy leads to a variation of the eccentricity of the orbit, 
and may turn it into an ionizing trajectory. In this case the Jung's recipe
\cite{Jung:JPA86-1345} is to feed back the electron which escapes to infinity
by doing a step backwards in time without collision with the core (i.e. a pure
Coulomb step with reverted velocity), and then re launch the electron towards
the molecule along the resulting asymptotic trajectory
(Fig.~\ref{fig:Principle}(b)). By doing this, one obtains trajectories
which cross infinitely many times the $r=r_0$ sphere. Poincar\'e surfaces of
section are described by a set of parameters of the molecular system at those 
instants where the electron
crosses outwards the sphere. To define these parameters (several sets are
possible) we first need to define precisely the reference frames and count
the number of independent parameters for the  determination of the dimension 
of this Poincar\'e surface of section.

\subsection{Reference frames}

\begin{figure}[!hp]
\setlength{\figwidth}{0.9\textwidth}
\centering
\OneWithLabel{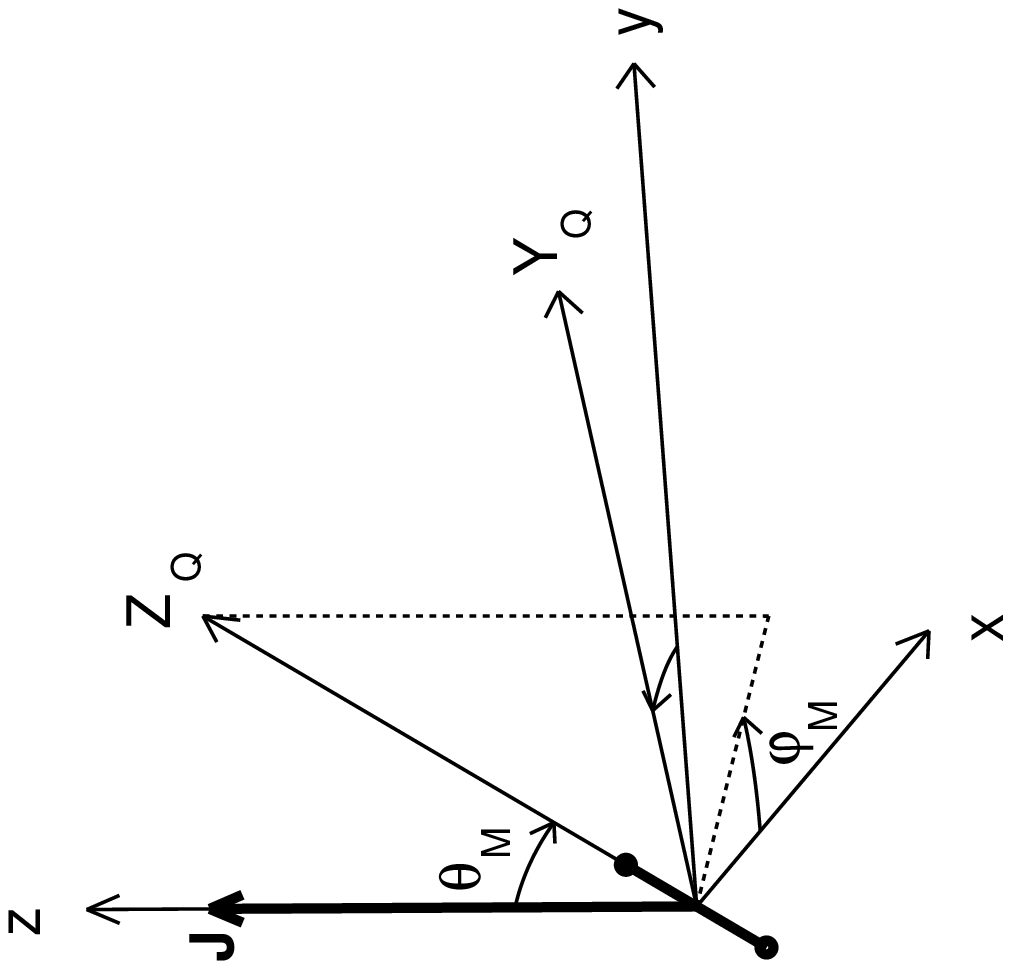}{a}
\OneWithLabel{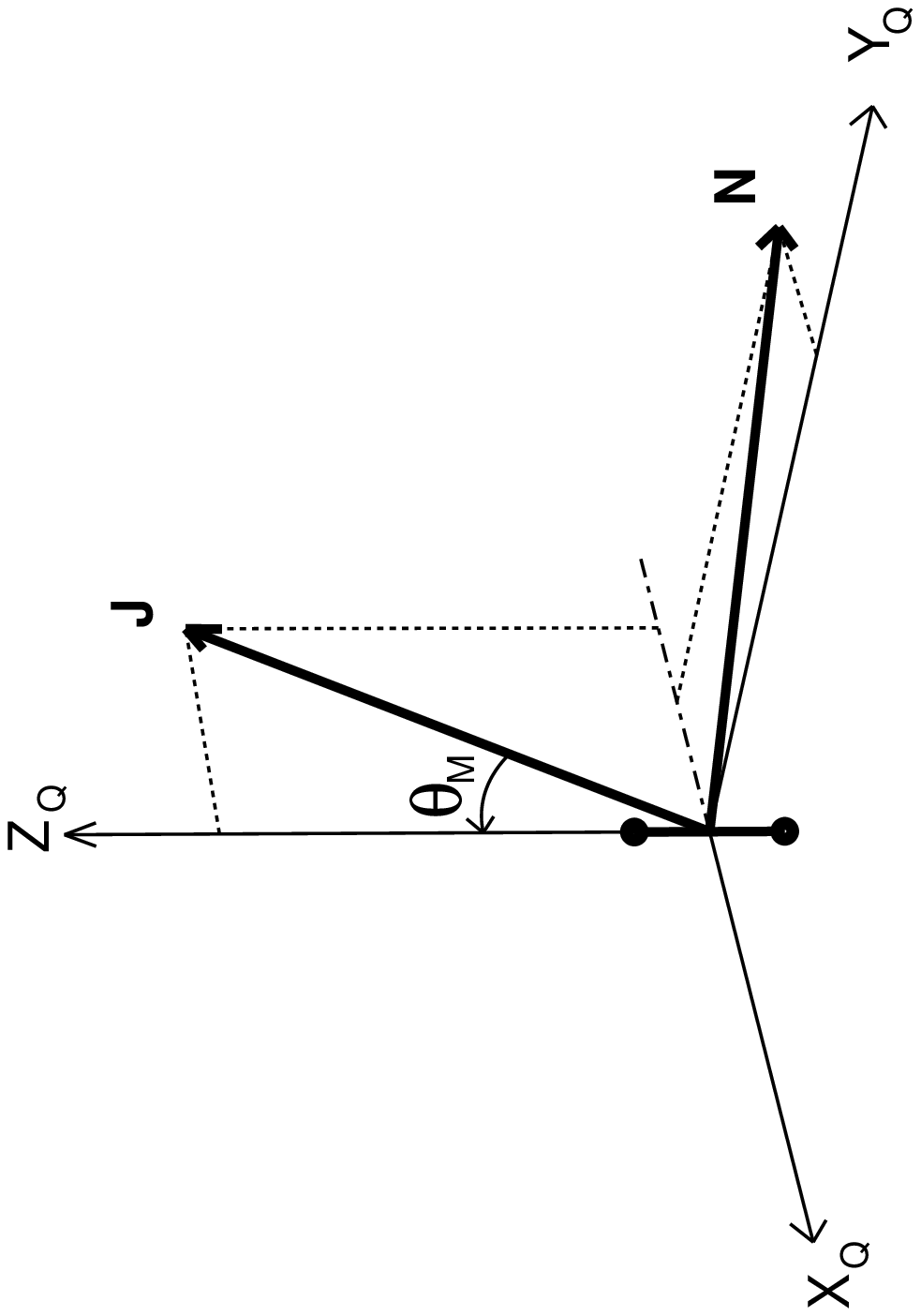}{b}
\OneWithLabel{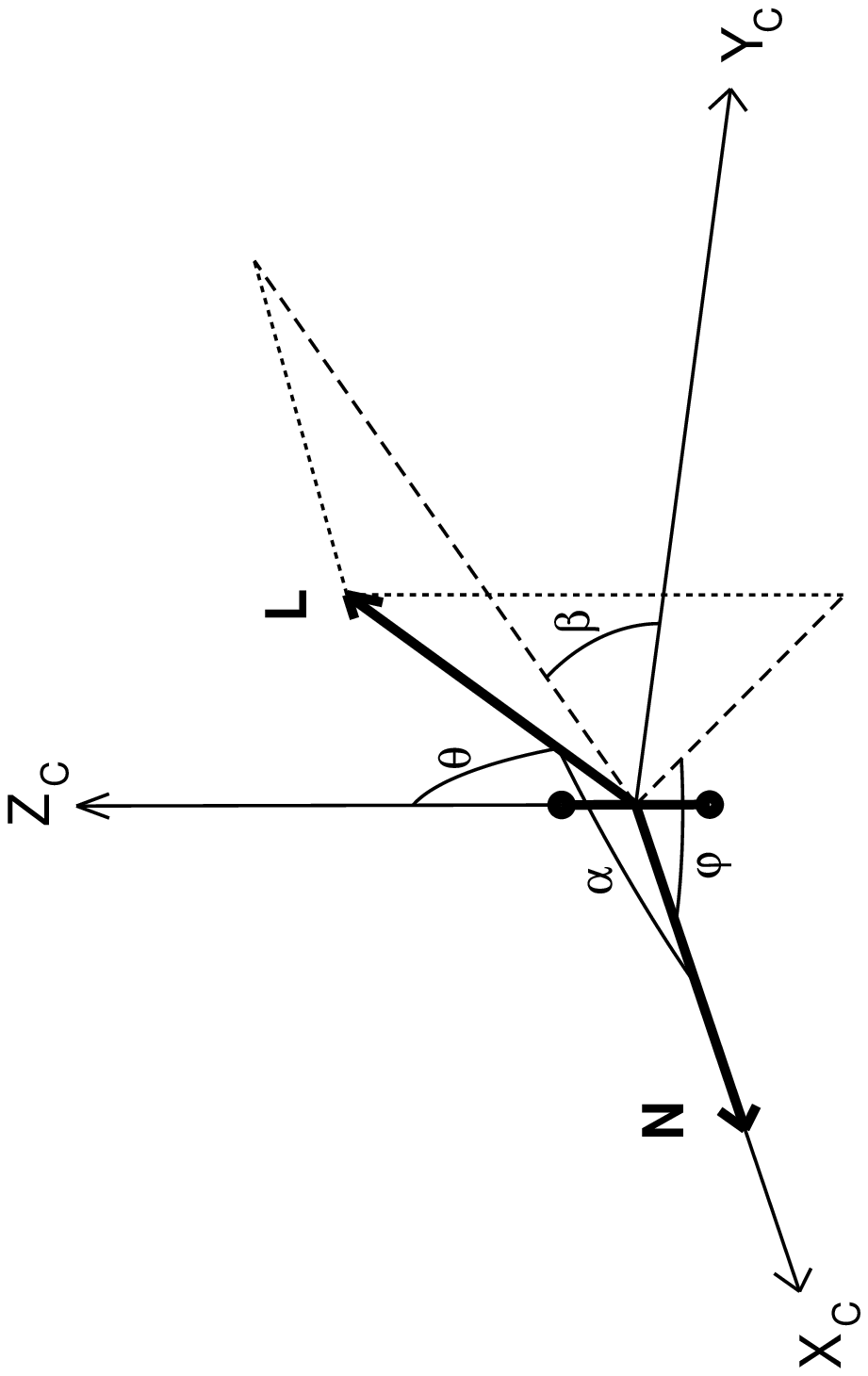}{c}
\caption{\label{fig:RefFrames}
Reference frames for the molecular system.
(a) Laboratory frame. Its $Oz$ axis is selected along the total angular
momentum $\vec{J}$
(b) ``Quantum'' molecular reference frame. Obtained from the laboratory frame
with a rotation of Euler angles $\varphi_M$, $\theta_M$, $0$. Its $OZ_Q$ axis
is along the molecular axis $\unitvec{M}$.
(c) ``Classical'' molecular reference frame. It is obtained from the
``quantum'' by a rotation around $\unitvec{M}$ which brings the $OX_C$ axis
along the angular momentum $\vec{N}$ of the core. This frame is physically the
most significant, and all other figures of this paper are drawn with this
convention.}
\end{figure}

In this work, we will use three reference frames, the laboratory frame and two
molecular reference frames, entitled ``quantum'' and ``classical'' reference
frames. They are shown in Fig.~\ref{fig:RefFrames}.

\begin{enumerate}
\renewcommand{\theenumi}{{\alph{enumi}}}
    \item Laboratory frame. Due to the global rotational invariance which 
entails
conservation of the total angular momentum $\vec{J}$, we can select, without
loss of generality, its $Oz$ axis along the total angular momentum
$\vec{J}$. The other two axis are arbitrary (but fixed). In this frame the axis
$\unitvec{M}$ of the core has (varying) polar angles $\theta_M$ and $\phi_M$.
    \item ``Quantum'' molecular reference frame. It is obtained  by a rotation
of the laboratory
frame by the Euler angles  $\theta_M$, $\varphi_M$, $0$ (as common in
quantum mechanics \cite{MessiahII64} we use the y convention for the Euler 
angles
\cite[Appendix B]{Goldstein80}). We entitled it ``quantum'' because it is the
frame used in the quantum theory of molecules. The third Euler angle is 
arbitrary for a diatomic molecule, and is sometimes selected as $\pi/2$
\cite{Landau:QM77} for normalization reasons, which do not concern us here.
Therefore, we prefer to stick to the conventions of the first paper by Fano
\cite{Fano:PRA70-253}. We restrict to the diatomic case for simplicity, a third
Euler angle would be necessary for a polyatomic core. In this frame $\vec{J}$
lies in the ``vertical'' plane $O{X_Q}{Y_Q}$. The angular momentum of the core
$\vec{N}$ for a diatomic molecule is perpendicular to the core axis
$\unitvec{M}$, because the moment of inertia of the core around it is zero. It
is thus in the ``horizontal'' plane $O{X_Q}{Y_Q}$.
    \item ``Classical'' molecular reference frame. This frame is obtained from
the ``quantum'' frame by a rotation around $O{Z_Q}=O{Z_C}$ such that the new
$O{X_C}$ axis is along $\vec{N}$. As will become obvious below, from the physical point of view this frame is the most appropriate for the visualization
of the results. As in previous publications
\cite{Lombardi:JCP88-3479,Leyvraz:PLA00-309,Dietz:JPA96-95,Lombardi:PRA93-3571,Leyvraz:PD01-169} we will use it in all figures of this paper.
\end{enumerate}

Formulas to go from one reference frame to the others are obtained from simple
geometric arguments based on the previous remarks, and are established in
Appendix~\ref{sec:QCFrames}.

\subsection{\label{sec:ParamCount}Count of parameters and the dimension of the
Poincar\'e surface of section}

The configuration space is described by five coordinates in the laboratory
frame. Two angles $\theta_M, \varphi_M$ describe the position of the axis of
the diatomic ionic core (a third would be necessary for a polyatomic core). Two
angles $\theta_e, \varphi_e$ and a radial distance $r$ describe the position of
the electron. Thus configuration space has a dimension of five and phase space
has a dimension of ten. Conservation of the total angular momentum $\vec{J}$
reduces the dimension of phase space by 4, two for the Poisson commuting
actions $J_z$ and $J^2$, and two for the associated angles. The first angle can
be selected as the polar angle $\varphi_J$ associated with $J_z$, that is the
angle of the projection of $\vec{J}$ onto the laboratory plane $Oxy$, which
is constant (with our choice of the laboratory axis it is in fact undetermined 
but
this is of no importance). The second can be a global rotation angle around
$\vec{J}$ associated to $J^2$, which is cyclic (but not constant): see e.g. the
analogous reduction to the $\{j_1,j_2,j_3,w_1,w_2,w_3\}$ action angle system
for the pure Kepler problem in \cite[Chap. 10-7]{Goldstein80}. 
Consequently, the dimension of
the reduced phase space in the general case is six, and a Poincar\'e
surface of section has dimension four.

The introduction of the auxiliary approximation that
the modulus $L$ of the angular momentum $\vec{L}$ of the electron is conserved
during the collision reduces the dimension of the phase space by two
($L^2$ and the associated angle), and thus the Poincar\'e surface of section
has dimension two, which enables graphical representations. This approximation
was used in our previous works
\cite{Lombardi:JCP88-3479,Leyvraz:PLA00-309,Dietz:JPA96-95,Lombardi:PRA93-3571,Leyvraz:PD01-169},
and is valid in the experimental works which motivated the theoretical
study \cite{Labastie:PRL84-1681,Bordas:JP85-27}. The reason for its validity
is that the
splitting between different electronic $L^2$ levels produced by the spherically
symmetric part of the potential of the core is much larger than the splitting
between sub levels of $L^2$, produced by its anisotropic part. 
In the experiments, $L$ takes low values, restricted to 0\ldots 2 by
a two steps laser excitation. Here, however we want to study the semi classical
limit, where the angular momenta take large values. There are two possibilties 
to increase angular momenta.
First we can increase them without changing other molecular
parameters. This is experimentally
possible since the application of R.F. fields \cite{Hulet:PRL83-1430} or of
combined electric and magnetic fields \cite{Delande:EL88-303} in atomic Rydberg
states enable to climb the $L$ ladder up to circular orbits. In this case the
constant $L$ approximation breaks down for higher $L$. A study of the 
consequences  
for this case has been performed in ref.~\cite{Benvenuto:PRL94-1818}. 
However, this is not
what we aim at, since it implies a complete change of the physics of the problem
at hand. We want to gain insight into the physics of the experimental works in
refs.~\cite{Labastie:PRL84-1681,Bordas:JP85-27}, by establishing its
semi-classical limit. Mathematically this amounts to letting $\hbar \to 0$ while
keeping all classical parameters constant. Practically, since $\hbar$ is a
constant of nature (taken as 1 in atomic units), this requires the 
increase of all classical parameters while keeping constant their
dimensionless ratios. For example, decreasing $\hbar$ by a factor of $m$ 
implies to multiplication of all momenta by $m$, angular $L$, $J$, $N$, 
and principal, $n$. The
Kepler period $T_e$ of the orbital motion of the electron, which is 
proportional to $n^3$ is
thus multiplied by $m^3$, the period $T_N$ of the rotational motion of the core
which is proportional to $I/N$ (where $I$ is the moment of inertia of 
the core) must be
multiplied by the same factor, so that $I$ scales as $m^4$. Similarly, one must
scale the ratio between the geometric size of the core and the De Broglie
wavelength, keeping constant the core's shape parameters (ratio anisotropic /
isotropic). This maintains constant the ratio of the classical periods
associated with $L$, due to the isotropic part of the short range potential, and
with $\Lambda = \vec{L} \cdot \unitvec{M}$, due to the torque produced by the
anisotropic part of this short range potential. This maintains the validity of
the approximation of constant $L$. Of course this needs proportionally larger
molecules, more difficult experimentally to deal with, but not completely out
of reach since recent experimental studies have found Rydberg states on
molecules as large as Benzene Argon complexes \cite{Neuhauser:PRL98-5089},
Diazacyclooctane (DABCO) and bis (benzene) chromium (BBC)
\cite{Even:JPC94-3472}.

Thus, for practical reasons, we shall continue to use the constant $L$
approximation in many formulae and all pictures presented in this article. Yet
we shall show, especially in the present and the next sections, that these
restrictions are not essential for the interpretation of MQDT as a QPM, and how
formulae can be generalized.

Finally notice that in ref.~\cite{Leyvraz:PD01-169}, we started with only seven
parameters. The three missing parameters were the three angles associated to
the three conserved momenta $J^2$, $J_z$ and $L^2$. This is legitimate in
classical mechanics since they can be computed afterwards by a mere quadrature.
For the comparison between classical and quantum mechanics, which is the main 
purpose of this work,
this is natural since these angles are totally undetermined
due to Heisenberg's uncertainty principle for well defined values of the
corresponding momenta.

\subsection{Parameterizations of the Poincar\'e surface of section}

We first discuss the case of constant $L$. For a proper parameterization of 
a two dimensional Poincar\'e surface of section, we need two conjugated
parameters, a momentum and an angle. We will not use the most obvious angles
$\theta_e$, $\varphi_e$ of the electron and their conjugated momenta, but
angles associated with angular momenta $L_{Z_Q}$ of the electron and $N$ of the
core because in the quantum MQDT description of the molecule,
these momenta enter naturally as (approximate) quantum numbers for the
limiting cases \cite{Fano:PRA70-253}. There are different choices, depending on
whether we work in the molecular or the laboratory frame.

\begin{enumerate}
    \item ``Quantum'' molecular frame. We use the projection $\Lambda =
L_{Z_Q}= \vec{L} \cdot \unitvec{M}$ of the electron angular momentum onto the
core axis, and the angle $\varphi_L^Q$, the azimuthal angle of the projection of
$\vec{L}$ into the horizontal plane $O{X_Q}{Y_Q}$, conjugated like the
$j_1,w_1$ pair in \cite[eq.~10-139]{Goldstein80}. This choice of parameters
implies that the Poincar\'e surface of section is a sphere on which every point
is defined by $\theta_L^Q = \arccos(\Lambda/L)$ and $\varphi_L^Q$, 
i.e. the direction of $\vec{L}$ in the molecular frame.
    \item Laboratory frame. We use the modulus $N$ of the angular momentum of
the core (not one of its projection onto laboratory frame). Intuitively, the
associated angle is the angle of rotation $\varphi_N$ of the molecular core
$\unitvec{M}$ around the direction of $\vec{N}$, and the only possible 
reference for this angle is given by the direction of the total angular
momentum $\vec{J}$. A proof of this intuition, namely a precise definition of
this angle, formulae relating the pairs \{~$L_{Z_Q}$, $\varphi_L^Q$~\} and
\{~$N$, $\varphi_N$~\}, and the proof that this transformation is canonical is
given in Appendix~\ref{sec:ClassQuantTransform}.
    \item ``Classical'' molecular frame. We use $L_{Z_C}=L_{Z_Q}$
(the two $Z$ axis coincide), and $\varphi_L^C$, the angle referred to $N$ in the
``horizontal'' plane (Fig.~\ref{fig:RefFrames}(c)). Notice that this pair is
\emph{not} conjugated, as discussed in
section~\ref{sec:SemiClassics-principle}.
The consequence is that the Poincar\'e map does not conserve the area of the
sphere with these coordinates, a small effect for the choice of parameters we
have used in this paper. We stick nevertheless to this choice for graphical
representations, because it gives more intuitive physical pictures as shown
below.
\end{enumerate}

Notice that with this choice of parameters, the direction of the plane of the
Kepler orbit is well determined in the molecular frame, since it is 
perpendicular to
$\vec{L}$, and its eccentricity is well determined by $L$ and the electron
energy $E_e$, but that the direction of its main axis in this plane, i.e. the
Laplace-Runge-Lenz vector $\vec{A}$ \cite{Goldstein80}, is not. As explained in
ref.~\cite{Lombardi:JCP88-3479}, the reason for this is that no component of
$\vec{A}$ , which defines this major axis, commutes with $L^2$ which is kept
fixed. Our counting of parameters implies that the direction of $\vec{A}$ 
is not an independent parameter. 

Introducing the approximation that $L$ is a constant has as a consequence that
the information about the main axis direction is hidden in the
angle associated to $L^2$. Then, the choice of action angle variables 
corresponds to that typically used in the pure bound Kepler
problem in celestial mechanics \cite{Goldstein80}. The actions are $j_1=L_z$,
$j_2=L$ and $j_3=\sqrt{-1/2E_e}$ (i.e. ``principal quantum number''). The
associated angles are the ``longitude of the ascending node'' $\Omega$
($=\varphi_L+\pi/2$ with our notations), the ``argument of the perihelion''
$\omega$, which is the direction of the main axis in the orbital plane (or the
direction of $\vec{A}$ in this plane) we are looking for, and the ``mean
anomaly'' which is proportional to time. Thus, the angle of the main axis of the
ellipse, which is constant for pure Kepler motion, i.e. for Rydberg molecules
when the electron is far away from the core, but varies during the
collision, is part of the map only, if the approximation $L^2$ constant is 
omitted.

\subsection{Description of the classical Poincar\'e map}

We now briefly describe the PM, both in the laboratory and in the molecular
frame. The dynamics is composed of two independent consecutive steps:

\begin{enumerate}
\renewcommand{\theenumi}{{\roman{enumi}}}
  \item collision step: when $r \ll r_0$ the electron feels the short range
anisotropic part of the potential. We suppose that this step is very short, so
that the core axis remains fixed (impulse approximation). It is thus
best described in the molecular frame. Due to the cylindrical symmetry of this
potential the projection $\Lambda=\vec{L} \cdot \unitvec{M}$ of $\vec{L}$ onto
the core axis $\unitvec{M}$ is conserved. With the auxiliary approximation that
$L^2$ is constant, this amounts to the collision being described by a
precession of $\vec{L}$ around $\unitvec{M}$ by an angle $\delta\varphi_L$.
Furthermore the invariance of the potential under reflection with respect to
any plane containing $\unitvec{M}$ implies that $\delta\varphi_L$ is odd with
respect to 
$\theta_L\to\pi-\theta_L$. The simplest choice is $\delta\varphi_L =
K\cos\theta_L$, which defines the strength $K$ of the interaction. Notice 
that this $\theta_L$ dependant rotation about the $OZ$ molecular axis is a
twist, not a global rotation of the sphere, implying that the dynamics generated
by the map can be chaotic \cite{Berry:CBDS83-171,LesHouches91}. 
Since $\vec{L}$ has changed, at least in direction if we make the
approximation $L^2$ constant, due to the  
the conservation of total angular momentum $\vec{J} = \vec{L} +
\vec{N}$, $\vec{N}$ changes both in direction and in
magnitude. This has far reaching consequences. The change of the magnitude
entails a change of the rotational energy of the core $E_N = B N^2$ ($2 B$ is
the reciprocal of the core moment of inertia $I$ according to usual
spectroscopic notation \cite{HerzbergI50}). Consequently, due to the 
conservation of the total energy $E$ the electron's energy
$E_e = E - E_N$ also changes. If $E_e$
becomes positive the molecule ionizes, if it remains negative the resulting
changes of $T_e$ and $T_N$ will change the parameters of the next Coulomb step.
While the ``quantum'' molecular reference frame is fixed during
the collision, in the ``classical'' molecular reference frame, 
its $OX_C$ axis, which is chosen along $\vec{N}$, will change 
(Fig.~\ref{fig:RefFrames}(c)). 
This produces an additional phaseshift 
$\delta\varphi_L^\prime$, which was called frame recoil
in ref.~\cite{Lombardi:JCP88-3479} where explicit formulae are given. It can
also be computed from eq.~(\ref{eq:deltaphiQversC}) of
Appendix~\ref{sec:ClassQuantTransform}, which gives the angular position of
$\vec{N}$ in the fixed ``quantum'' molecular frame, by taking the difference of
positions before and after the $\delta\varphi_L$ kick.
  \item Coulomb (or free rotation) step: when $r \gg r_0$ the electron only
feels the spherically symmetric $-1/r$ part of the potential. It moves on a
Kepler orbit. Its angular momentum $\vec{L}$ and its Laplace-Runge-Lenz vector
$\vec{A}$ are constant in the laboratory $Oxyz$ reference frame. Meanwhile the
molecular core rotates at constant speed around its angular momentum $\vec{N}$
which is perpendicular to its axis $\unitvec{M}$. When seen in the
``classical'' molecular frame, with $OZ$ along $\unitvec{M}$ and $OX_C$ along
$\vec{N}$ (Fig.~\ref{fig:RefFrames}(c)), $\vec{L}$ (and $\vec{A}$) seem to
rotate the opposite way (clockwise) around $OX_C$. The total angle of rotation
is $\delta\beta = -2\pi T_e/T_N$, {\it i.e.} the ratio of the periods of the
electron orbit and of the core rotation. We have $T_e = 2 \pi (-2 E_e)^{3/2}$
a.u. and $T_N = 2 \pi / 2 B N$ a.u..
\end{enumerate}

The crucial point of this model is the exchange of energy between the electron
and the core during the collision step. If we neglected it the model
would coincide with with a 
``kicked spin'' model \cite{Nakamura:PRL86-5,Haake:ZPB87-381}, which cannot
lead to ionization.

For the free rotation step we prefer to visualize the results in  
the ``classical'' molecular frame. Indeed in the ``quantum''
molecular frame the free rotation is about a direction of $\vec{N}$ in the
molecular $O{X_Q}{Y_Q}$ plane (see Fig.~\ref{fig:RefFrames}(b)), which
varies from step to step. We will visualize
ionization on the Poincar\'e sphere in the ``classical'' frame by noting that
positive electron energies $E_e$ correspond to points where the total energy
$E$ is greater than the rotational energy $E_N$. Using $\vec{J}^2 =
\left( \vec{L} + \vec{N} \right)^2$, this leads to the inequality
\begin{equation}\label{eq:IonisationCap}
E_N=BN^2=B\left(-L\cos\alpha+\sqrt{J^2-L^2\sin^2\alpha}\right)^2<E
\end{equation}
This condition only depends on the angle $\alpha$ between $\vec{L}$ and
$\vec{N}$ and accordingly the
positive electron energy region corresponds to a cap around the positive
$\vec{N}$ axis, which is fixed in this frame (Fig.~\ref{fig:RefFrames}(c)). 
If we relax the condition $L^2$
constant, we have a four dimensional PM and formulae for the collision and the
recoil are modified, but eq.~(\ref{eq:IonisationCap}) remains valid.

Practically, there are two possibilities to compute the classical PM. 
In all our
previous works, we considered it entirely in the ``classical'' molecular
reference frame, by computing in turn free rotation, collision
$\delta\varphi_L$, frame recoil, new values of $N$, $E_N$, $T_N$, $E_e$, $T_e$
and plotting the results in the ``classical'' reference frame by using 
equations given in Appendix~\ref{sec:ClassQuantTransform}. 
In the present work we alternate between laboratory frame and ``quantum''
reference frame, in a way summarized in the first three columns of
table~\ref{table:PM} (the last two will be filled below in
sections~\ref{sec:quantum} and~\ref{sec:SemiClassics}). 
We choose this more complex, and thus slower, procedure, because
as outlined in section~\ref{sec:SemiClassics}, the coordinates 
in the
``classical'' reference frame are not canonically conjugate. This is of no harm
for purely classical computations, but forbids the semi-classical analysis. 
Of course we have checked that both representations give the same results.

\begin{table}[!htb]
\caption{\label{table:PM} Poincar\'e Map: summary. Molecular coordinates are in
the ``quantum'' reference frame.}
\begin{tabular}{c|c|c|c|l}
Coords. & Action & Classical & MQDT & Semi Classical\\
\hline&&&&\\
$L_Z^\prime \varphi_L^\prime$&&&&\\
&M.$\to$L.
&eqs.(\ref{eq:N_Q},\ref{eq:phiN})
& $\matr{U}_{N^\prime\Lambda^\prime }^{(LJ)}$
& $F_3(L_Z^\prime,\varphi_{N^\prime})$\quad eq.(\ref{eq:F2LQ})
\\
$N^\prime \varphi_{N^\prime}$&&&&\\
&Free r.
&$\delta\varphi_N =
\left\{
\begin{array}{l}
2 \pi \frac{T_e^\prime}{T_N^\prime}\\
-2 \pi \frac{\partial \nu_{N^\prime}}{\partial N^\prime}
\end{array}
\right.$
&$\rme^{2 \rmi\pi\nu_{N^\prime}}$
&$F_3(N^\prime,\varphi_{N})= N^\prime \varphi_{N} + 2\pi \nu_{N^\prime}$\\
$N^\prime \varphi_N$&&&&\\
&L.$\to$M.
&eqs.(\ref{eq:LzQ},\ref{eq:phiLQ})
&$\matr{U}_{\ \Lambda N^\prime}^{\dag(LJ)}$
&$F_2(\varphi_N,L_Z)$\quad eq.(\ref{eq:F2LQ})
\\
$L_Z \varphi_L^{\prime\prime}$&&&&\\
&Coll.
&$\delta\varphi_L =
\left\{
\begin{array}{l}
2K\cos\theta_L\\
-2\pi\frac{\partial\mu_{\Lambda}}{\partial\Lambda}
\end{array}
\right.$
&$\rme^{2 \rmi\pi\mu_{\Lambda}}$
&$F_3(L_Z,\varphi_L)=L_Z \varphi_L + 2 \pi \mu_{L_Z}$\\
$L_Z \varphi_L$&&&
\end{tabular}
\end{table}

Finally we will describe how to construct the combined Poincar\'e--Jung 
scattering map.
The JSM is defined such that the electron which escapes to infinity is fed back
by doing a step backwards in time without collision with the core (i.e. a pure
Coulomb step with reverted velocity), and then re launching the electron
towards the molecule along the resulting asymptotic trajectory for the next
collision (Fig.~\ref{fig:Principle}(b)). Notice that $\vec{L}$ and $\vec{A}$
are constant in the \emph{laboratory} frame during such a pure Coulomb step. For
the $L^2$ constant case this amounts to starting from the same point on the
Poincar\'e sphere in the \emph{molecular} frame if $E_e>0$ after a collision,
i.e. to skip the three first steps in Table~\ref{table:PM}. Indeed the rotation
of the core is neglected during the impulse collision step, and during journeys
to and from infinity of the electron these rotations cancel since they are made
for $r>r_0$, where the short range potential is negligible. In the general case
the ``argument of the perihelion'' $\omega$ is also conserved. The proper JSM
is a map from ionized to ionized states, the PM a map from bound to bound
states, but there is obviously no technical problem to iterate a combined map
from unbound to bound states and vice versa by selecting the JSM recipe to feed
back ionizing trajectories. This seemingly strange feed back was suggested and
discussed in ref.~\cite{Jung:JPA86-1345}. We shall show that it is indeed a
good representation of the results in quantum mechanics under semi-classical
conditions.

\section{\label{sec:quantum} MQDT as a generalized quantum Poincar\'e map}

\subsection{\label{sec:Quantum-usual} The usual MQDT theory}

MQDT for molecules is a well established, time honored theory
\cite{Fano:PRA70-253,Fano:JOSA75-979,Greene:AAMP85-51,Jungen96}.
We will outline here only what is
needed for our purposes. The theory closely parallels the preceding classical
ideas, which originated from the classical limit of MQDT, but which were
implicitly present in Fano's seminal paper \cite{Fano:PRA70-253}, albeit with a
fully quantum way of writing down the theory. Wave functions have different
forms for $r\lesssim r_0$ and $r\gtrsim r_0$ and they are matched at $r = r_0$.
Going from $r=0$ to $r=\infty$ we have
\begin{enumerate}
\renewcommand{\theenumi}{{\roman{enumi}}}
    \item Collision step: $r\lesssim r_0$.
The electron's motion is rigidly coupled to the molecular axis $\unitvec{M}$.
Its speed exceeds by far the speed of the motions of the nuclei of the
molecular ion and the Born-Oppenheimer factorization of the molecular wave
function applies (Hund's case (b) coupling type according to molecular
spectroscopy nomenclature\cite{HerzbergI50}). Due to the cylindrical symmetry
of the potential the projection $\Lambda = \vec{L} \cdot \unitvec{M}$ of the
electron's angular momentum onto the molecular axis $\unitvec{M}$ is a constant
of the motion, whereas the squared angular momentum $\vec{N}^2$ of the
molecular ion rotation does not have a definite value. We may or may not have a
definite value of $L^2$ depending on the use of the auxiliary approximation
$L^2=\mathrm{const}$.
The angular part of the wave function is thus an eigenfunction of $\Lambda$ (in
addition to $J$ and $M_J$) denoted by $X^{(J,\Lambda )}_{M_J}
(\theta_e^\prime,\varphi_e^\prime,\unitvec{M})$ \cite{Fano:PRA70-253}, where
($\theta_e^\prime,\varphi_e^\prime$) are the angular coordinates of the Rydberg
electron in the ``quantum'' \emph{molecular} reference frame $O{X_Q}{Y_Q}{Z_Q}$
defined in Fig.~\ref{fig:RefFrames}(b) and Appendix~\ref{sec:QCFrames}. For
$r\sim r_0$, i.e. outside of the molecular core, where the potential is
approximately pure Coulomb, the radial part of the Rydberg electron wave
function is a linear combination of two independent Coulomb functions. We
choose the one regular at $r=0$, $s(E_e,r)$, as well as one irregular at $r=0$,
$c(E_e,r)$, suitably selected for its asymptotic properties when $r\to\infty$
\cite{Seaton:RPG83-167}.
The normalized total wave function reads \cite{Fano:PRA70-253} as
\begin{equation}\label{eq:CollisionWF}
     \Psi_\Lambda (\theta_e^\prime ,\varphi_e^\prime, \unitvec{M} ;r) =
X^{(L,J,\Lambda)}_{M_J} (\theta_e^\prime ,\varphi_e^\prime
 ,\unitvec{M}) \ \left(
s(E_e ,r) \cos (\pi\mu_\Lambda ) + c(E_e ,r) \sin (\pi\mu_\Lambda ) \right).
\end{equation}
This implicitely defines the ``quantum defects'' $\mu_\Lambda$.
Notice that $E_e$ depends
on $\Lambda$. If the subsidiary approximation $L^2$ constant is not used, the
$X$'s are not eigenstates of $L^2$, but can be developed in such a basis,
implying a summation over $L$ and an extra label $L$ on the $\mu$'s and the
$E_e$'s. The quantum defects can in principle be computed by ab initio
electronic molecular calculations between $r=0$ and $r=r_0$, but we will take
them here as semi empirical parameters, as is frequently done. We will assume
that they are independent of energy, in a restricted range above or below zero
electron energy. This is justified by the fact that in the inner region the
change of the total energy is negligible as compared to the electron kinetic 
energy in the
Coulomb well. The wave function in the collision region should not depend
significantly on such small variations of the total energy. For large distances
there is a difference, leading to bound or unbound states. We will neglect
refinements used in actual molecular calculations to compute levels with
somewhat lower principal quantum numbers, for which one allows quantum defects
to vary slowly with $E_e$.
 \item Coulomb step: $r\gtrsim r_0$.
The electron is coupled to the isotropic part of the Coulomb potential only,
and cannot exchange angular momentum with the core. The core angular momentum
$N$ and the electron angular momentum $L$ are thus separately conserved, as is
$\vec{J}=\vec{L}+\vec{N}$. The angular part of the wave function
$\Phi^{(L,J,N)}_{M_J}(\theta_e,\varphi_e,\unitvec{M})$ corresponds to the
coupling by a Clebsch-Gordan coefficient of an electron wave function labeled
by $L$ and of a molecular core wave function labeled by $N$ (Hund's case (d)
coupling type according to molecular spectroscopy
nomenclature\cite{HerzbergI50}). Here ($\theta_e,\varphi_e$) are the electron
coordinates in the \emph{laboratory} reference frame $Oxyz$. The radial part of 
the wave
function of the Rydberg electron is again a linear combination of regular and
irregular Coulomb wave functions. The total wave function reads
\begin{equation}\label{eq:CoulombWF}
  \Psi_N(\theta_e ,\varphi_e, \unitvec{M};r)=\Phi^{(L,J,N)}_{M_J}(\theta_e,
\varphi_e , \unitvec{M})\left( s(E_e ,r)\,
c_N+c(E_e ,r)\,d_N\right),
\end{equation}
where $E_e=E-B N (N+1)$ can be positive or negative and $c_N$ and $d_N$ are to
be determined from asymptotic conditions. Here again an additional summation 
over $L$ is
necessary when omitting the auxiliary approximation that $L^2$ be conserved.
\end{enumerate}

For $r\sim r_0$ we have one and the same wave function developed into two
different bases (\ref{eq:CollisionWF}) and (\ref{eq:CoulombWF}). The electron
angular parts correspond to a change from the molecular frame to the laboratory
frame. The orthogonal matrix $\matr{U}$ which performs this change
\cite{Fano:PRA70-253},
\begin{equation}\label{eq:Udef}
    X^{(L,J,\Lambda)}_{M_J}=\sum_N\Phi^{(L,J,N)}_{M_J}
\matr{U}_{N\Lambda}^{(LJ)},
\end{equation}
is proportional to a Clebsch-Gordan coefficient:
\begin{equation}\label{eq:Uclebsch}
    \matr{U}_{N\Lambda }^{(LJ)}=\langle L-\Lambda J\Lambda\vert L J N 0\rangle
(-1)^{J-N+\Lambda }(2-
\delta_{\Lambda 0})^{\frac12},
\end{equation}
if we use the definition of rotational wave functions given in
Appendix~\ref{sec:QCFrames}.

We transform a general wave function $\Psi = \sum_\Lambda \Psi_\Lambda
A_\Lambda$ from the molecular to the laboratory frame by combining
(\ref{eq:CollisionWF}) and (\ref{eq:Udef}). The matching of the wave functions
(\ref{eq:CollisionWF}) and (\ref{eq:CoulombWF}) yields
\begin{eqnarray}\label{eq:TotalWF}
  \Psi (\theta_e ,\varphi_e ,\unitvec{M};r) &=& \sum_{N} \Phi^{(L,J,N)}_{M_J}
(\theta_e , \varphi_e ,\unitvec{M}) \Psi_N(r) \nonumber\\
  \Psi_N(r)&=&\sum_\Lambda \matr{U}_{N\Lambda} \left( s(E_e ,r) \cos
(\pi\mu_\Lambda ) + c(E_e,r) \sin (\pi\mu_\Lambda )
\right) A_\Lambda
\end{eqnarray}
taking into account that the implicit $E_e$
dependance of the radial wave functions on $\Lambda$ in
eq.~(\ref{eq:CollisionWF}) and on $N$ in eq.~(\ref{eq:CoulombWF}) is negligible
\emph{near $r\sim r_0$}. This would be true neither for $r\ll r_0$ nor for
$r\gg r_0$.

The coefficients $A_\Lambda$ in eq.~(\ref{eq:TotalWF}) 
are obtained from the asymptotic
properties of the wave function for $r\to\infty$. These properties in
turn depend on whether the electronic energy in each asymptotic channel labeled
by $N$, namely $E_e=E-B N (N+1)$, is positive (open channels) or negative
(closed channels). In any case we define a principal quantum number $\nu_N$
(non integer) by $E_e=-1/2\nu_N^2$. For closed channels $\nu_N$ is real, for
open channels it is imaginary.

In a \emph{closed} channel, for the motion to be bounded, the exponentially
increasing part of the Coulomb wave function must be zero, which implies
that the radial part $\Psi_N(r)$ of the wave function in
eq.~(\ref{eq:TotalWF}) must be proportional to a suitable combination of
regular and irregular Coulomb wave functions which is regular when
$r\to\infty$, i.e. of the form: \cite{Seaton:RPG83-167,Hartree:PCPS28-426}:
\begin{equation}\label{eq:AsyClosed}
    \Psi_N(r)\propto -\cos(\pi\nu_N)\ s(E_e,r) + \sin(\pi\nu_N)\ c(E_e,r),
\end{equation}
Combining equations (\ref{eq:TotalWF}) and (\ref{eq:AsyClosed}) yields
\begin{equation}\label{eq:EqClosed}
  \sum_\Lambda \matr{U}_{N\Lambda} \sin(\pi(\nu_N+\mu_\Lambda))A_\Lambda = 0
\end{equation}

In an \emph{open} channel the Coulomb wave function may be decomposed into
incoming and outgoing Coulomb waves $\varphi^\pm$ which have the asymptotic
behavior \cite{Seaton:RPG83-167}:
\begin{align}\label{eq:DefPhi}
  & \varphi^\pm(r) = \frac{1}{\sqrt{2}}\left( c \pm is\right)\sim (\pi
k)^{-\frac12}\exp\left( \pm \rmi\zeta
\right) &\mathrm{for}\ r\to\infty \\
  & \zeta = k r -\frac12 L\pi +\frac1k \ln (2kr )+\mathrm{arg}(\Gamma
(L+1-\rmi k)), &k=\frac{\rmi}{\nu_N }. \nonumber
\end{align}
Scattering is defined with respect to pure Coulomb scattering, that is, the
eigenphases $\tau$ are measured relative to the Coulomb phases $\zeta$. The
wave function will have the correct asymptotic behavior in the limit
$r\to\infty$ for the asymptotic channel $N$ if its radial part is of the form:
\begin{equation}\label{eq:AsyOpen}
    \Psi_N(r) \propto \sum_{N^\prime} C_{N^\prime}
\left(\varphi^-(\nu_{N};r)\delta_{N N^\prime} - \varphi^+
(\nu_{N};r) \matr{S}_{N N^\prime}\right),
\end{equation}
where the sum runs \emph{only} over all \emph{open} channels, and the
$C_{N^\prime}$ are constants to be related to the $A_\Lambda$. This defines the
scattering matrix $\matr{S}$. It is interesting to consider the matrix
$\matr{T}$ which diagonalizes $\matr{S}$, i.e.
\begin{equation}\label{eq:S=TtauTt}
   \matr{S}_{N N^\prime} = \sum_\ell \matr{T}_{N \ell} \exp(2\rmi\pi\tau_\ell)
\transpose{\matr{T}}_{\ell N^\prime}
\end{equation}

Using eq.~(\ref{eq:DefPhi}), we obtain:
\begin{eqnarray}\label{eq:PsiNl}
  \Psi_N(r) & \propto & \sum_l \Psi_{N\ell}(r) \nonumber \\
  \Psi_{N\ell}(r) &=& \left( \frac{\sqrt{2}}{\rmi} \sum_{N^\prime} C_{N^\prime}
\transpose{\matr{T}}_{\ell N^\prime}
\exp(-\rmi\pi\tau_\ell) \right) \times \nonumber\\
 & & \matr{T}_{N\ell} \left( s(E_e ,r) \cos(\pi\tau_\ell) + c(E_e,r)
\sin(\pi\tau_\ell ) \right)
\end{eqnarray}

\emph{Near $r\sim r_0$} eq.~(\ref{eq:TotalWF}) should have this form and we
obtain, for each value of $\ell$, two equations:

\begin{eqnarray}
0 & = & \sum_\Lambda \matr{U}_{N\Lambda} \sin(\pi(\mu_\Lambda-\tau_\ell))
A_\Lambda(\ell) \label{eq:EqOpen}\\
\matr{T}_{N \ell} & \propto & \sum_\Lambda \matr{U}_{N\Lambda}
\cos(\pi(\mu_\Lambda-\tau_\ell)) A_\Lambda(\ell)
\label{eq:EqT}
\end{eqnarray}

Equations (\ref{eq:EqClosed}) and (\ref{eq:EqOpen}) can be combined to
\begin{equation}\label{eq:SA=0}
    \matS\vert A(\ell)\rangle = 0
\end{equation}
with
\begin{equation}\label{eq:defS}
  \matS=
  \left(
  \begin{array}{lcl}
     \matr{U}_{N\Lambda} \sin(\pi(\mu_\Lambda-\tau_\ell)) &;& N\,\mathrm{open}
\\
     \matr{U}_{N\Lambda} \sin(\pi(\mu_\Lambda+\nu_N)) &;& N\,\mathrm{closed} \\
  \end{array}
  \right)
\end{equation}
where $\mathcal{S}$ stands for sine, not to be confused with the $S$ matrix in
eq.~(\ref{eq:S=TtauTt}).

\subsection{\label{sec:Quantum-QPM} MQDT as a Quantum Poincar\'e map}

We now seek an alternative along the lines sketched in
ref.~\cite{Leyvraz:PLA00-309} for the case of all channels closed. The matrix
$\matS$ defined in eq.~(\ref{eq:defS}) is the imaginary part of a \emph{complex
unitary} matrix
\begin{equation}\label{eq:defE}
\matE = \matC + \rmi \matS =
\left(
\begin{array}{lcl}
  \matr{U}_{N\Lambda}\exp(\rmi\pi(\mu_\Lambda-\tau_\ell))&;& N\,\mathrm{open}
\\
  \matr{U}_{N\Lambda} \exp(\rmi\pi(\mu_\Lambda+\nu_N))&;& N\,\mathrm{closed} \\
\end{array}
\right)
\end{equation}

Splitting real and imaginary parts in the unitarity relation
\begin{equation}\label{eq:Eunit}
    \matE^\dag \matE = (\transpose{\matC}-\rmi\transpose{\matS}) (\matC + \rmi
\matS) = \Id
\end{equation}
gives
\begin{equation}
 \transpose{\matC}\matC+\transpose{\matS}\matS=\Id \label{eq:CtC+StS=1}
\end{equation}
\begin{equation}
 \transpose{\matC}\matS=\transpose{\matS}\matC \label{eq:CtS=StC}
\end{equation}
Equation (\ref{eq:CtS=StC}) implies that if $\vert A(\ell)\rangle$ is an
eigenvector of $\matS$ with eigenvalue zero, then $\vert B(\ell)\rangle =
\matC\vert A(\ell)\rangle$ is an eigenvector of $\transpose\matS$ with
eigenvalue zero, i.e. a solution of the transpose of eq.~(\ref{eq:SA=0}):
\begin{equation}\label{eq:StB=0}
    \transpose{\matS} \vert B(\ell)\rangle = 0
\end{equation}
Writing down the components of $\vert B(\ell)\rangle$,
\begin{equation}\label{eq:B=CA}
    B_N(\ell) = \sum_\Lambda \matC_{N \Lambda} A_\Lambda(\ell)
\end{equation}
we see that the matrix $\matC$ relates the eigenfunction for channel $\ell$
in the molecular reference frame labeled by $\Lambda$ to that in the laboratory
reference frame labeled by $N$. Note that eq.~(\ref{eq:B=CA}) is the same
as eq.~(\ref{eq:EqT}), except that the matrix $\matr{T}$ is defined for
\emph{open} channels only, while $B$ is defined for \emph{all} channels.

Next, consider the \emph{symmetric complex unitary} matrix:
\begin{equation}\label{eq:EtE}
  \begin{array}{rcl}
   \transpose{\matE} \matE &=& (\transpose{\matC}+\rmi\transpose{\matS}) (\matC
+ \rmi \matS) \\
    &=&(\Id -2 \transpose{\matS} \matS) + 2 \rmi \transpose{\matC} \matS,
  \end{array}
\end{equation}
using eqs.~(\ref{eq:CtC+StS=1},\ref{eq:CtS=StC}). This implies that if
$\matS\vert A_\Lambda\rangle=0$, then
\begin{equation}\label{eq:EtE=1}
    \transpose{\matE} \matE \vert A_\Lambda\rangle = \Id \vert
A_\Lambda\rangle,
\end{equation}
i.e. the solutions of eq.~(\ref{eq:SA=0}) are eigenvectors of
$\transpose{\matE}\matE$ with eigenvalue $1$.

Similarly the $\vert B_N\rangle$ are eigenvectors of $\matE\transpose{\matE}$
with eigenvalue $1$
\begin{equation}\label{eq:EEt=1}
    \matE \transpose{\matE} \vert B_N\rangle = \Id \vert B_N\rangle
\end{equation}

Eqs. (\ref{eq:EtE=1},\ref{eq:EEt=1}) imply the determinant equation
\begin{equation}\label{eq:det=1}
    \det(1-\transpose{\matE}\matE)=0.
\end{equation}
Comparing this result with the equation for the $T(E_n)$ matrix defined by
Bogomolny \cite{Bogomolny:CAMP90-67,Bogomolny:N92-805,Bogomolny:C92-5} was
essential for the interpretation of MQDT as a QPM in
ref.~\cite{Leyvraz:PLA00-309}. We will examine this relation more closely in
the next sections.

For scattering systems we have to go a little further, because we have to
search for the phase shifts at any energy rather than for eigenvalues and
eigenfunctions. For this situation we shall derive another useful form of
these equations.
Remember that in this case vectors and matrices in
eqs.~(\ref{eq:EtE=1},\ref{eq:EEt=1}) depend on the channel number $\ell$ and 
that
the $B_N(\ell)$ are identical to the eigenvectors $T_{N\ell}$ of the $S$
matrix. We are looking for a set of equations written for open channels only, 
whereas
eq.~(\ref{eq:EtE=1}) is written for all channels, open and closed. For this
purpose we write eq.~(\ref{eq:EEt=1}) in bloc diagonal form, splitting the $N$
channels into open channels indexed by $o$ and closed channels indexed by
$c$. This is possible for elements indexed by the asymptotic quantum number
$N$, e.g. $\matE\transpose{\matE}$, not for those indexed by the Born
Oppenheimer quantum number $\Lambda$, e.g. $\transpose{\matE}\matE$. We obtain
\begin{eqnarray}
  (\matE\transpose{\matE})_{oo} B_o + (\matE\transpose{\matE})_{oc} B_c &=& B_o
\label{eq:Splitoc1}\\
  (\matE\transpose{\matE})_{co} B_o + (\matE\transpose{\matE})_{cc} B_c &=& B_c
\label{eq:Splitoc2}
\end{eqnarray}
Eliminating $B_c$ gives
\begin{equation}\label{eq:SenEEt}
    \left( (\matE\transpose{\matE})_{oo} + (\matE\transpose{\matE})_{oc} \left(
\Id_{cc} - (\matE\transpose{\matE})_{cc}
\right)^{-1} (\matE\transpose{\matE})_{co} \right) B_o = B_o
\end{equation}

Notice then that we can rewrite eq.~(\ref{eq:defE}) as
\begin{equation}\label{eq:E=nuUmu}
    \matE = \exp(\rmi\pi\matr{\nu})\ \matr{U} \exp(\rmi\pi\matr{\mu})
\end{equation}
where $\matr{\nu}$ is the diagonal matrix in the laboratory reference frame
with diagonal elements
\begin{equation}\label{eq:defMatNu}
    \matr{\nu}_{NN} =
    \left\{
    \begin{array}{lcl}
       -\tau_\ell &;& N\,\textrm{open}\\
       +\nu_N  &;& N\,\textrm{closed}
    \end{array}
    \right.
\end{equation}
and $\matr{\mu}$ is the diagonal matrix in the molecular reference frame with
diagonal elements $\mu_\Lambda$.

Then, defining
\begin{equation}\label{eq:defCalU}
    \matU = \matr{U} \exp(2\rmi\pi\matr{\mu}) \transpose{\matr{U}}
\end{equation}
Equation (\ref{eq:SenEEt}) can be rewritten as
\begin{equation}\label{eq:SenU}
    \left( \matU_{oo} + \matU_{oc} \left( \exp(-2\rmi\pi\matr{\nu})_{cc} -
\matU_{cc} \right)^{-1} \matU_{co} \right)
B_o(\ell) = \exp(2\rmi\pi\tau_\ell) B_o(\ell)
\end{equation}
The matrix within the left large parenthesis no longer depends on $\tau_\ell$;
it is defined in open channels, and for each $\ell$ has the same eigenvalue
(eq.~(\ref{eq:S=TtauTt})) and the same eigenvector (eqs.~(\ref{eq:EqT}) and
(\ref{eq:B=CA})) as the $S$ matrix. We thus obtain the expression
\begin{equation}\label{eq:Soo}
    \matr{S} = \matU_{oo} + \matU_{oc} \left( \exp(-2\rmi\pi\matr{\nu})_{cc} -
\matU_{cc} \right)^{-1}
\matU_{co}
\end{equation}
for the $S$ matrix.

\subsection{\label{sec:Quantum-Computation} Computational advantages}

The usual MQDT (sec.~\ref{sec:Quantum-usual}) uses the equation $\det\matS=0$
(eq.~\ref{eq:defS}) which can be rewritten as a polynomial in
$\tan(\pi\tau_\ell)$ whose degree equals the number of open channels. Each root
gives a well defined $\tau$ because, according to eq.~(\ref{eq:S=TtauTt}), the
eigenphases of the $\matr{S}$ matrix can take values between 0 and 1 (or -0.5
and 0.5) only. Depending on the choice of the total energy there are the
possibilities that all channels are closed (there is then only
one value of the set $\{A_\Lambda\}$) or open. In the last case one obtains
as obvious solutions the values $\tau_\ell=\mu_\Lambda$ (see
(eq.~\ref{eq:defS}). 
This shows that the $\mu_\Lambda$ are exactly the usual collision phase shifts
when the energy is high enough for all channels to be open. In the general case
equation (~\ref{eq:defS}) is not solved by a diagonalization algorithm. The 
usual way to
find a solution is
\begin{itemize}
    \item All channels closed: No $\tau_\ell$.
All matrix elements depend on energy $E$ through the $\nu_N$. One varies $E$
and locates the zeros of the determinant by a root searching algorithm.
    \item Some channels closed, some open.
One fixes a given value of $E$ and searches the values of $\tau_\ell$ for which
the determinant vanishes. The search may be done by computing the determinant
and varying $\tau_\ell$ like in the preceding case, or by writing the
polynomial in $\tan\tau_\ell$ and computing its roots. In any case it is a root
searching algorithm.
\end{itemize}

As explained in ref.~\cite{Leyvraz:PLA00-309}, where illuminating figures are
given, the practical problem with such a standard approach is that when zeros
are very close the determinant crosses twice or more often the zero line at
extremely close values, and nearly tangent to the zero line.  This makes it very
difficult to avoid the missing of some pairs of zeroes. Moreover, in such a 
situation it is very difficult to recover the wavefunctions individually as 
they switch
between nearly orthogonal states for a very small change of 
$E$ or $\tau_\ell$. This situation is quite common in the case of a nearly 
integrable phase
space, where there are systematic near degeneracies associated to the lack of
level repulsion due to approximately conserved quantum numbers. The problem
of missing levels 
becomes particularly severe for the high values of angular momenta we use to
study the semi classical limit.

In our QPM formulation (sec.~\ref{sec:Quantum-QPM}) for bound systems the
relevant practical aspect \cite{Leyvraz:PLA00-309} was that we could
diagonalize the unitary matrix $\transpose{\matE}\matE$ by standard techniques.
Then the equation is solved by a search of zeros of the eigenphases, which move
monotonously as a function of energy. Though it still requires a root searching
algorithm, it is much more efficient for near degenerate levels than the
standard MQDT equation $\det\matS=0$ and may be accelerated by linear 
interpolation. In the case of approximate double
degeneracies, the diagonalization algorithm always reliably provides a good 
pair of orthogonal wavefunctions.

Similarly in the ionized case the computation of the $S$ matrix with
eqs.~(\ref{eq:SenU},\ref{eq:Soo}) is reduced to a standard diagonalization of a
\emph {complex symmetric} matrix (for any \emph{real energy}), with an
auxiliary inversion of a matrix in the closed channel space. The components of 
the
resulting wave function on the open channels ($B_o(\ell)\equiv T_{o\ell}$) are
obtained by the diagonalization and the components on the closed channels are
then deduced with eq.~(\ref{eq:Splitoc2}). An other way to use this equation is
to look for the zeros of the matrix to invert in the \emph{complex energy}
plane, which gives poles of the $\matr{S}$ matrix.
We will use both techniques in the next sections.

\section{\label{sec:Results}Comparison between classical and quantum evolution
on the surface of section}

\subsection{Principles}

One of the basic tools we use to study properties of our molecular system is to
compare the PM with Husimi \cite{Husimi:PPSJ40-264} or Wigner
\cite{Wigner:PR32-749} plots of eigenfunctions, as well as the time evolution
of such quantum distributions in phase space with that of a 
swarm of classical trajectories. In such plots we represent the angular part of
the wave function, ignoring its radial part. In other words we represent both
the classical and the quantum evolution on the surface of section, thus
relating the PM and the QPM.

First we recall that we can perform the comparison of such objects in two 
different frames, the
molecular or Born-Oppenheimer frame and the laboratory frame. The molecular
frame wave functions have coefficients labeled by $\Lambda$, the laboratory
frame wave functions have coefficients labeled by $N$. The coefficients of a
given wave function in those two reference frames are related by
eq.~(\ref{eq:B=CA}). According to eq.~(\ref{eq:defE}) the matrix $\matC$ which
performs the transformation is composed of the matrix $\matr{U}$, which
transforms between the molecular and the laboratory frames and of phase shifts
in the two frames. We will explain the meaning of these phase shifts at the end
of this section. Note that, according to
\cite[eq.~(A2.1)]{Edmonds74}, the asymptotic value of $\matr{U}$ for large
angular momenta reduces to a rotation of $\pi/2$ around $OY$. $\matr{U}$ thus
basically performs a rotation between two orthogonal polar axis $\unitvec{M}$ 
and $\vec{N}$.
We shall focus on four key points in an electron orbit: perigee,
apogee, and the two crossings of the surface of the sphere of radius $r_0$, the border between the free
motion and the collision region. For hyperbolic motion the apogee does not
exist. Wave functions for all these four points are naturally contained in the 
MQDT. Perigee coefficients are the $A_\Lambda$ introduced in
eq.~(\ref{eq:TotalWF}). They are naturally expressed in the laboratory frame. In
the case where all channels are bound, apogee coefficients are the components
$B_N$ of the eigenvector $\vert B\rangle$ of $\transpose{\matS}$ introduced in
eqs.~(\ref{eq:StB=0},\ref{eq:B=CA}). They are naturally expressed in the
laboratory frame, but may be expressed in the molecular frame by multiplication
with the matrix $\transpose{\matr{U}}$. The coefficients at the crossing of the
sphere of radius  
$r_0$ before and after collision are the angular parts of incoming
$\varphi^-(r)$ and outgoing $\varphi^+(r)$ in eq.~(\ref{eq:AsyOpen}). They are
defined for unbound motion, but the basis for the validity of MQDT is that at
$r_0$ the inner part of the wave function is approximately the same
for bound and unbound
motion. So even for bound motion one can speak of ``infinitely far from the
core'' motion at $r_0$. This was a key point for the derivation of the
classical limit of MQDT in ref.~\cite{Lombardi:JCP88-3479}. The incoming and
outgoing angular parts are naturally defined in the laboratory frame, but
inserting eq.~(\ref{eq:DefPhi}) into eq.~(\ref{eq:TotalWF}) we see that they
are respectively $\rme^{\rmi \pi \matr{\mu}} A_\Lambda$ and $\rme^{-\rmi \pi
\matr{\mu}} A_\Lambda$ in the molecular frame.

These considerations are clearly displayed by combining 
eqs.~(\ref{eq:EtE=1}) and (\ref{eq:E=nuUmu}):

\begin{eqnarray}\label{eq:1mEET}
0 &=& \det(\Id-\transpose{\matE}\,\, \matE)\\
  &=& \det(\Id-\rme^{\rmi\pi\matr\mu} \transpose{\matr{U}}
\rme^{\rmi\pi\matr\nu} \ \ \ \rme^{\rmi\pi\matr\nu} \matr{U}
\rme^{\rmi\pi\matr\mu}),\nonumber
\end{eqnarray}

As written this is the determinant of a matrix in the molecular frame. It
represents a QPM between perigee and perigee, and the associated eigenfunctions
are the $\vert A_\Lambda \rangle$. These are the angular part of the perigee
wave function. They depend on the energy $E$ and (if there are open channels)
on the phase shift $\tau_\ell$ which cause the determinant to vanish. The
second line of this equation reads, from right to left: apply half a collision
($e^{i\pi\matr\mu}$), rotate the axis from $OZ$ to $OX$ by $\matr{U}$, apply
half a free rotation ($e^{i\pi\matr\nu}$), you are at the apogee. Then apply
half a free rotation, rotate from $OX$ to $OZ$, apply half a collision and you
are back to the perigee.

One may cyclically interchange the matrices in the product entering the
determinant. Then eigenfunctions of the associated matrix are transformed
accordingly. Moving the matrices step by step one place from left to right this
gives first the QPM for the outgoing wave, $\vert
\rme^{\rmi\pi\matr\mu_\Lambda} A_\Lambda \rangle$ in the molecular frame, then
the same QPM in the laboratory frame, next the QPM between apogee and apogee in
the laboratory frame {\it etc}. This provides the explanation for the phase 
shifts
included in eq.~(\ref{eq:B=CA}): going from $A_\Lambda$ to $B_N$ involves not
only a change of the basis, but also half a collision, and half a free rotation.

\subsection{\label{sec:Closed} All channels closed}

\begin{figure}[!ht]
\centering
\setlength{\figwidth}{\textwidth}
\TwoByTwo{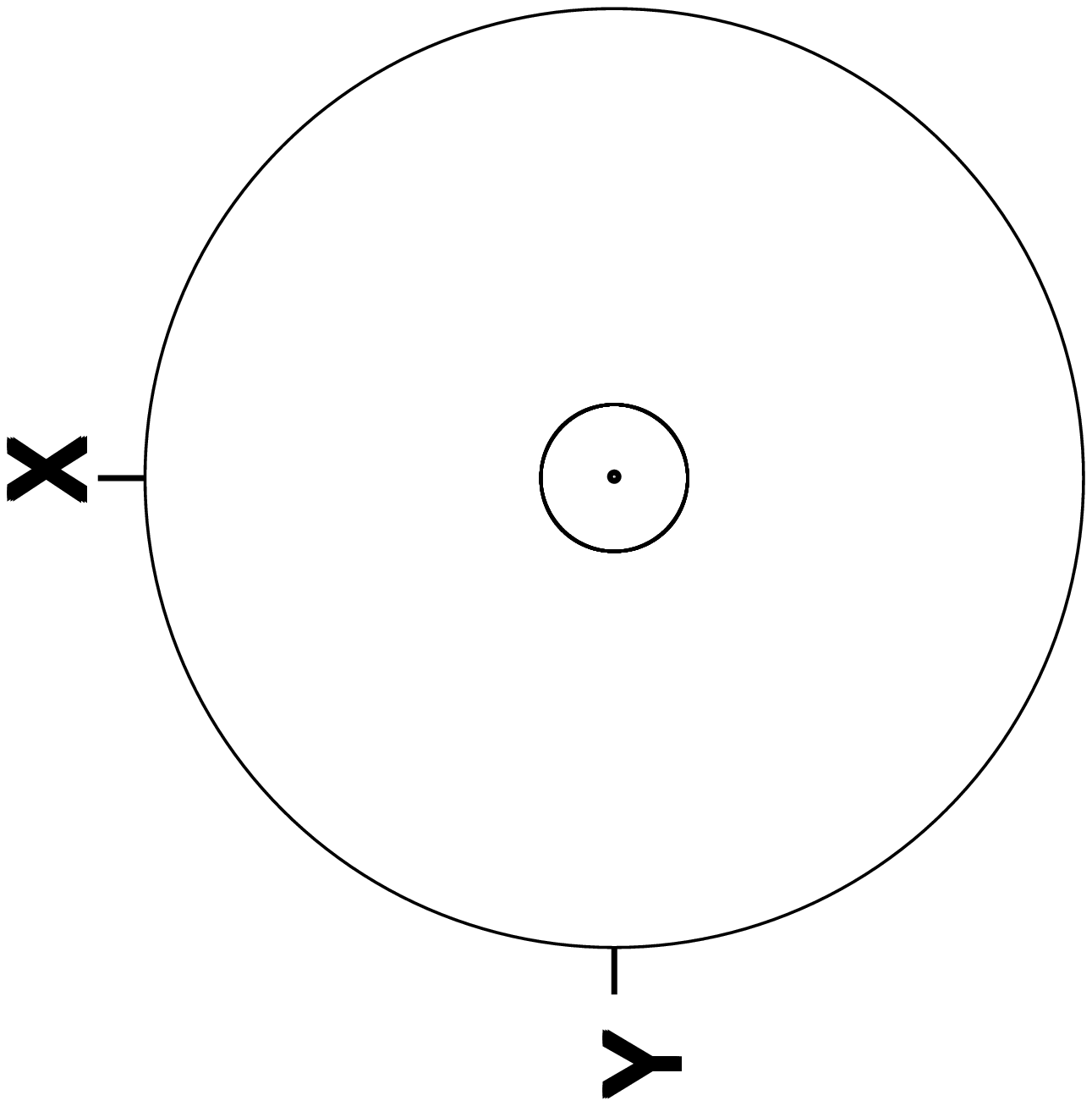}{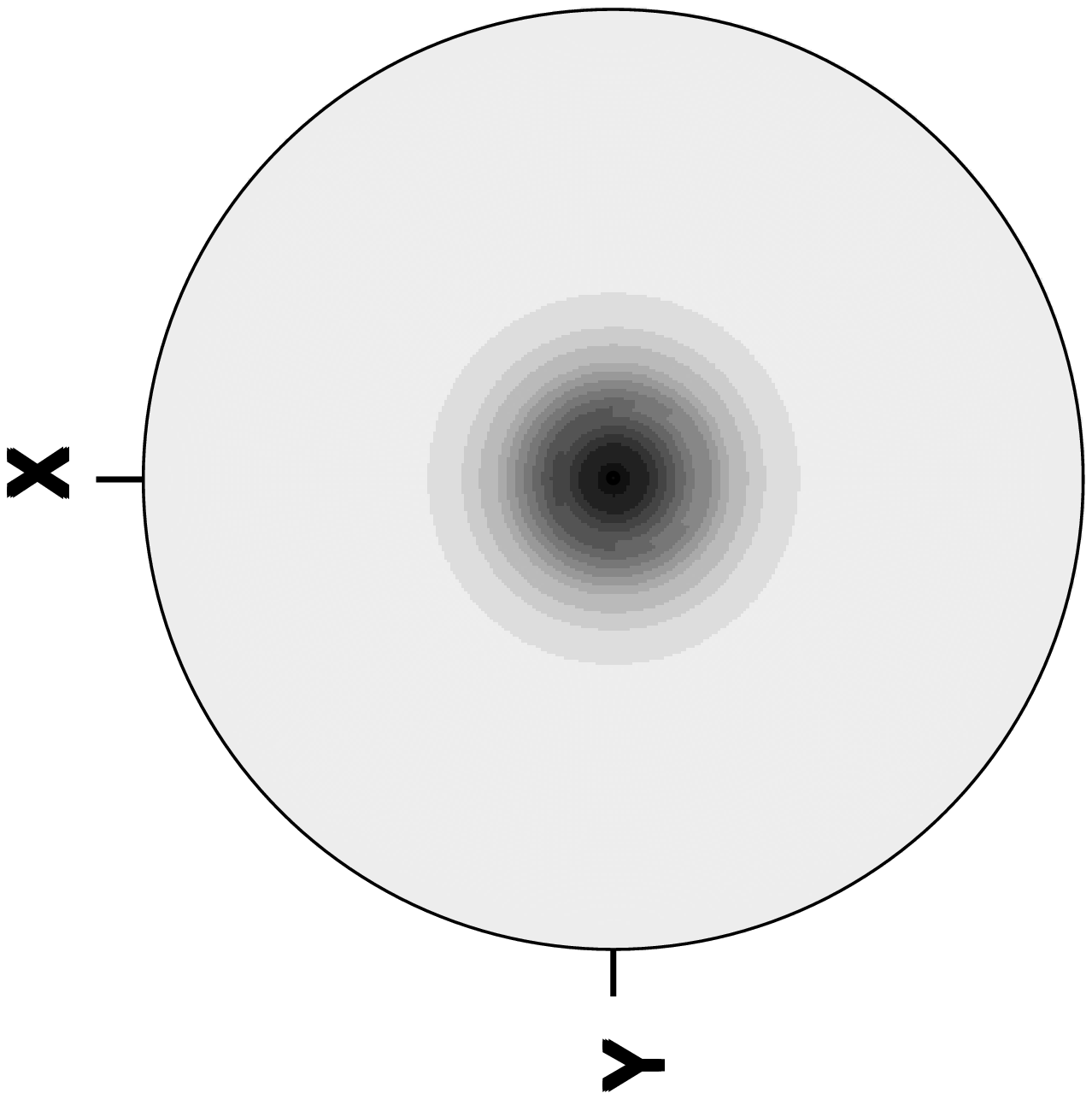}{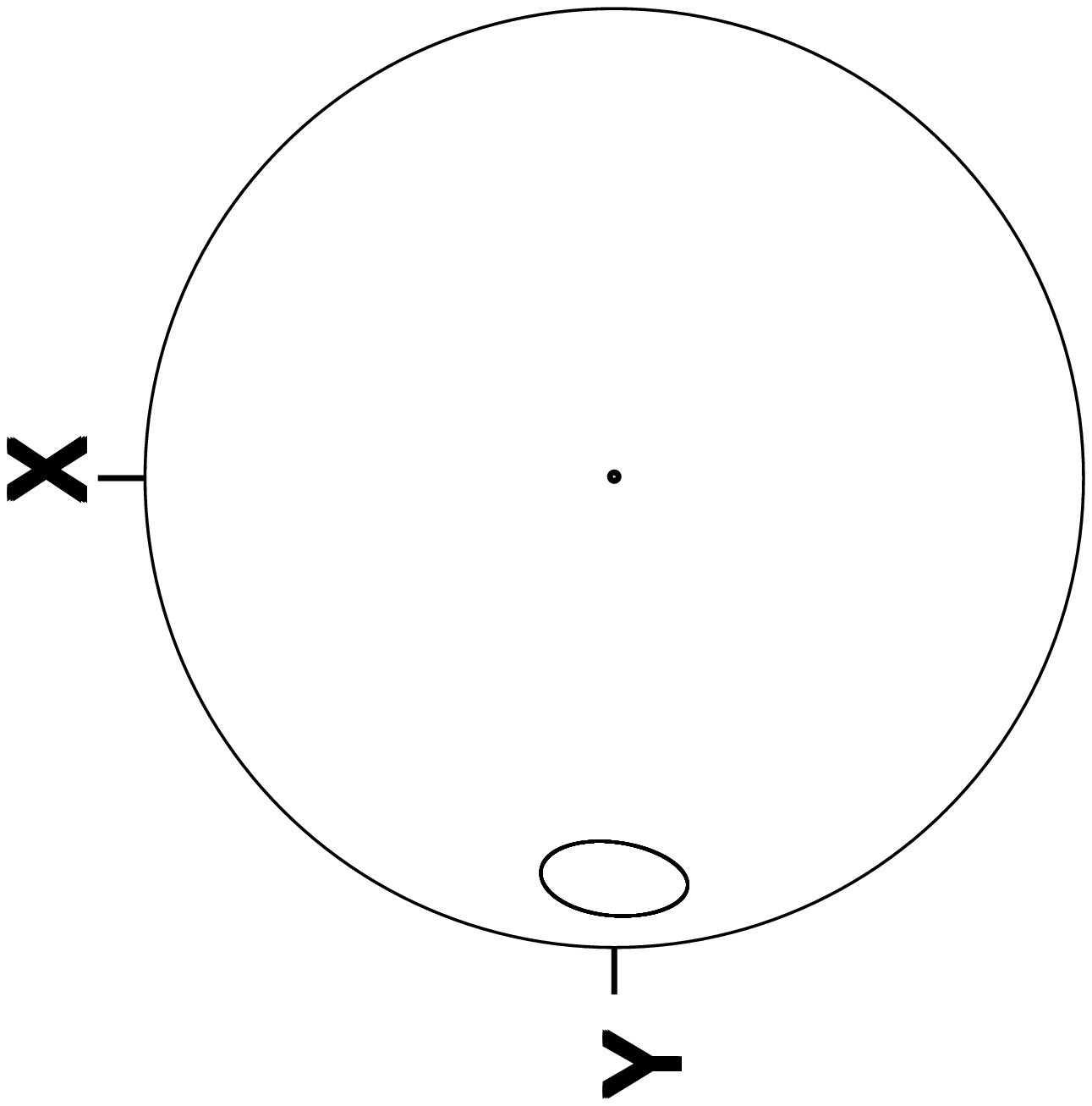}{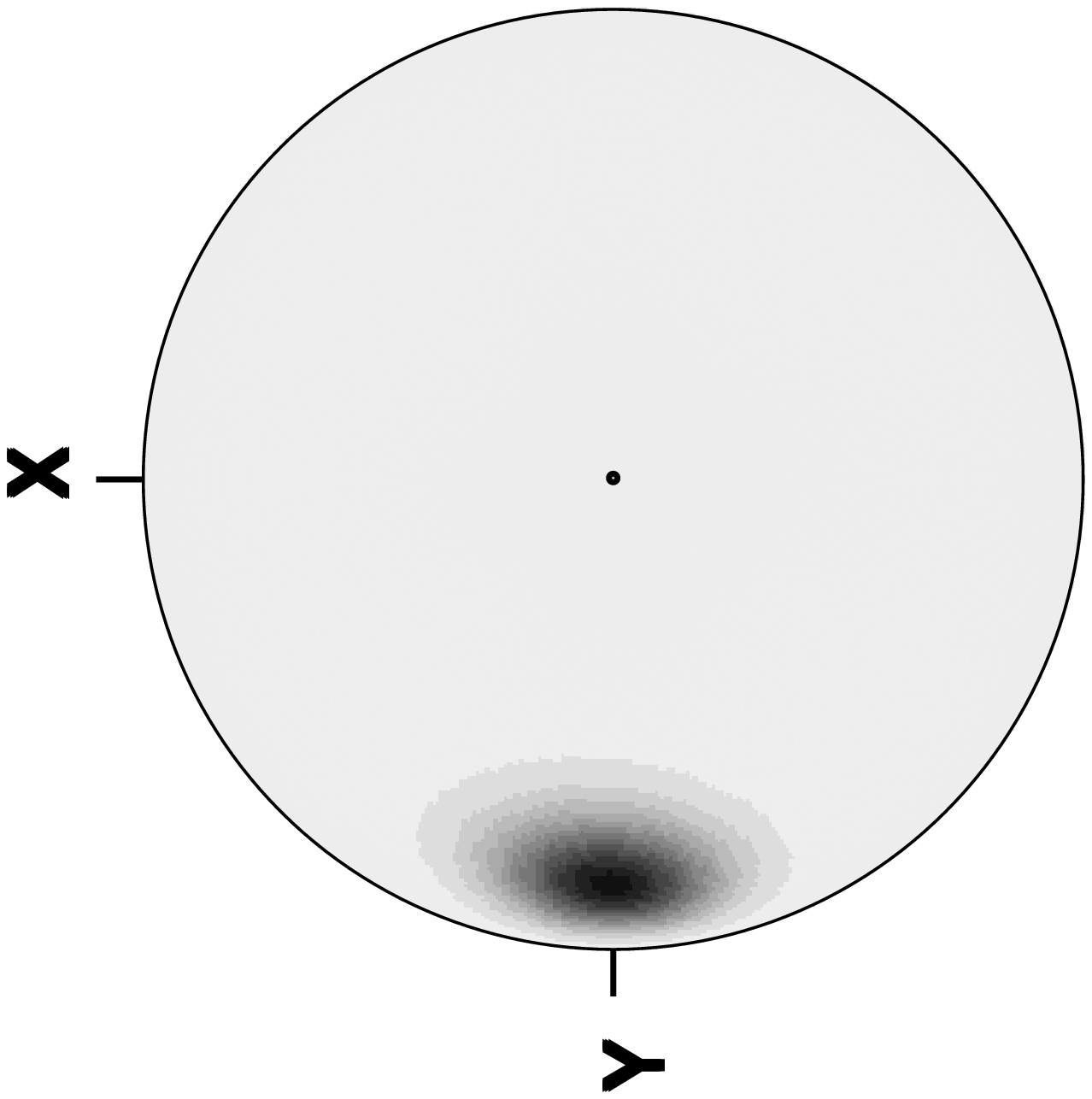}{a}{b}
\TwoByTwo{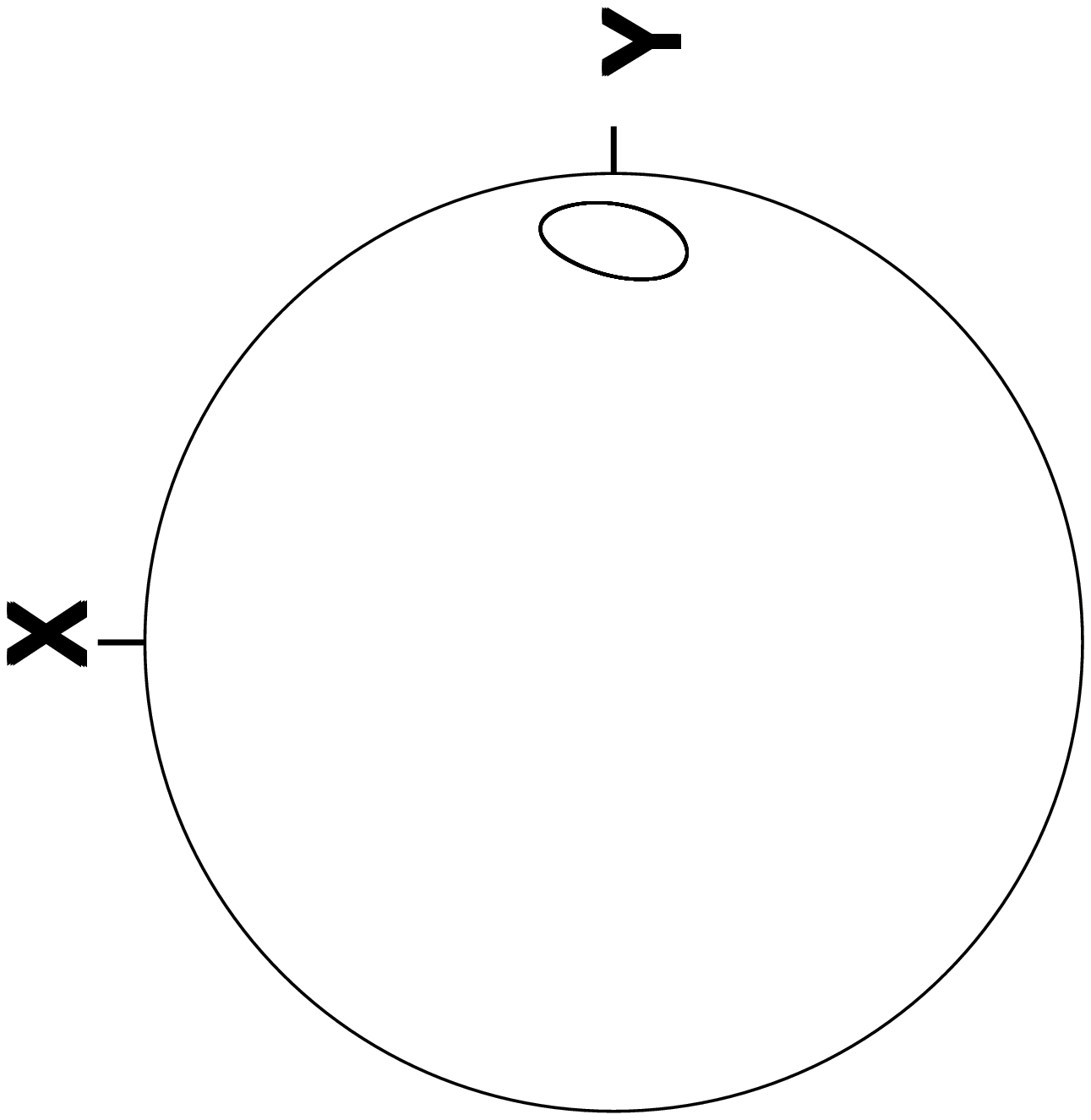}{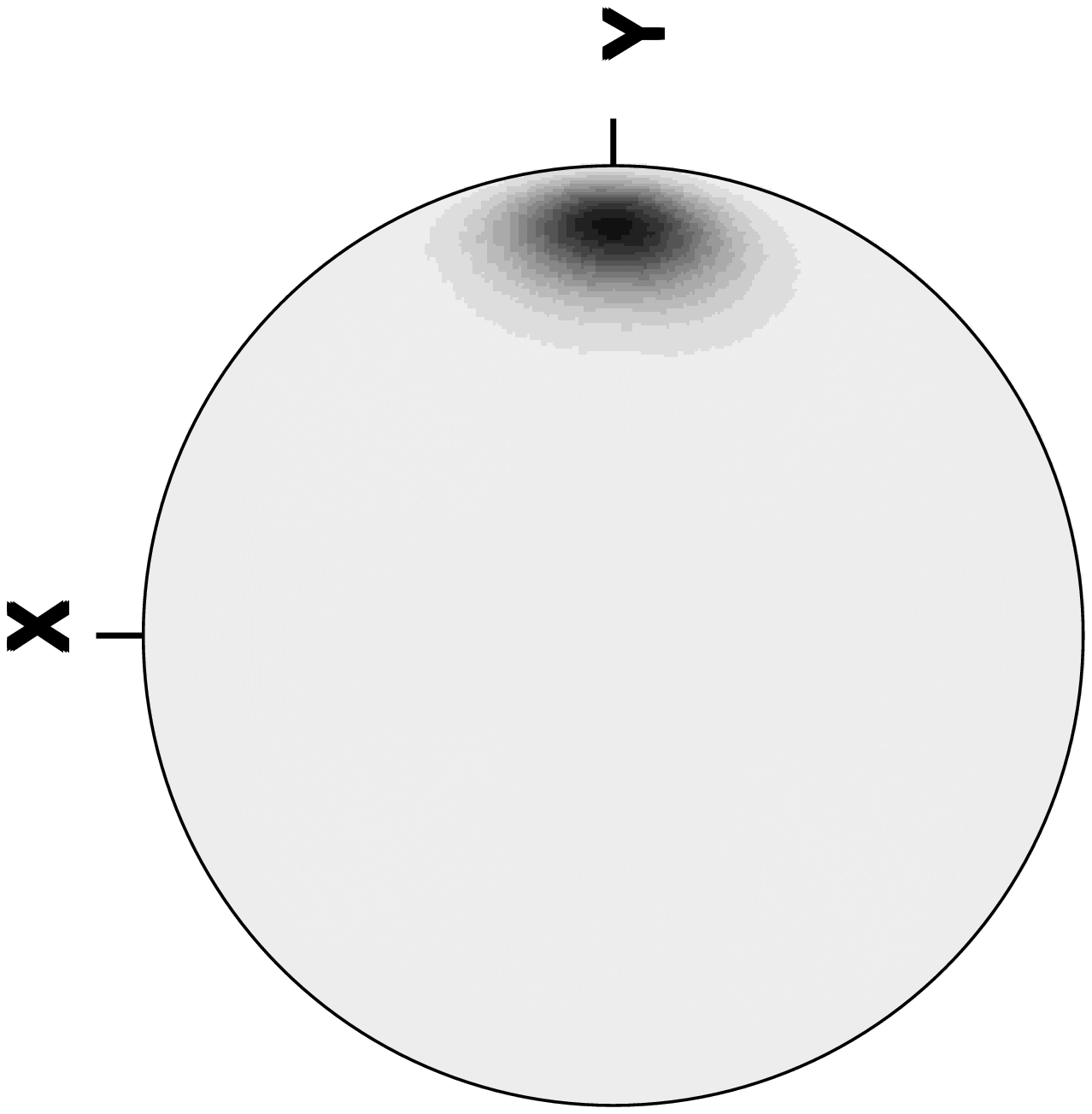}{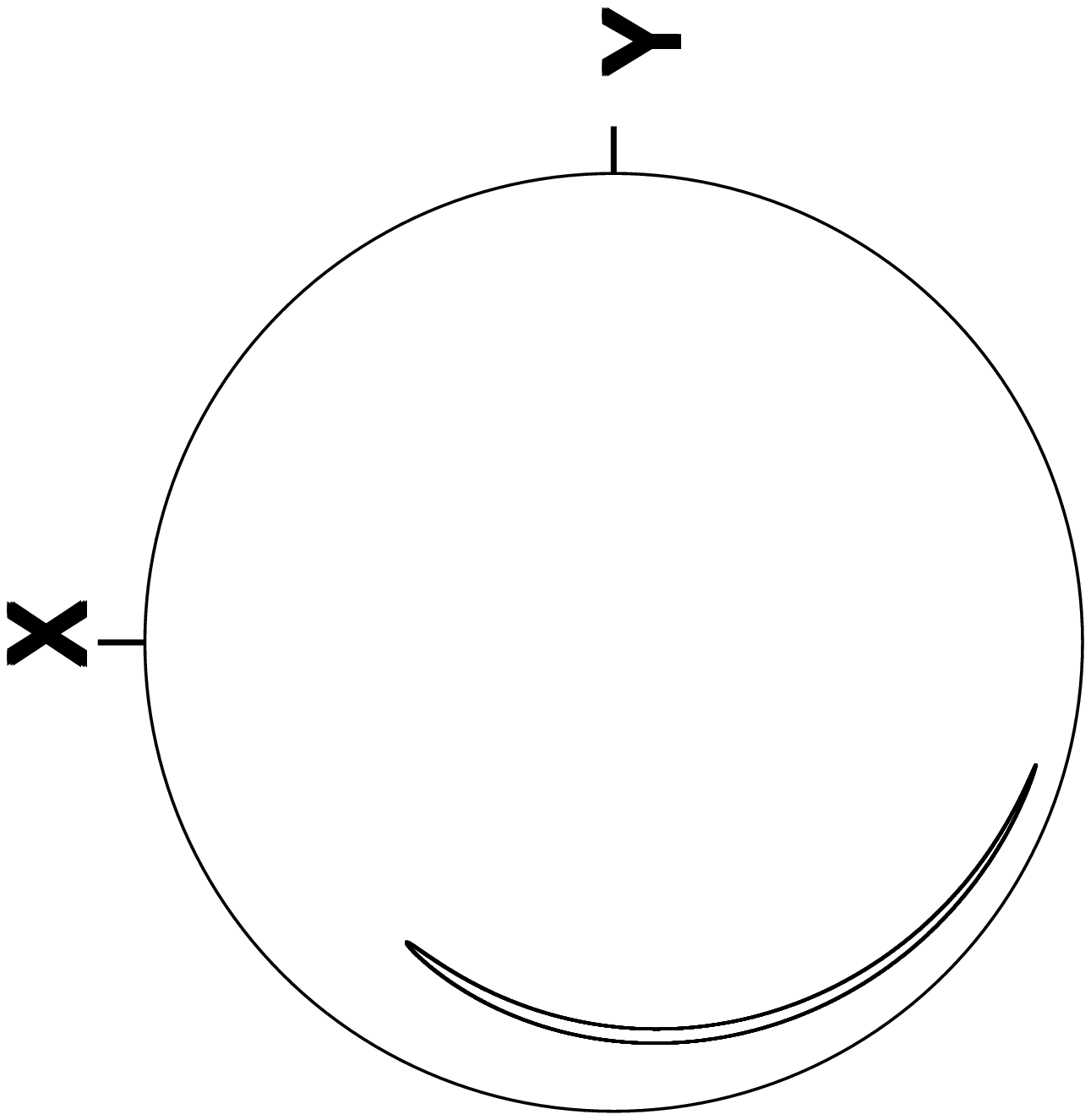}{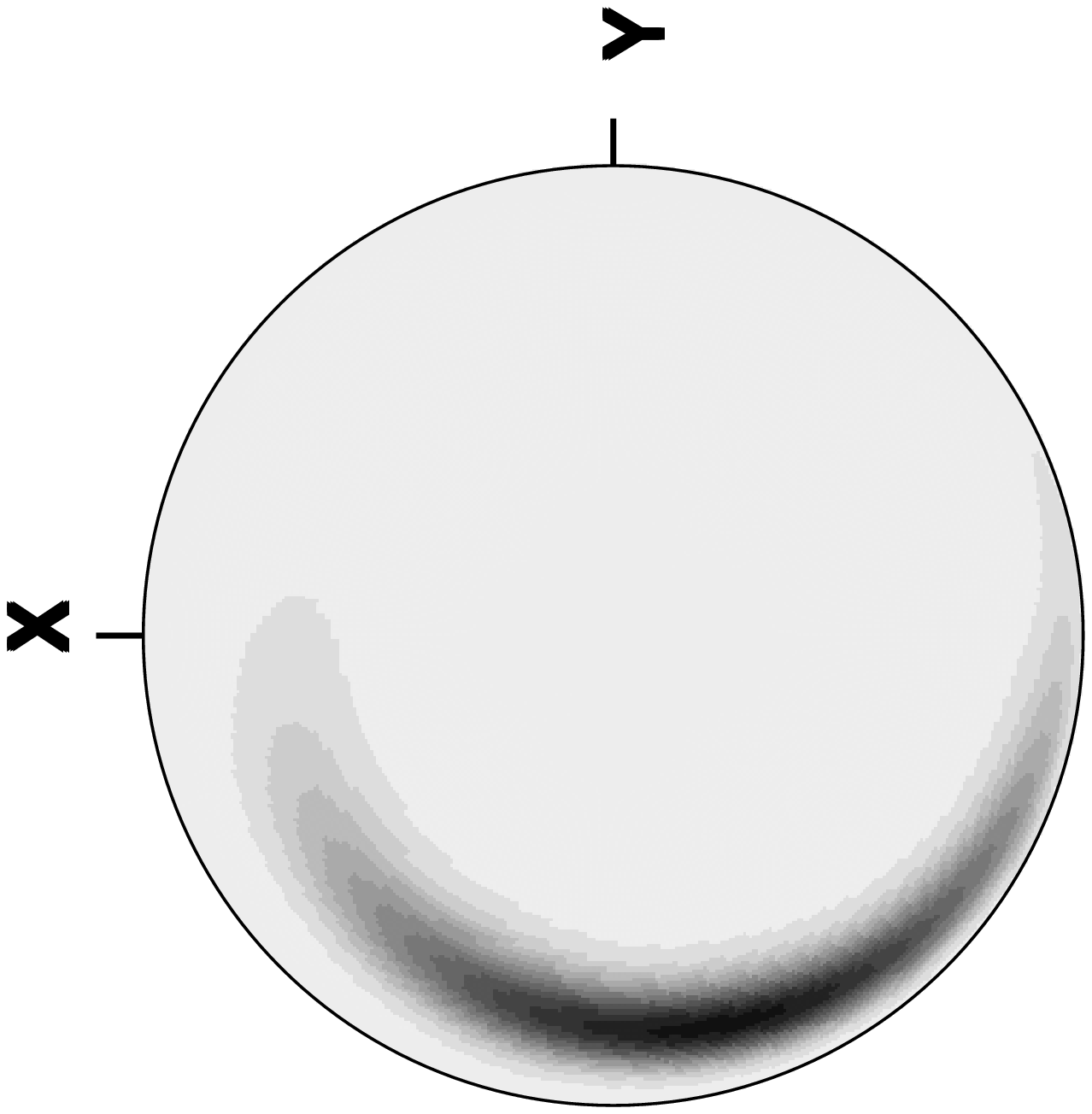}{c}{d}
\TwoByTwo{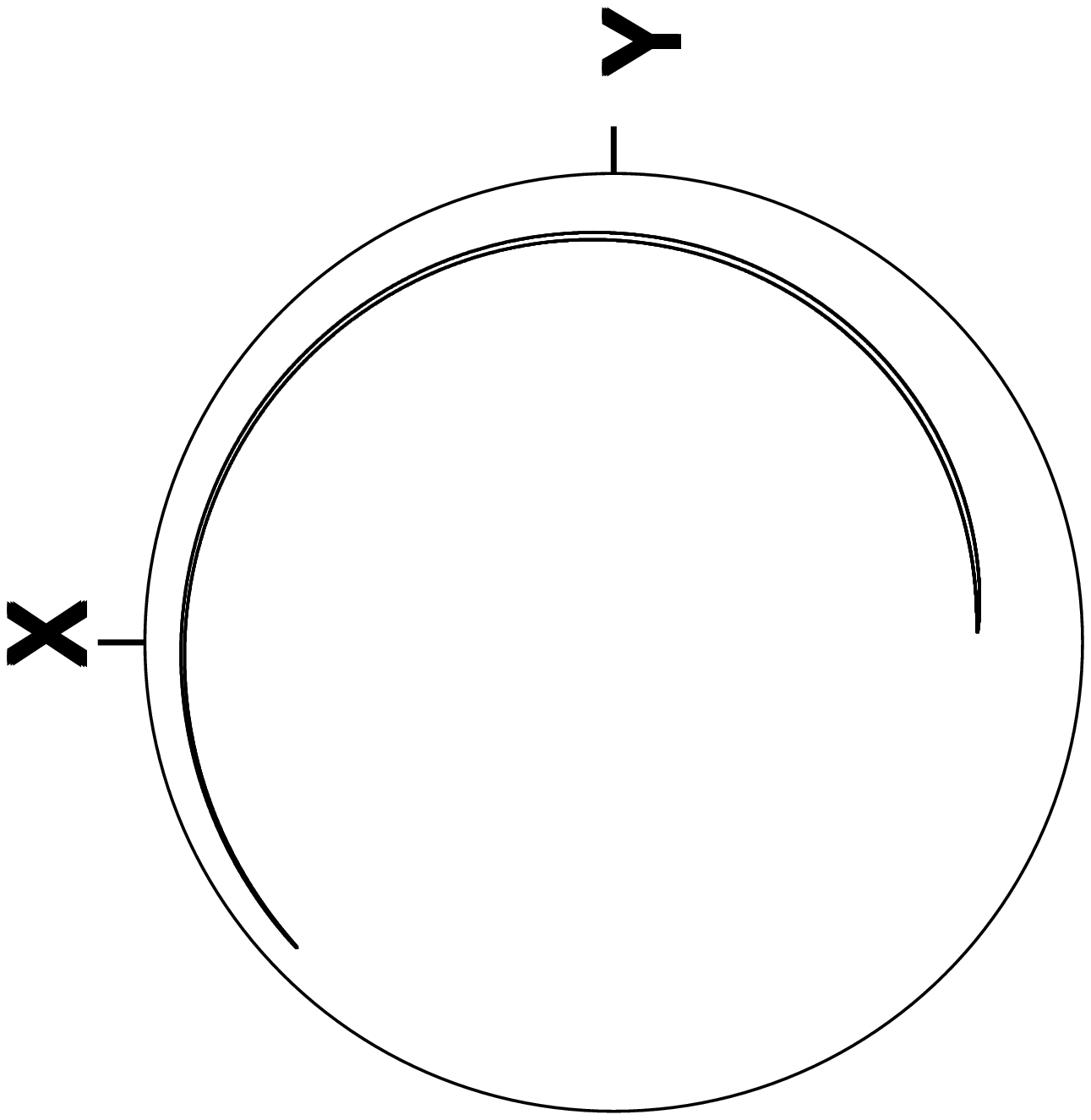}{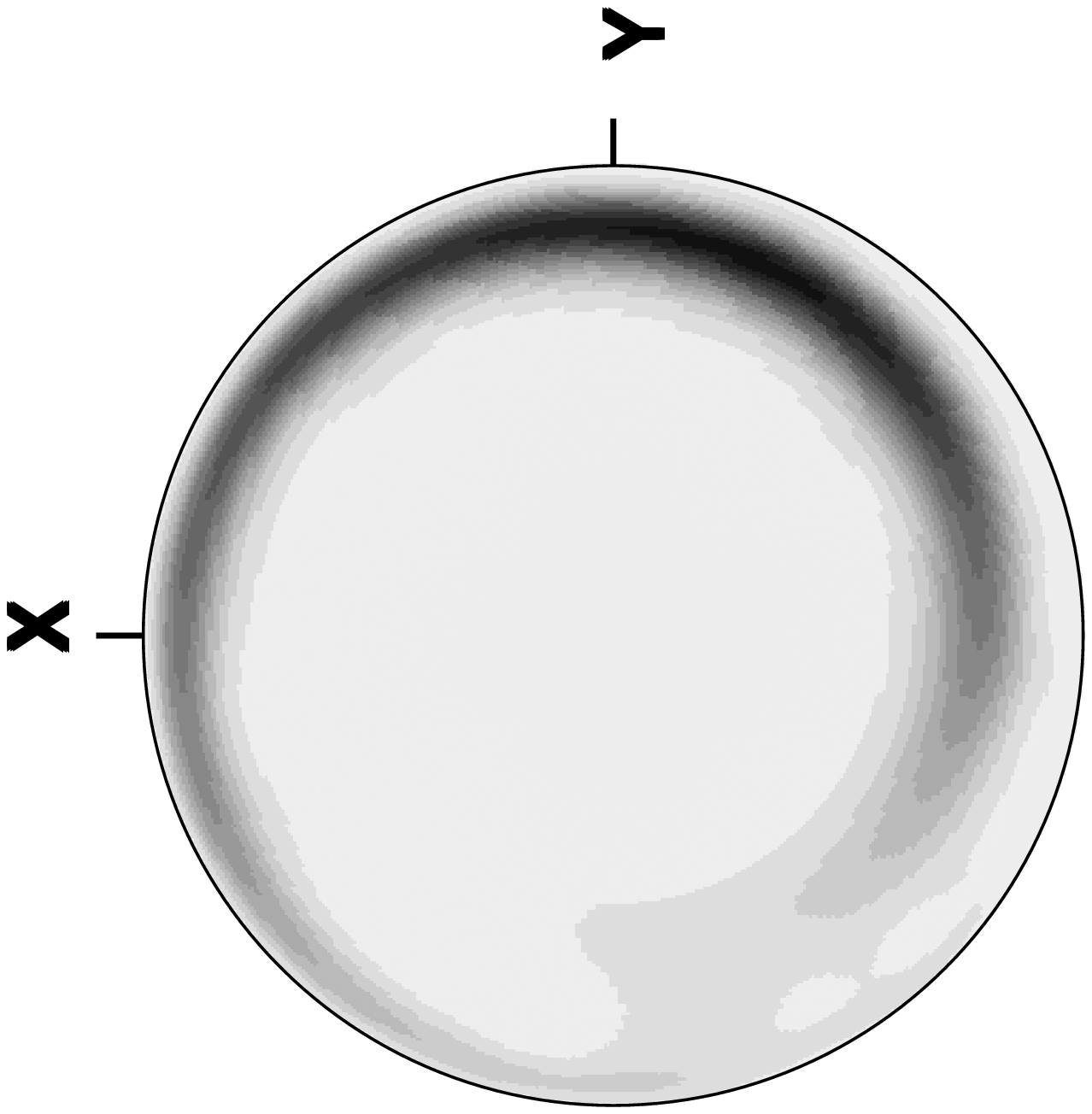}{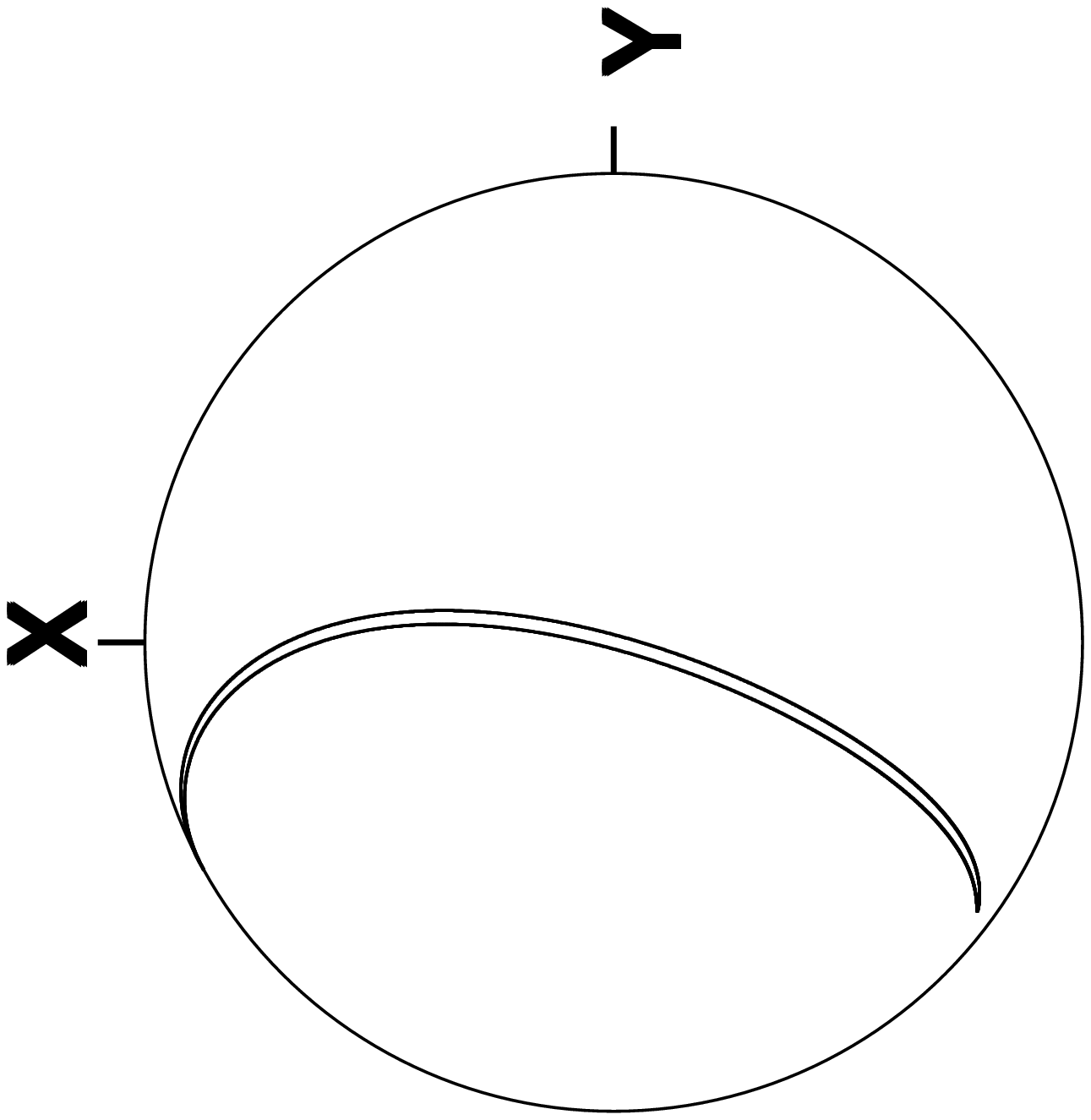}{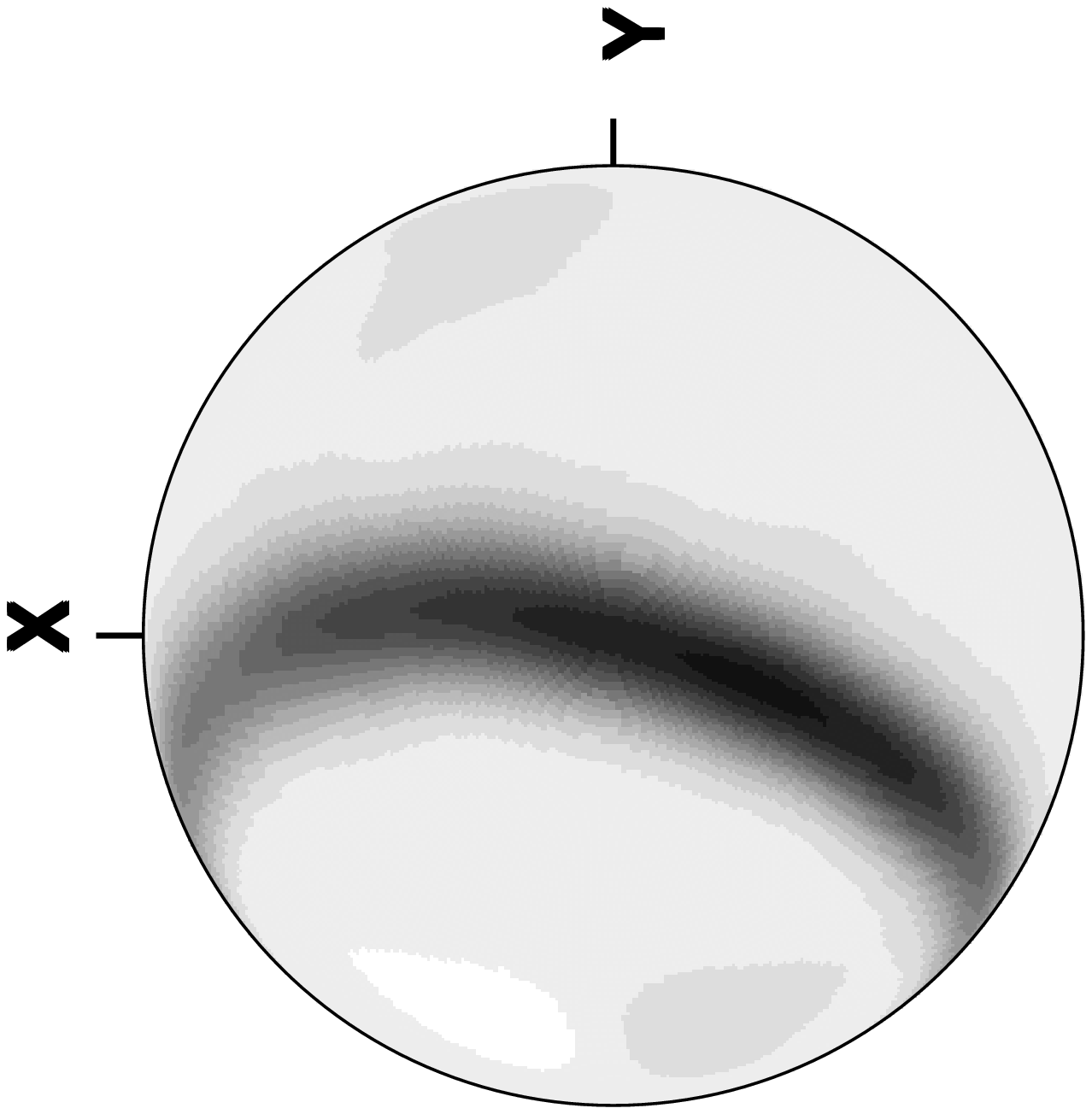}{e}{f}
\caption{\label{fig:RotCirc}
Comparison between Classical and Quantal time evolution. All channels closed.
Parameters $\nu_J=875-880\,\mathrm{a.u.}$, $\mu_{40}=39.8765$, corresponding to
$T_e/T_N=1/3$, $K=-3.98765\,\pi$. (a): Start. (b)-(d): Application of $\matE$.
Half a free rotation ($\rme^{\rmi \pi \matr\nu}$), $1/6$ of a turn clockwise
around $OX$ leads from (b) to (c)  (notice the flip from top view to bottom
view), $U$ changes axis from  $OX$ to $OZ$, half a collision ($\rme^{\rmi \pi
\matr\mu}$), stretch around $OZ$ leads from (c) to (d). (d)-(f): Application of
$\transpose{\matE}$. Half a stretch around $OZ$, $\transpose{\matr{U}}$, half a
rotation around $OX$.}
\end{figure}

That this is the correct interpretation is illustrated in 
Fig.~\ref{fig:RotCirc},
where we compare the time evolution of a (quantum mechanical) Husimi 
distribution and
a corresponding classical swarm of trajectories for a bound system.  The
parameters correspond to a mean energy for which $T_e/T_N=1/3$, so that a free
rotation is $1/3$ of a turn, and a strong coupling $K=-3.98765\,\pi$ giving a
strong stretch around $OZ$, nearly $\pm$ two full turns at the $\pm OZ$ poles
(and zero at the equator). The remaining parameters are chosen as 
$L=40$, $J=200$,
$2B=2.5\,10^{-12} \, \mathrm{a.u.}$. The angular momenta are large as compared
to those accessible in current molecular experiments, especially $L$ which  
usually is limited to
the range 0-2 in two step laser excitations. They correspond to semi classical
scaling of the experimental situation
\cite{Labastie:PRL84-1681,Bordas:JP85-27}, as explained in
section~\ref{sec:ParamCount}.

The quantum time evolution is computed by first obtaining a set of $2L+1$
consecutive eigenenergies $E_i$ and the corresponding angular wave functions
$A_\Lambda^i$ for increasing principal quantum number $\nu_J$ by one
(or a multiple of such sets to increase statistical accuracy
of the results). An initial radial wave function is then expanded in this
eigenbasis. The time evolution results by applying $\exp(\rmi E_i t)$ to the
$A_\Lambda^i$. Notice that it is applied only to the angular part, not to the
radial part of the wave functions. But for integer times, which correspond to a
full orbit because we express energies in terms of a mean $\nu$, the radial
part is the same, so the corresponding radial overlap is one. This procedure is
consistent with the QPM point of view. What we neglect in doing so, is the
spreading with time of the radial wave packet. This is consistent with a semi
classical picture. For a full quantum computation we would need to choose a
radial distribution $f(r)$ peaked around the corresponding radius $r$, perigee,
apogee, $r_0$, and to develop this initial state,
angular and radial part, in an eigenbasis. The initial quantum distribution is
the Husimi distribution of a $\vert L L \rangle$ state. The corresponding
classical distribution is a circle around $OZ$ with radius given by
$\cos\theta_L = L/\sqrt{L(L+1)}$. In Fig.~\ref{fig:RotCirc} the top pair (a)
equally corresponds to incoming, perigee and outgoing waves because the
collision is a rotation around $OZ$ and the distribution is rotationally
invariant around this axis. The next pair (b) is the apogee. Quantum
mechanically we cannot compute it from the previous distribution
because we can compute only the
evolution with an integral number of turns. It is thus obtained by a rotation
of the initial distribution around $OX$ by $1/6$ of a turn. But then everything
is computed from the four initial distributions by applying the above method.
One sees perfectly the evolution: $1/6$ of a turn around $OX$ (c), half stretch
around $OZ$ (d), half stretch along $OZ$ (e), $1/6$ of a turn around $OX$ (f)
etc. The quantum distribution follows perfectly the classical one in such short
times. For longer times they begin to differ due to the finite value of
$\hbar$.

\subsection{First Channel Open}

\begin{figure}[!htb]
  \centering
  \includegraphics[angle=270,width=0.5\textwidth]{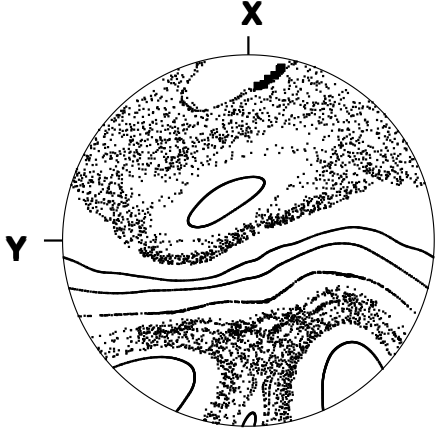}
  \caption{\label{fig:Poin1stOpen}
  First channel opening. Classical PM. Since Poincar\'e maps have a symmetry
$\mathrm{C_2}$ around $OX$, only half the sphere is needed. We use a
stereographic projection from $+OZ$ of the lower part of the sphere. The open
part is a small cap near the $+OX$ axis. Large black squares at its border are
ionizing trajectories.}
\end{figure}

To begin with, we select an energy just above the first threshold. As for the
rest of this section we select $L=50$, $J=100$, $2B=10^{-10}\,\mathrm{a.u.}$,
so $N$ varies between 50 and 150. As compared to the bound case, the ratio
$L/J$ has been increased from $1/5$ to $1/2$ to increase the transfer of energy
between electron and core. The PM is given in
Fig.~\ref{fig:Poin1stOpen}. The ionization region corresponds to a very small 
white cap
around the $+OX$ axis. The large black squares at its border are the points of
ionization of some trajectories.

\begin{figure}[!ht]
\centering
\includegraphics[angle=270,width=\figwidth]{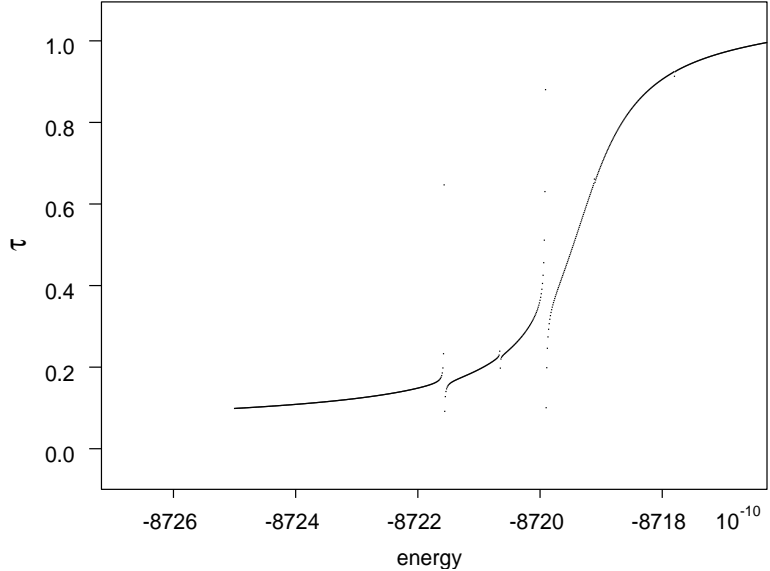}
\caption{\label{fig:Tau1stOpen}
First channel opening. Phase shift.  It appears at the first ionization
threshold, which is negative ($-8726\,10^{-10}$), simulating a stray space
charge: see text. There is only one phase shift $\tau$, because there is only
one channel open, in this energy range. It increases by one for resonances
between the open channel and some closed channels. Coupling $\mu_{50}=4$.}
\end{figure}

The value of the phase shift $\tau$ of the $S$ matrix as a function of energy
is shown in Fig.~\ref{fig:Tau1stOpen}: for each energy there is only one open
channel, thus one value of $\tau$.

\begin{figure}[!ht]
\centering
\setlength{\figwidth}{\textwidth}
\ThreeUnRelated{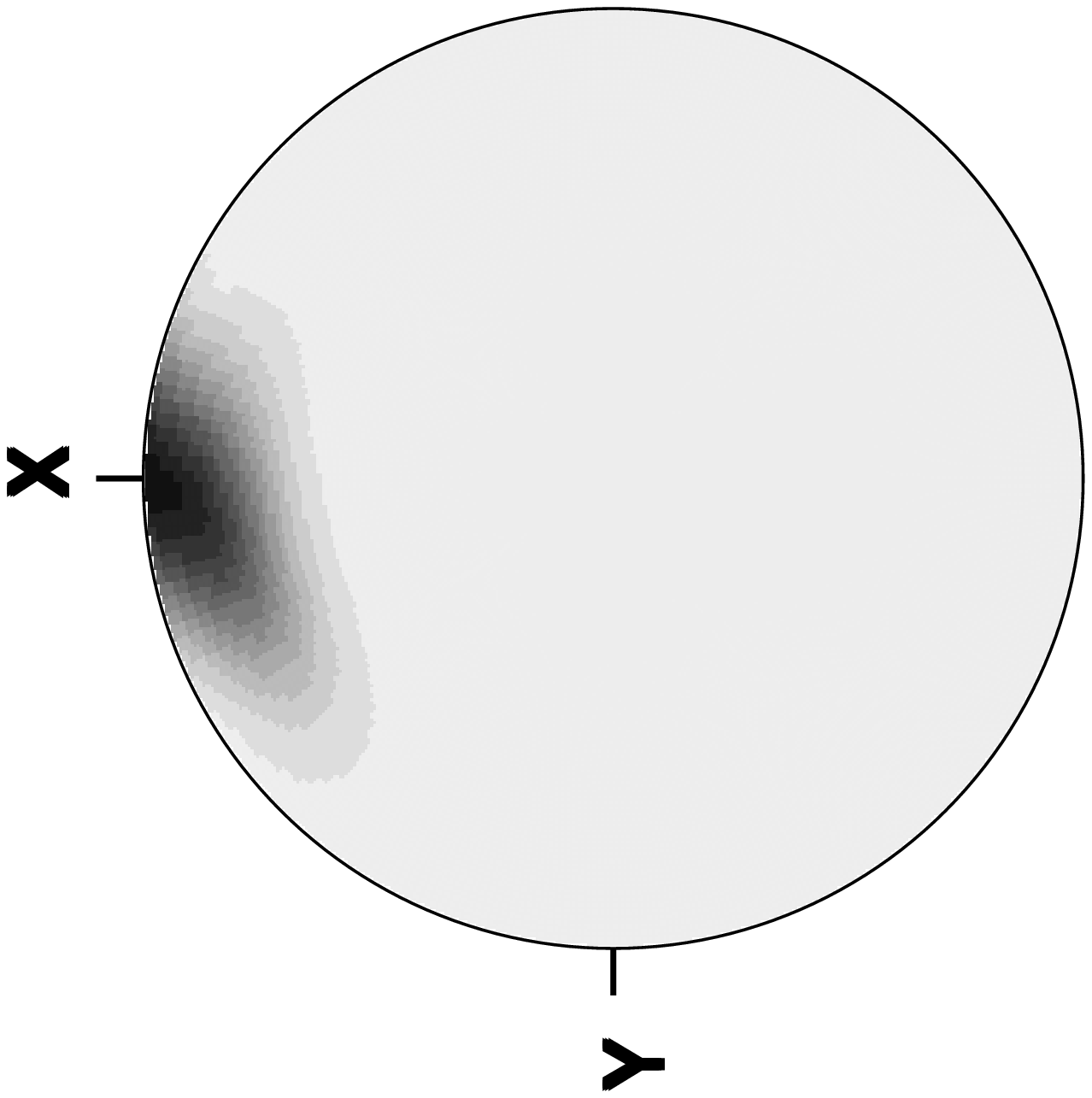}{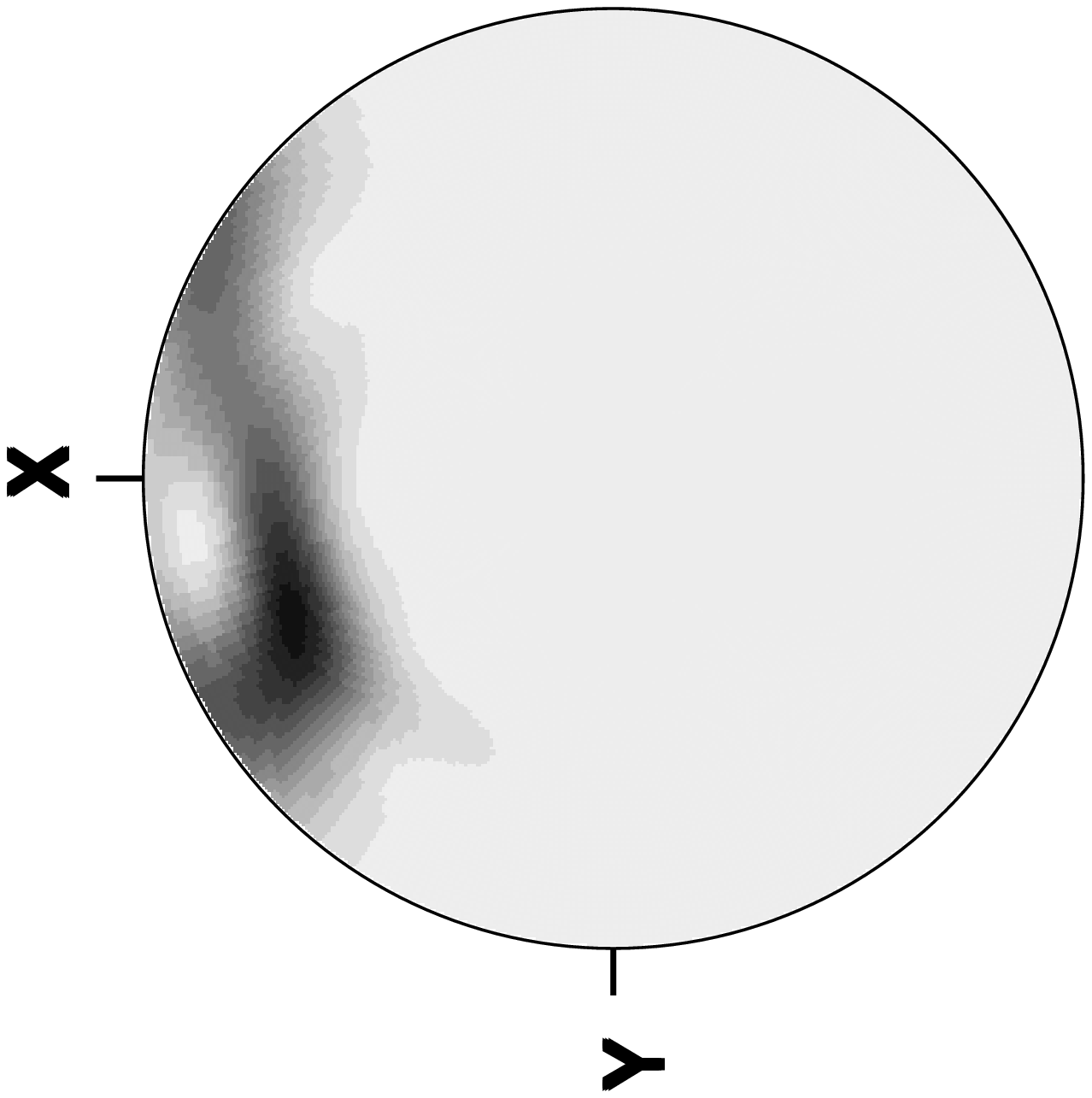}{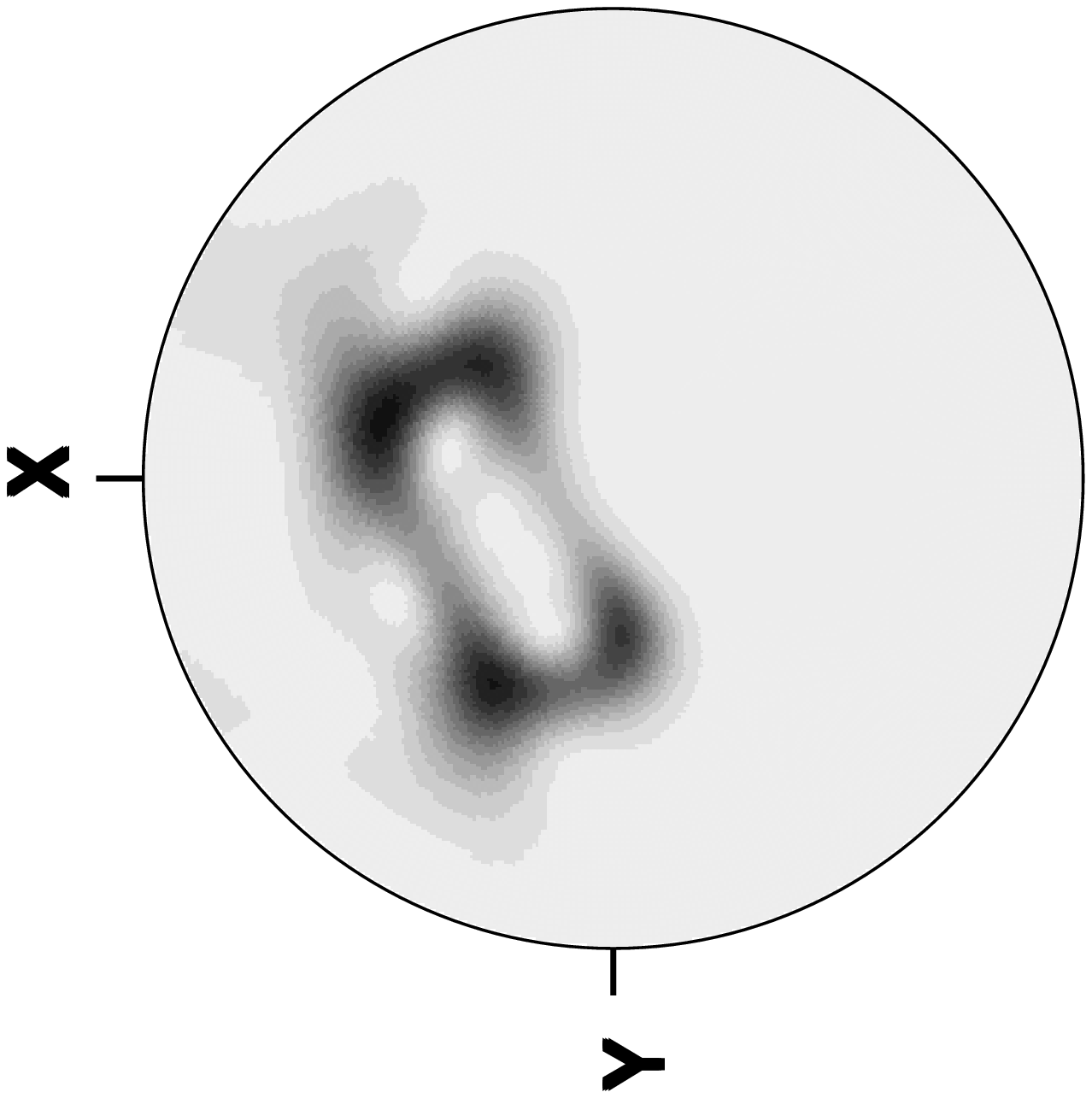}{a}{b}{c}
\ThreeUnRelated{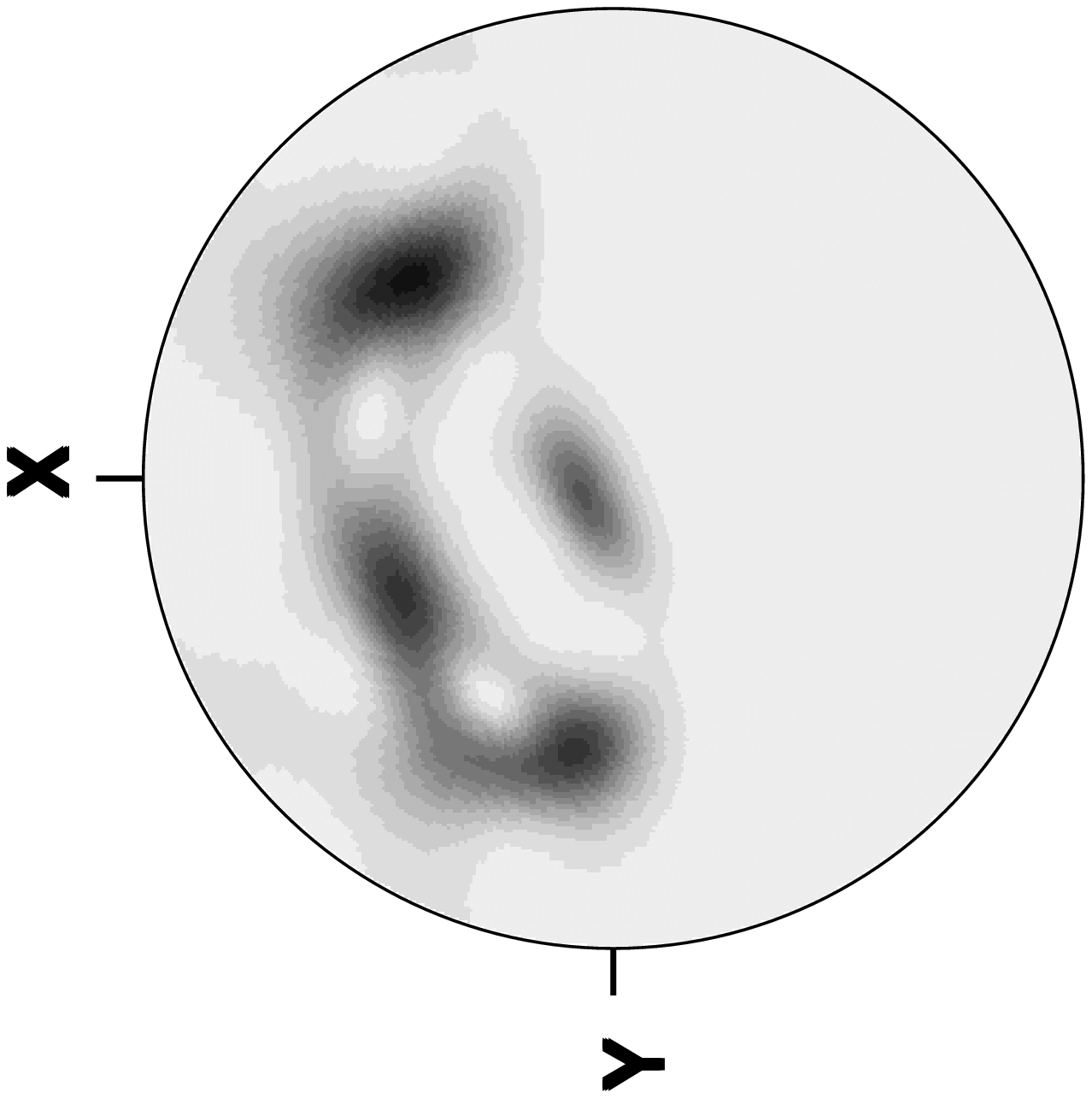}{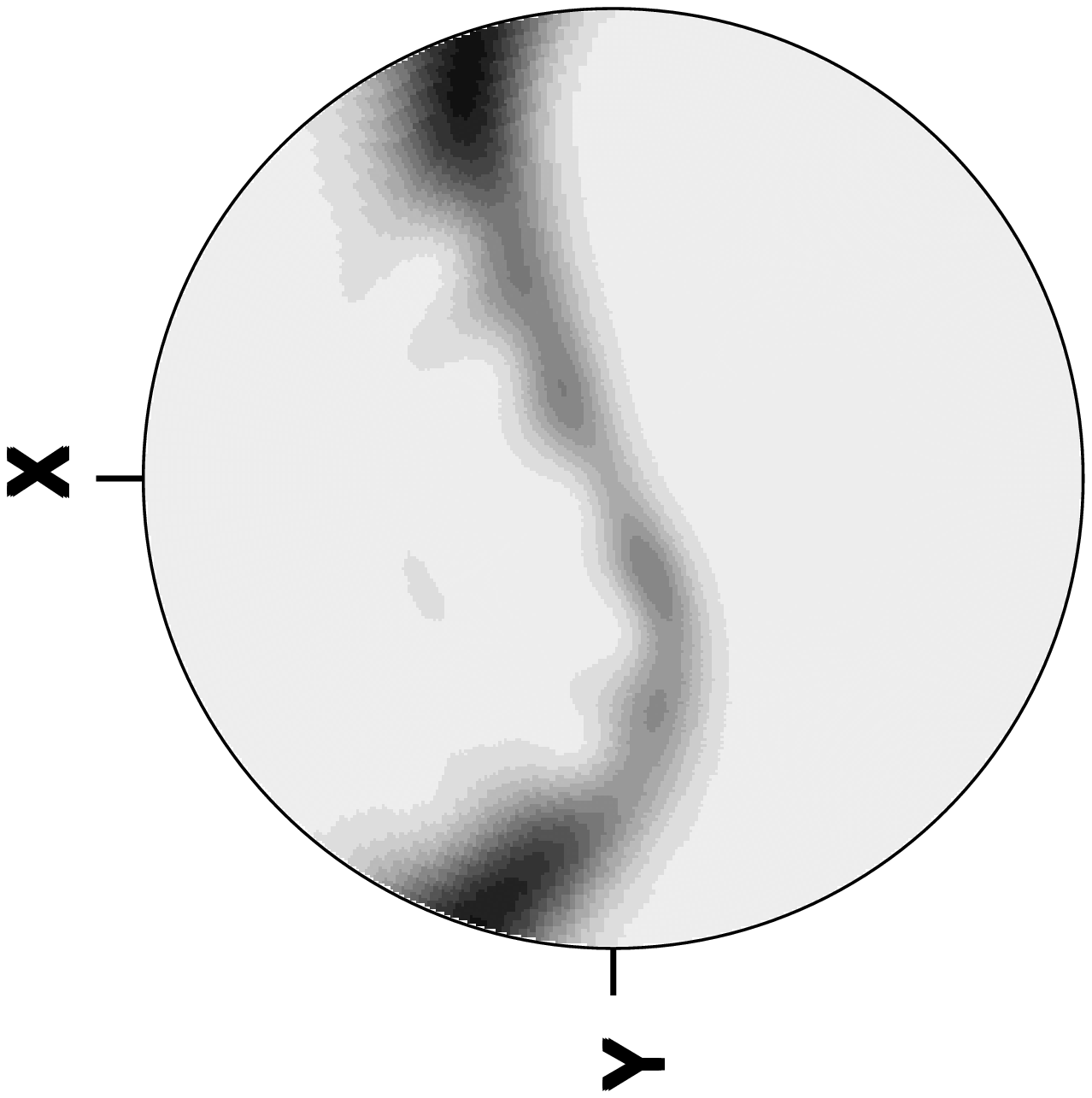}{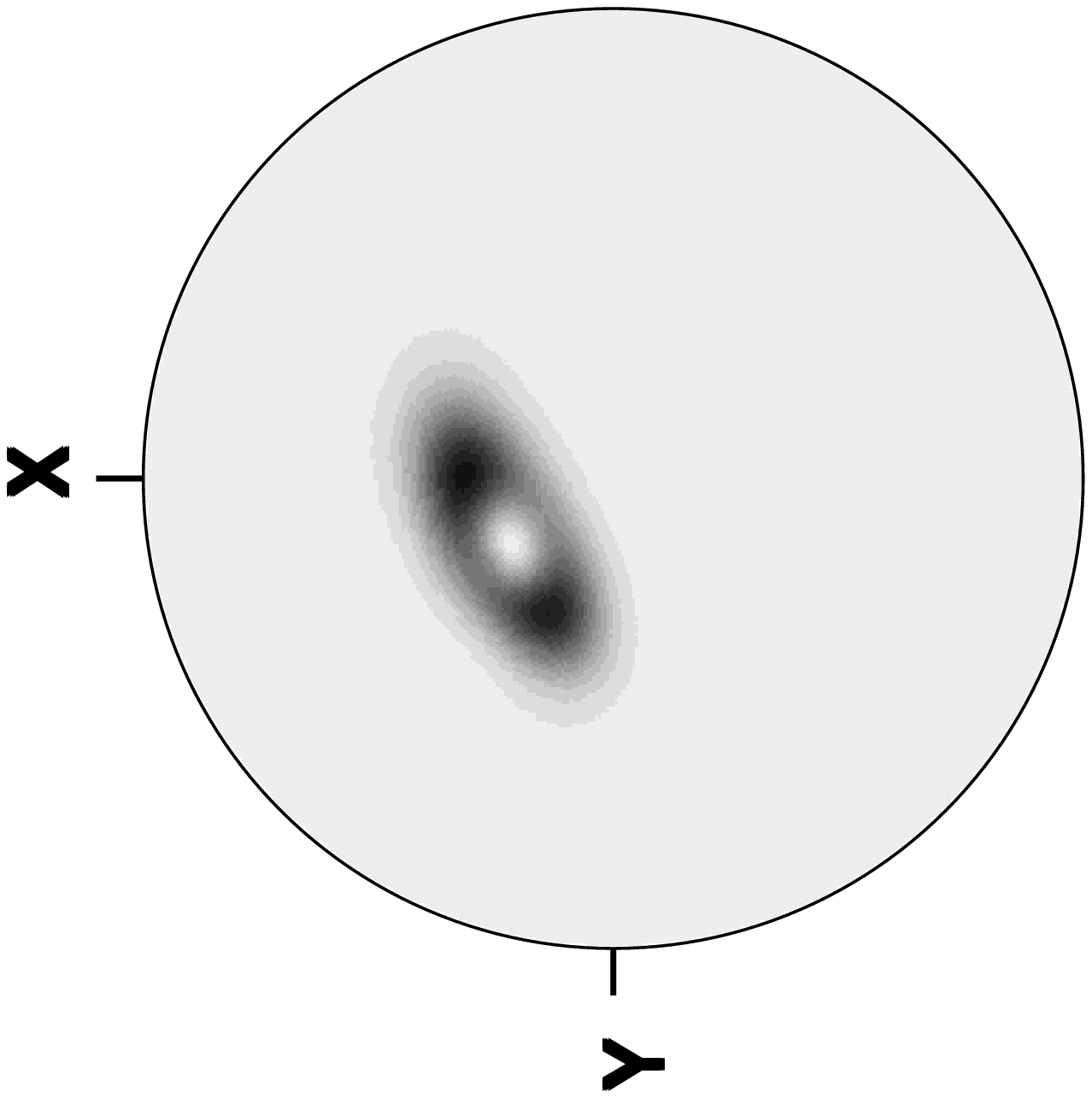}{d}{e}{f}
\ThreeUnRelated{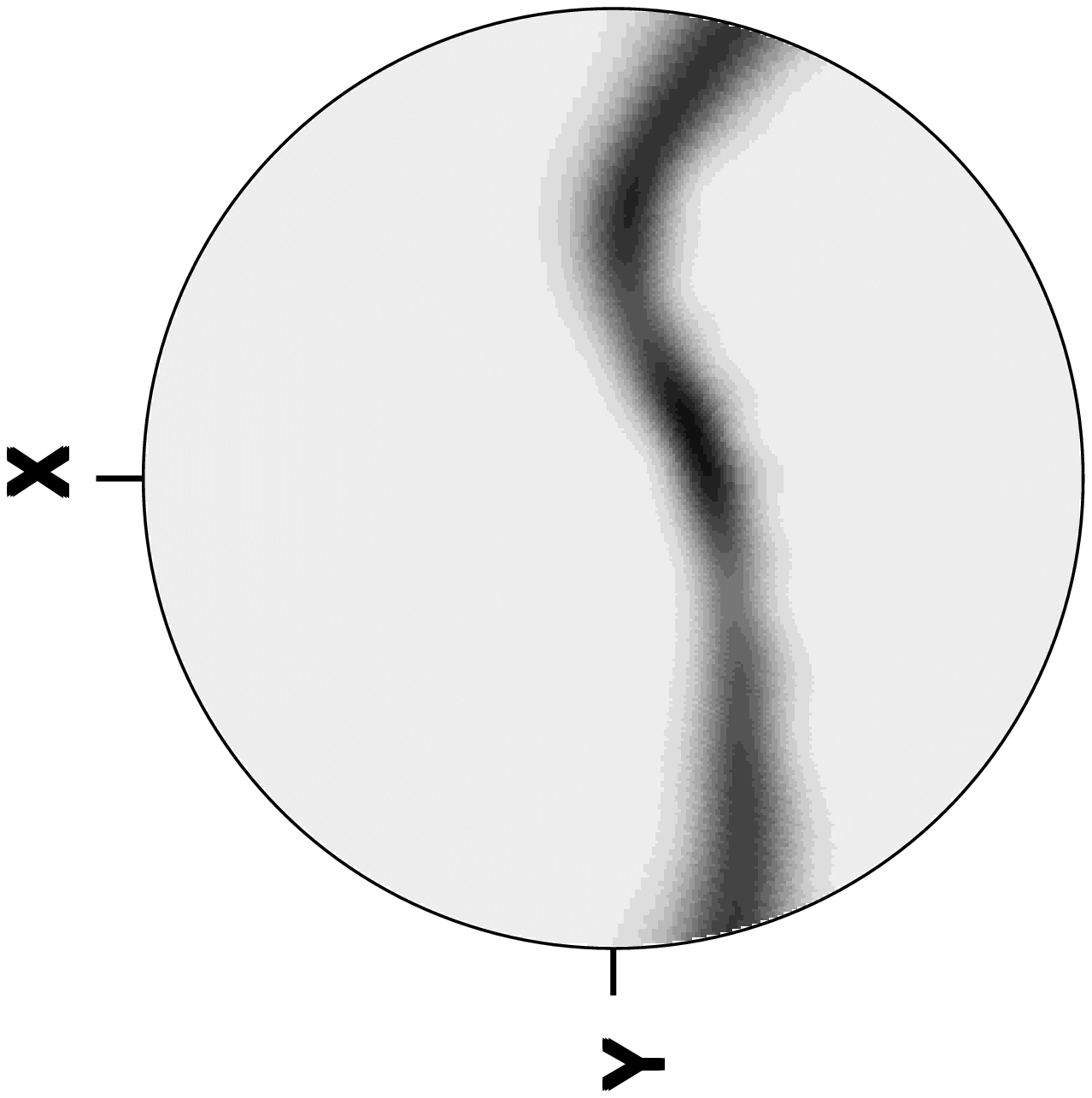}{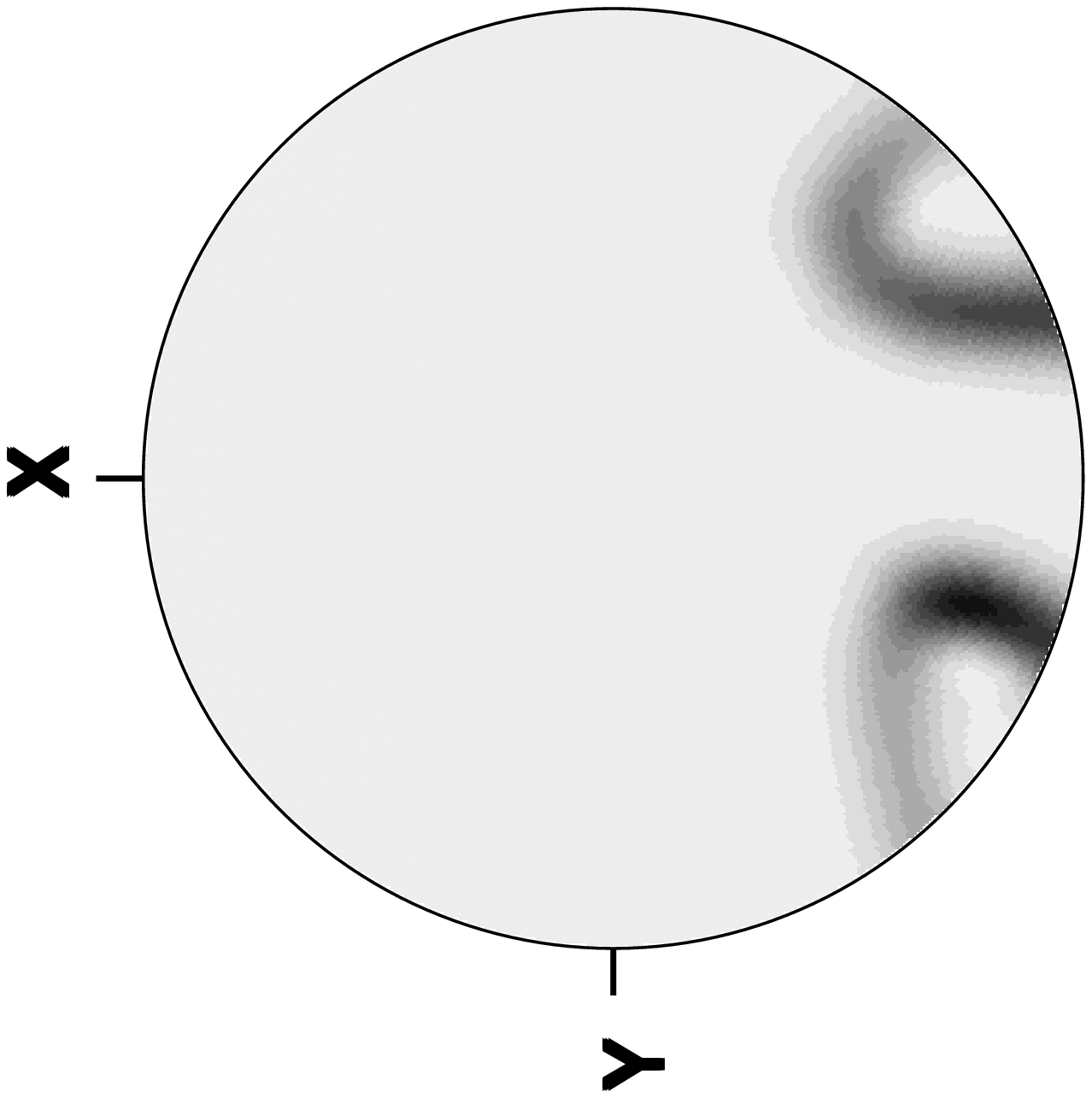}{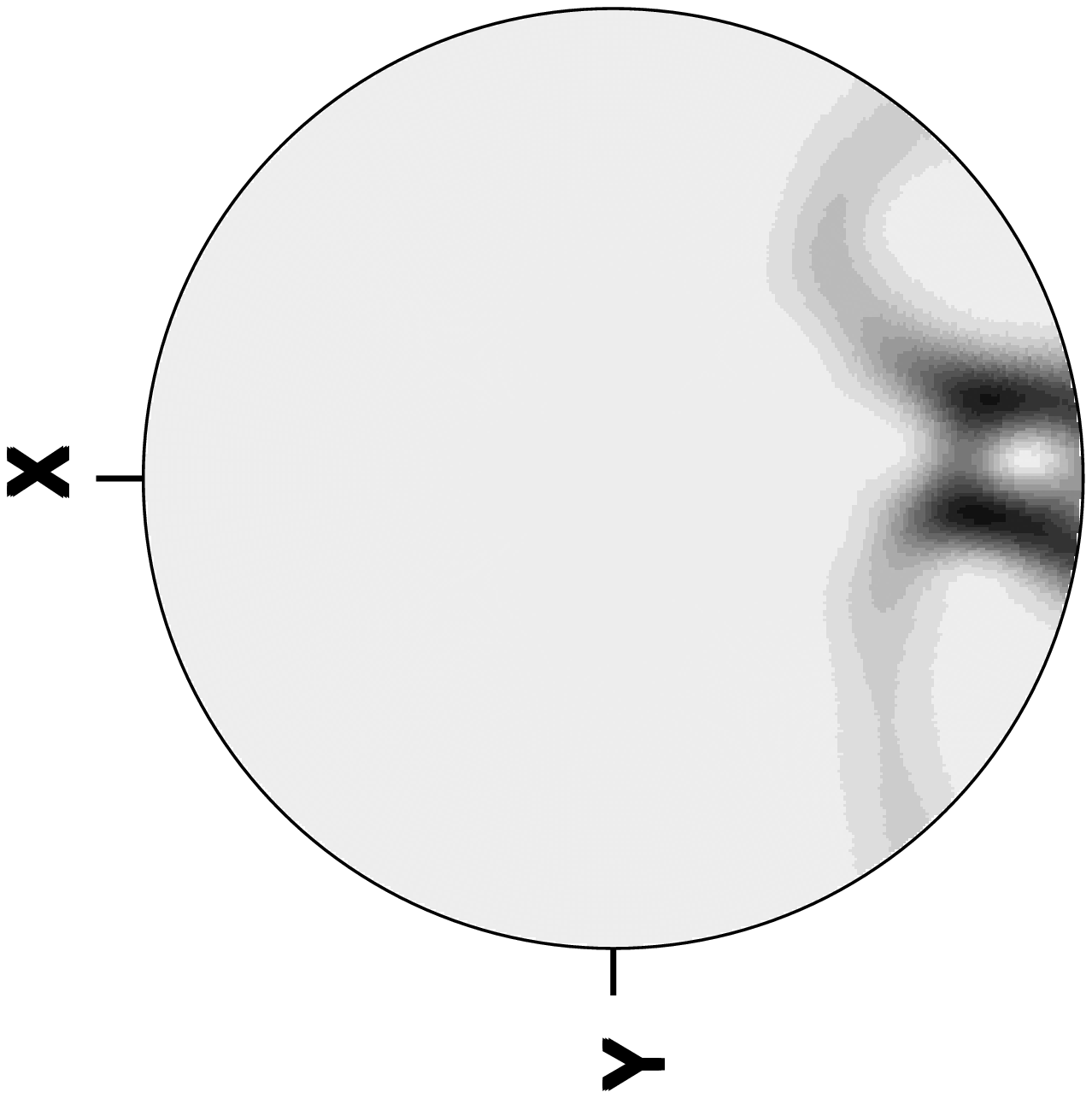}{g}{h}{i}
\caption{\label{fig:Husi1stOpen}
First Channel Opening. Husimi plots of resonances. (a) Wave function of the
open channel out of resonances. It is quantized on the open part of the sphere.
(b): very large resonance which spans all Fig.~\protect\ref{fig:Tau1stOpen}, it
has a large overlap with the open part of the sphere. (c)-(e) three narrow
resonances seen in this figure (in order of decreasing widthes) at
$-8.71991\,10^{-7}$ a.u. (width $1.11\,10^{-12}$ a.u.), $-8.72156\,10^{-7}$
a.u. (width $2.26\,10^{-13}$ a.u.) and $-8.72065\,10^{-7}$ a.u. (width
$3.15\,10^{-14}$ a.u.): they are located in the chaotic region connected to the
open part. (f) the narrowest resonance seen in this figure, as only a one point
glitch at $-8.71780\,10^{-7}$ a.u. (width $2.67\,10^{-15}$ a.u.): located on a
regular island with some overlap with the chaotic sea. (g)-(h) extremely narrow
resonances located in the regular regions: their widths are $2\,10^{-21}$ and
$1.5\,10^{-29}$ a.u.. (i) an extremely narrow resonance located in the second
chaotic sea between them. Its width is $1\,10^{-27}$, intermediate between the
two previous ones, because it is separated from the open part by a regular
region.}
\end{figure}

The Husimi plot of the wave function at the first channel opening  
(see Fig.~\ref{fig:Husi1stOpen}(a)) is located in the open part of the sphere.
Resonances for nearly bound states appear as increments of $\tau$ by one. The
widths of the resonances vary widely. The physical reason for this is easily
displayed by Husimi plots for the resonances observed in 
Fig.~\ref{fig:Tau1stOpen}. The
very broad resonance which spans all of Fig.~\ref{fig:Tau1stOpen} is seen to be
localized in the chaotic region, overlapping widely with the open part of the
sphere: Fig.~\ref{fig:Husi1stOpen}(b). We see three other resonances of
decreasing width, which are also localized in the chaotic sea, but farther from
the open part, as shown in Fig.~\ref{fig:Husi1stOpen}(c)-(e). Note that these
resonances live on the border of an integrable island, and are thus of similar
type as states computed in ref.~\cite{Casati:PRL99-524}.
The state \ref{fig:Husi1stOpen}(f) is even narrower as it lives on or around an
integrable island within this chaotic zone. It is seen as a one point glitch in
Fig.~\ref{fig:Tau1stOpen}. In addition there are much narrower resonances which
cannot be found easily by the previous method. They are localized on or beyond
the large regular region in the middle of the sphere, and have exceedingly
small width, due to an exponentially small overlap of the Husimi function with
the open part of the sphere.

Using the traditional method which looks for a zero value of a determinant we
have been able to locate them by the following trick. We can set artificially a
lower than zero threshold for ionization both in classical and in quantum
mechanics: this would correspond to the existence of a long range screening
potential, which might be due to stray electric fields or a space charge. In
the classical simulation, this corresponds to considering an electron which
after collision has an electronic energy higher than a given negative threshold
$E_\mathrm{thresh}$ as ionized and not returning. In quantum mechanics, since
the only difference between open and closed channels is the replacement of
$\nu_N$ by $-\tau$, this corresponds to performing this replacement for a
finite value of $\nu_N$, which corresponds to this $E_\mathrm{thresh}$. We have
selected such a case: see the negative energy for opening of the first channel
in Fig.~\ref{fig:Tau1stOpen}. If we do not apply this trick, the whole sphere
is closed for the same value of the parameters. We can thus use MQDT 
for closed channels to compute energies and eigenfunctions of all channels. We
select by inspection the wave functions quantized in the regular regions and
run the open channel MQDT with a very small step to locate the
resonances. Thus, we have been able to locate resonances of widths $2\,10^{-21}$
and $1.5\,10^{-29}$ a.u. located respectively in the regular regions near the
equator and the bottom of the sphere, and also a resonance of intermediate
width $1\,10^{-27}$ located in the second chaotic sea between these two regular
regions. In any case we have checked that the Husimi plots for the closed level
and the resonance are essentially equal, which provides a good check of our
interpretation.

But this situation is quite unsatisfactory in the general case of truly
positive energy, where we cannot use this trick to locate such narrow
resonances. Fortunately the method of locating the poles of the $\matr{S}$
matrix in the complex plane indicated at the end of section~\ref{sec:quantum}
is much more powerful. We can use the contour integration to locate the
resonances by dichotomy in the complex plane, and finally use a root searching
algorithm to compute them accurately once they have been approximately located.
To check this method we have determined for example that the resonance at
$-8.71991\,10^{-7}$ a.u. (Figs.~\ref{fig:Tau1stOpen} and
\ref{fig:Husi1stOpen}(c)) corresponds to a pole at complex energy value
$1.11\,10^{-12}$ a.u., consistent with the previous method, and that the
resonance at $-8.71780\,10^{-7}$, seen as a one point glitch in
Fig.~\ref{fig:Tau1stOpen} corresponds to a pole at complex energy value of
$2.67\,10^{-15}$ a.u. (Fig.~\ref{fig:Husi1stOpen}(f)).

\subsection{Many Open Channels}

We now increase the total energy to $\sim 4.50\,10^{-7}$ a.u., such that 
slightly more
than a half of the sphere is open. Ionization occurs in a cap around $OX$ with
$\cos\alpha\gtrsim -0.1$. We select two values of the coupling, a weak
$\mu_{50}=0.4$ and a strong $\mu_{50}=9.81$. The remaining parameters are 
chosen as $L=50$,
$J=100$, $2B=10^{-10}\,\mathrm{a.u.}$ as before.

\subsubsection{Weak coupling}

\begin{figure}[!htb]
\centering
\setlength{\figwidth}{\textwidth}
\TwoUnRelated{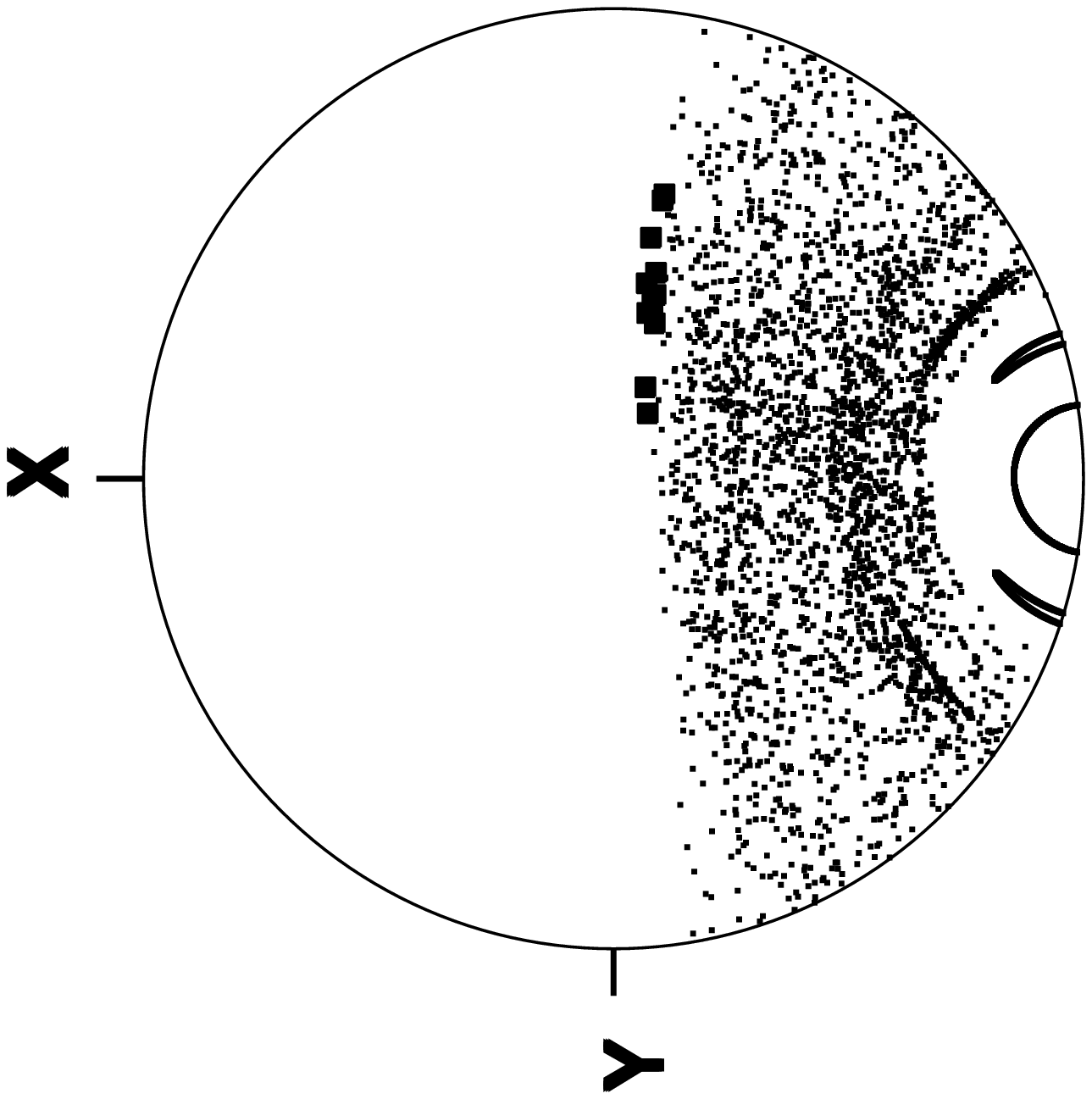}{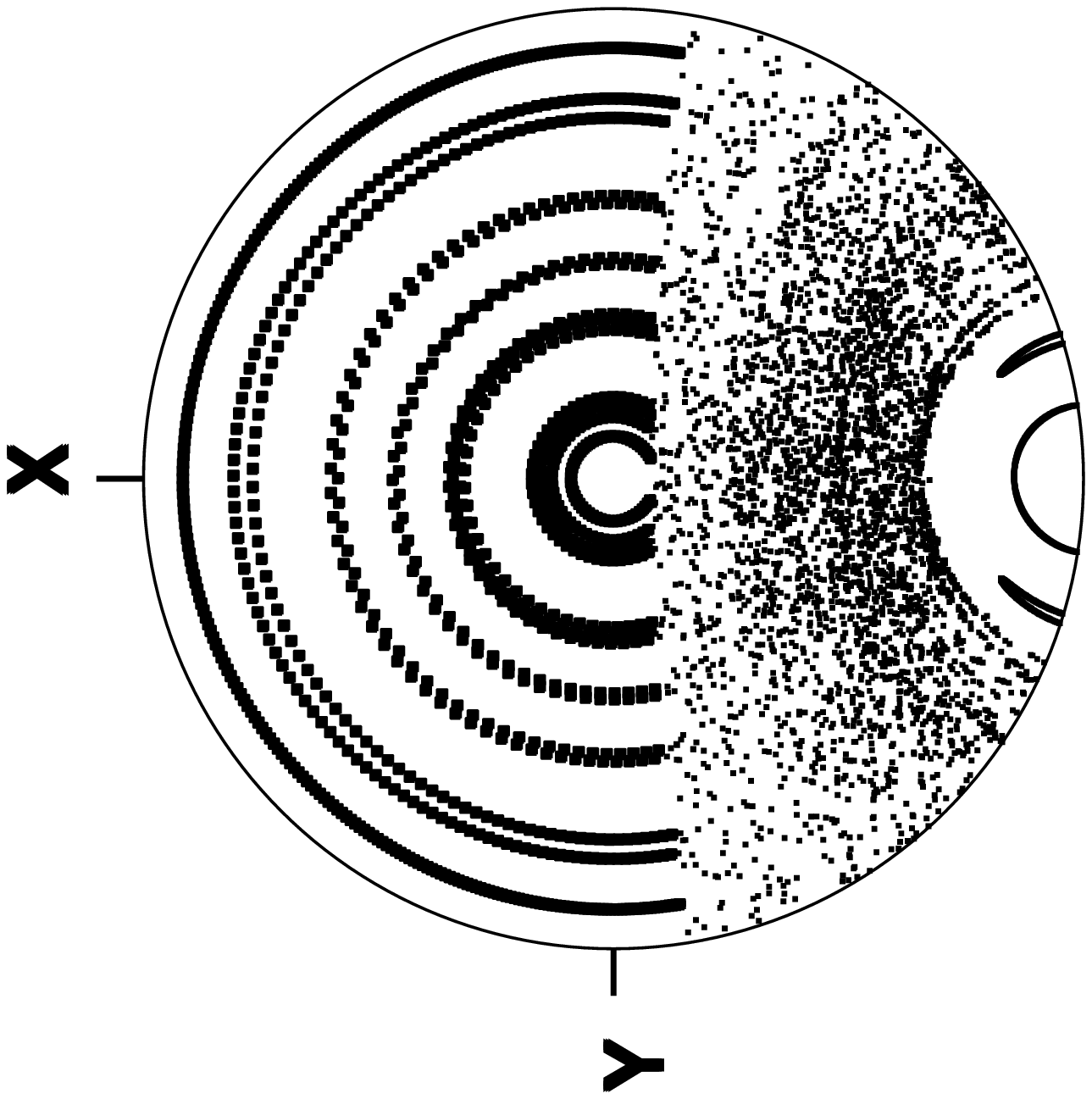}{a}{b}
\caption{\label{fig:PoinJungWeak}
Weak coupling. (a) Ordinary Poincar\'e Map. Iteration stops when the electron
ionizes (black squares)
(b) Combined Poincar\'e Jung map. When the electron ionizes iteration is
continued with the Jung recipe (Fig.~\ref{fig:Principle}(b)). It follows
circles on the open part of the sphere until it enters back into the closed
part.}
\end{figure}

Fig.~\ref{fig:PoinJungWeak}(a) displays the ordinary PM. Iteration stops when
the electron ionizes (black squares) and a new trajectory is launched in the
closed part of the sphere. Fig.~\ref{fig:PoinJungWeak}(b) shows the combined
PM+JSM. Except for the regular islands near the $-OX$ axis, it was obtained by
running a single trajectory, i.e. using the PM for negative electron energies,
and the JSM for positive energies. When the system is in the bound chaotic part
it diffuses slowly towards the back or the front of the sphere. This diffusion
mechanism was studied carefully in ref.~\cite{Leyvraz:PD01-169}. When it
ionizes, the JSM generates a set of one bounce events, tracing a semi
circle around $OZ$ in the open region,
until it enters the bound region at the end of the semi circle and starts anew
to diffuse. Due to the small value of $\mu$ the number of steps needed to 
complete the semi circle is very
large. The trajectory thus behaves like one injected very close to a parabolic
manifold.

\begin{figure}[!htb]
\centering
\includegraphics*[angle=270,width=\figwidth]{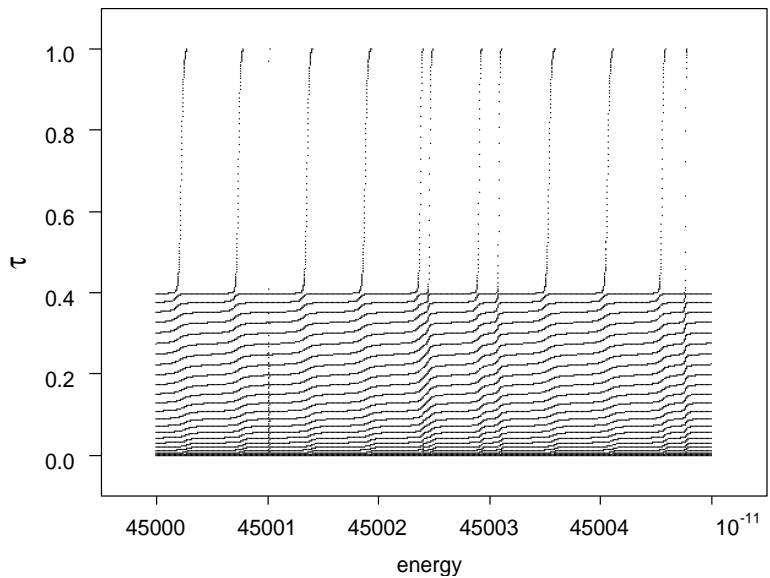}
\caption{\label{fig:Tauwk}
Phase shifts. Weak coupling $\mu_{50}=0.4$}
\end{figure}

\begin{figure}[!htb]
\centering
\TwoUnRelated{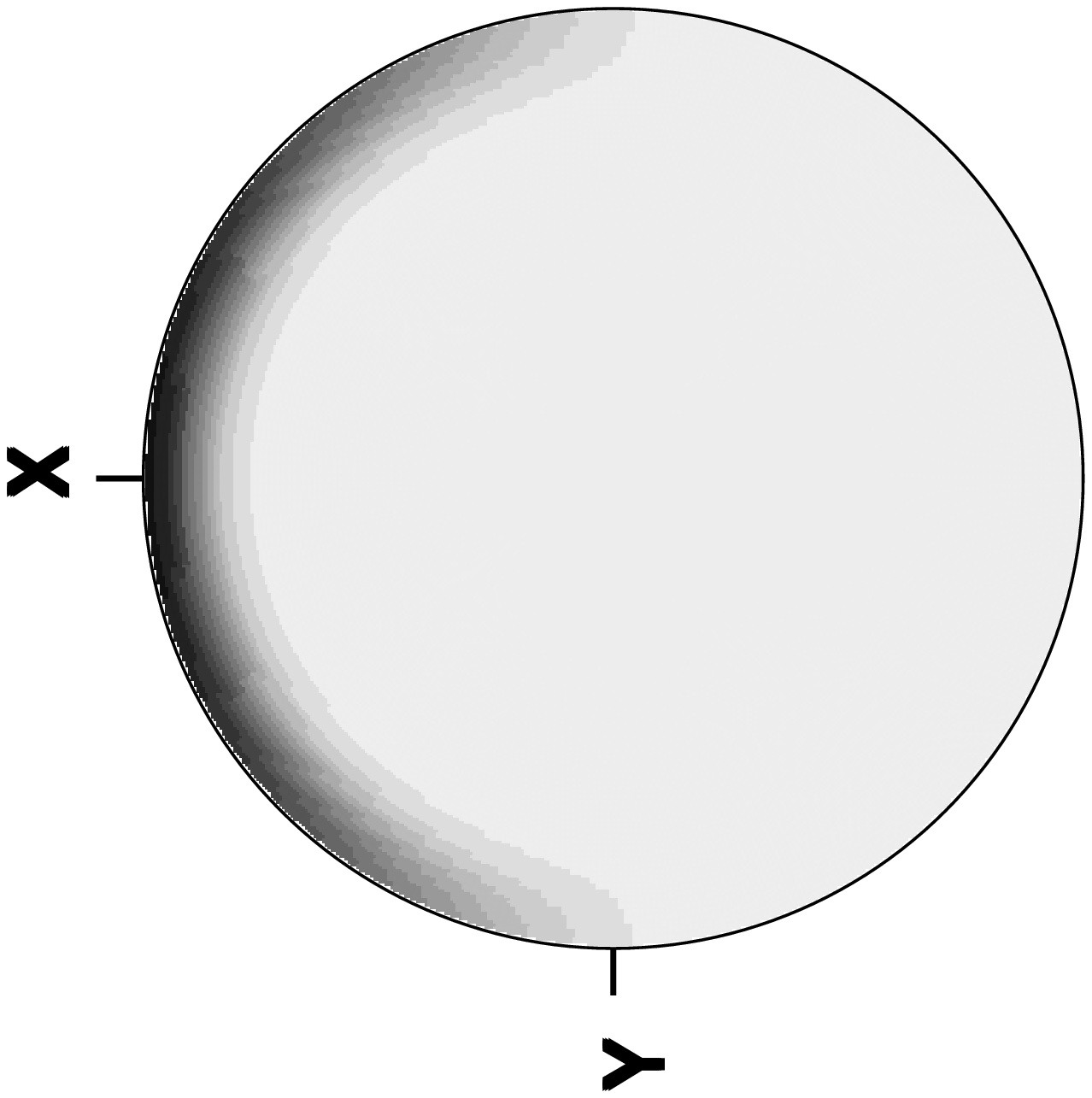}{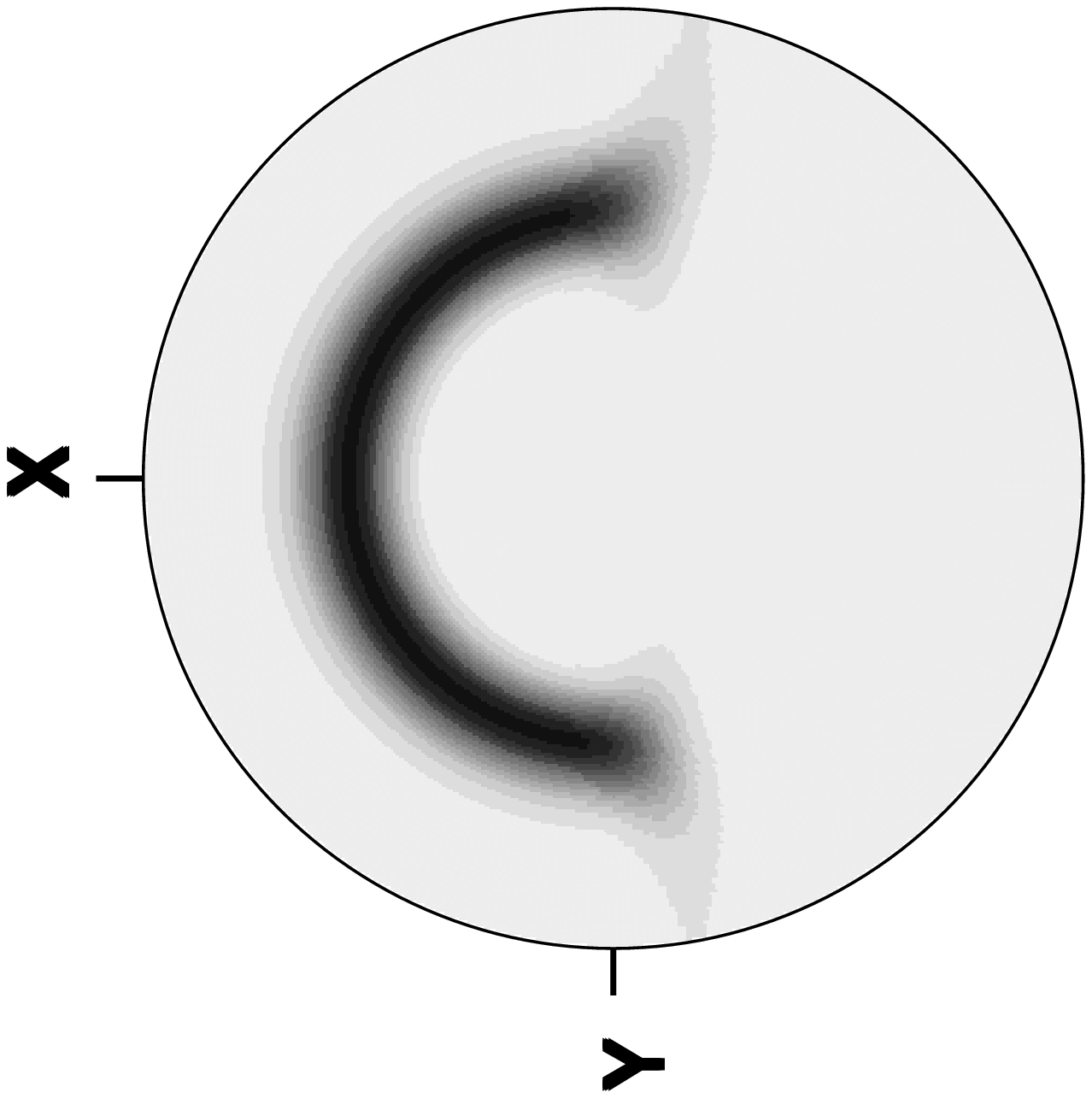}{a}{b}
\TwoUnRelated{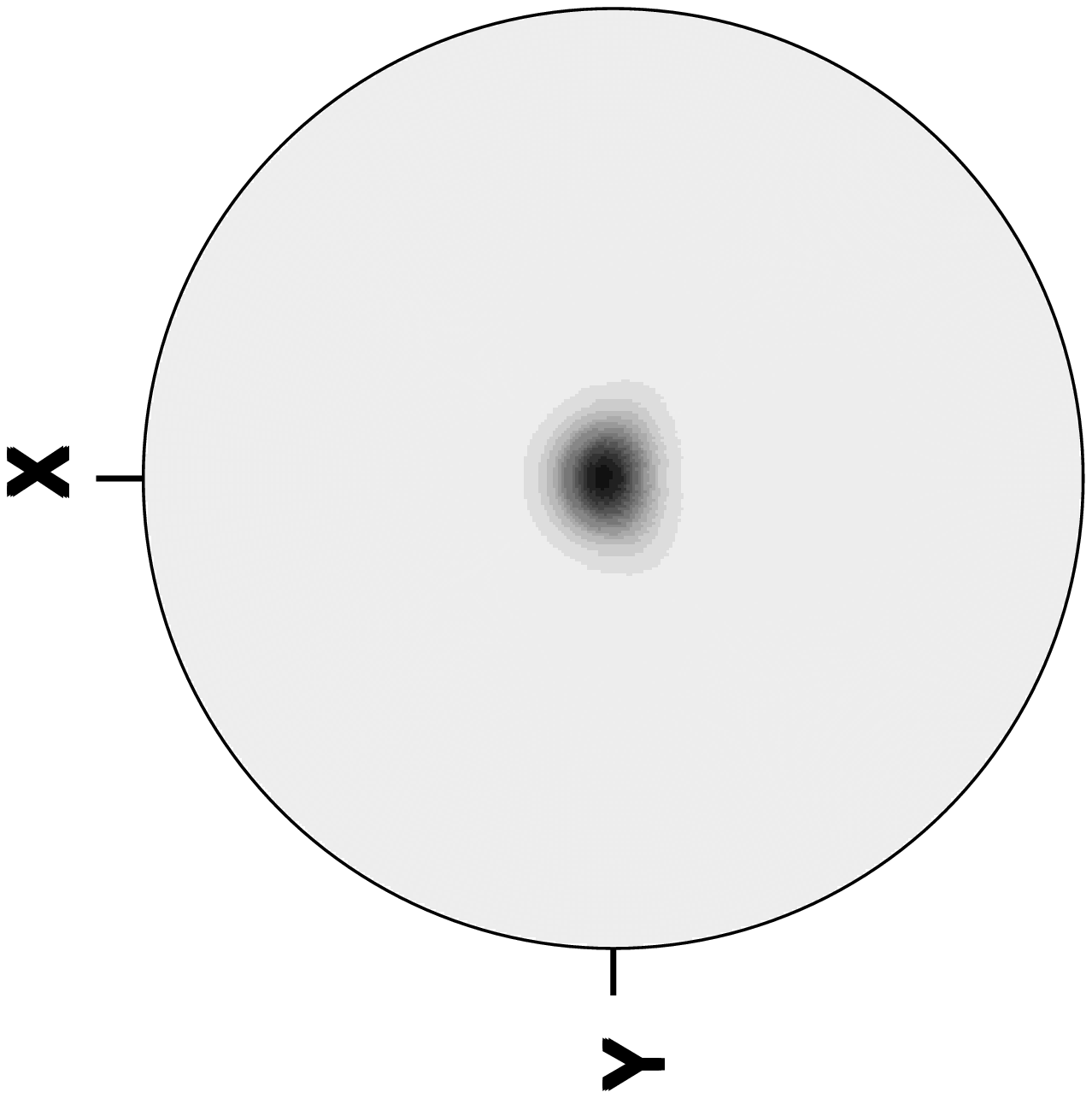}{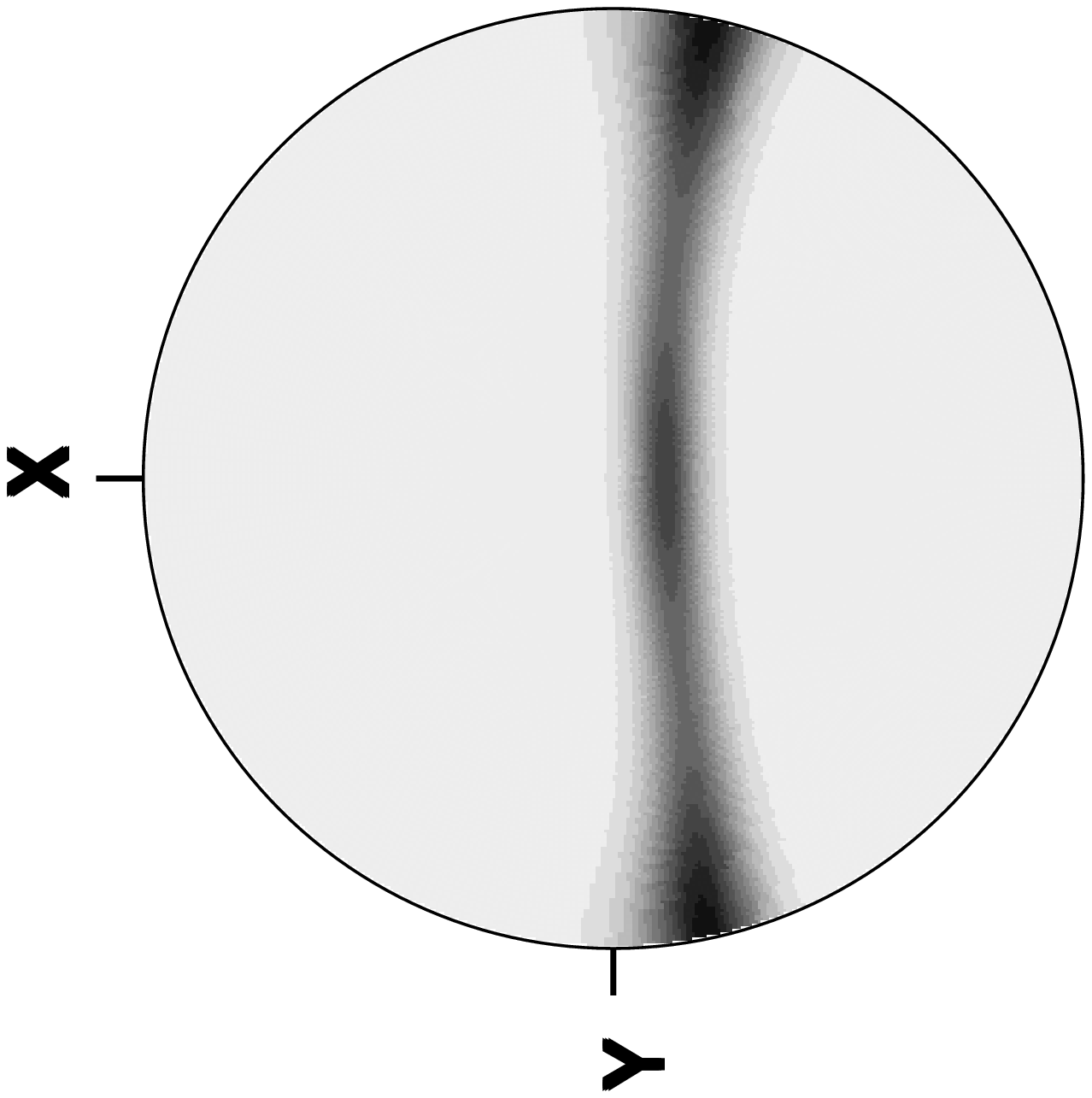}{c}{d}
\caption{\label{fig:JungMapWeak}
Husimi plot of the MQDT eigenfunctions. Weak coupling. (a-c) Open channels are
quantized along semi-circles which pertain to the Jung Map. (d) All resonances
seen in Fig.~\protect\ref{fig:Tauwk} are quantized near the border of the
separation between open and closed parts of the sphere.}
\end{figure}

Quantum mechanically Fig.~\ref{fig:Tauwk} displays the phase shifts for the
eigenchannels. There are 23 channels open at this energy. We readily see that
it displays a set of resonances.

Some Husimi plots of the eigenfunctions of the open channels are displayed in
Fig.~\ref{fig:JungMapWeak}(a-c). They follow closely the trajectories of the
JSM. In a previous letter we noted \cite{Dietz:JPA96-95}, that parabolic
manifolds in a JSM may lead to eigenstates of the $S$ matrix localized near
this manifold (see also below). As we mentioned above the semicircles behave
quite similarly, and as we do find states localized near these semi circles
this again proves the applicability of the JSM to the description of  
the ionizing part of the
spectrum in a semiclassical limit. Fig.~\ref{fig:JungMapWeak}(d) displays the
Husimi plot of the first resonance seen in Fig.~\ref{fig:Tauwk}. It is
quantized along the border of the regions of bound and unbound motion, 
yielding a large overlap
with the open part resulting in a  large width of the resonance. All 
resonances seen
in Fig.~\ref{fig:Tauwk} have similar Husimi plots. Functions lying further from
the open channels would again correspond to extremely narrow resonances, which
cannot easily be seen in the eigenphases with the present resolution.

\subsubsection{Strong coupling}

\begin{figure}[!htb]
\centering
\setlength{\figwidth}{\textwidth}
\TwoByTwo{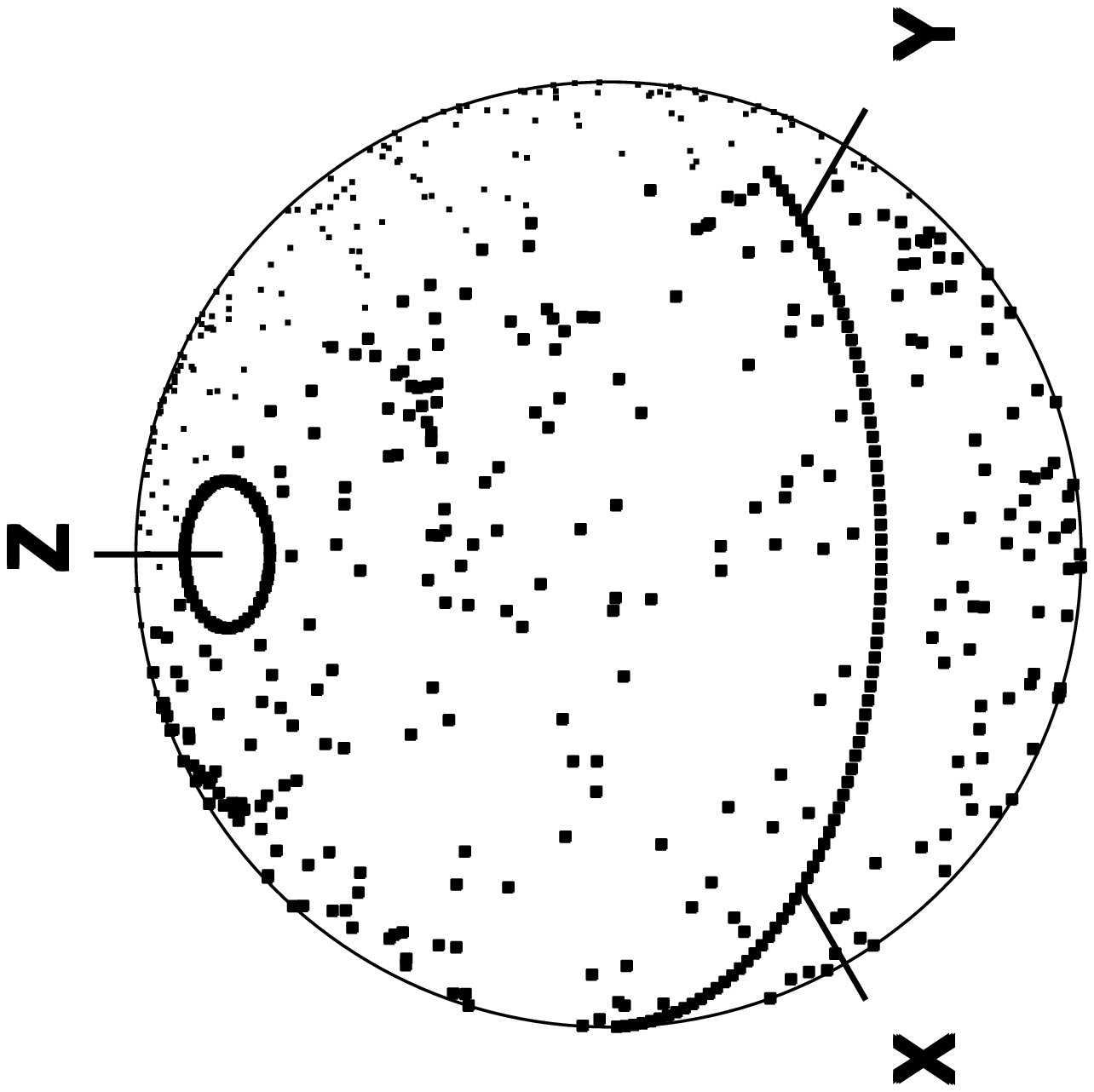}{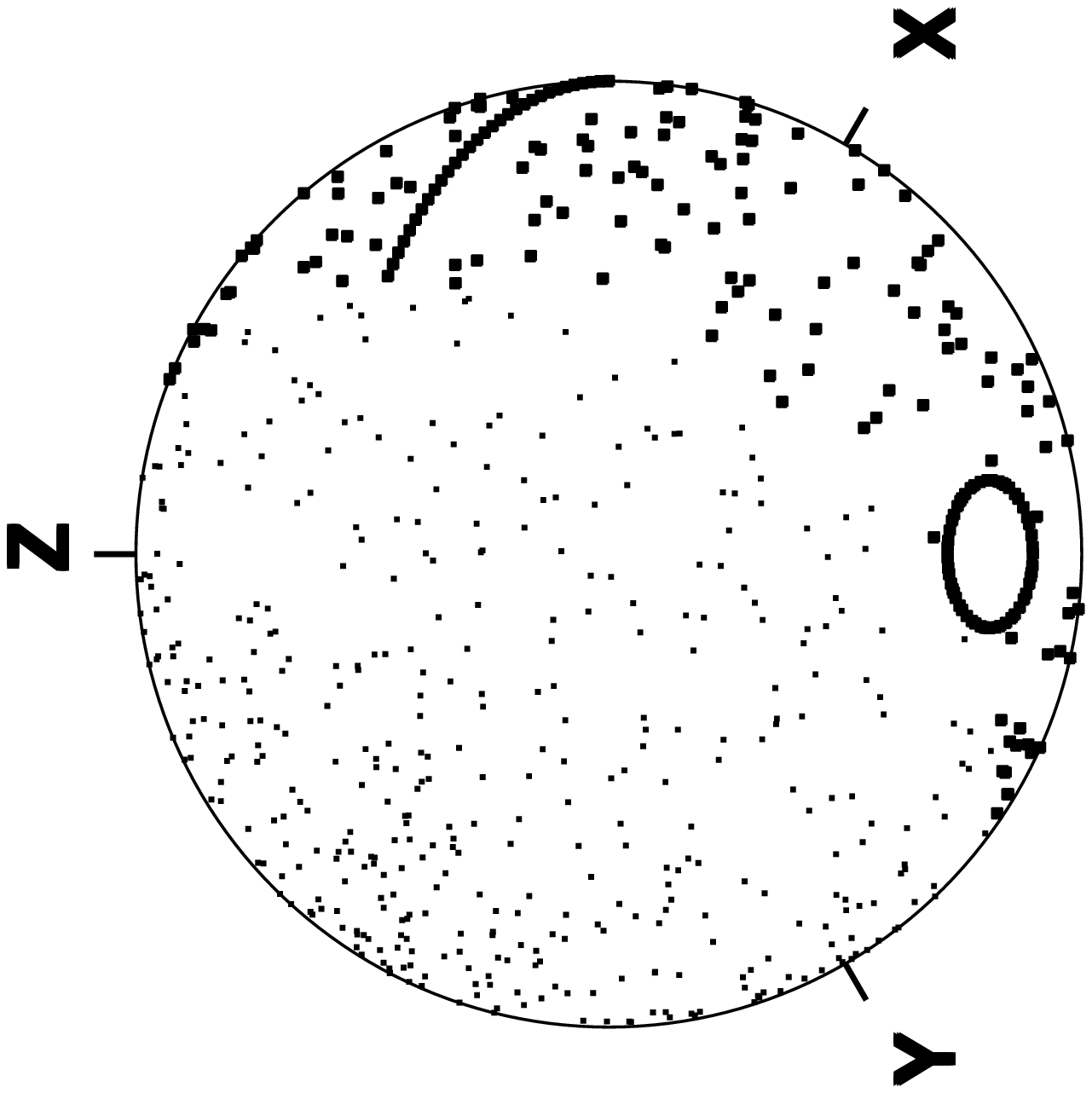}{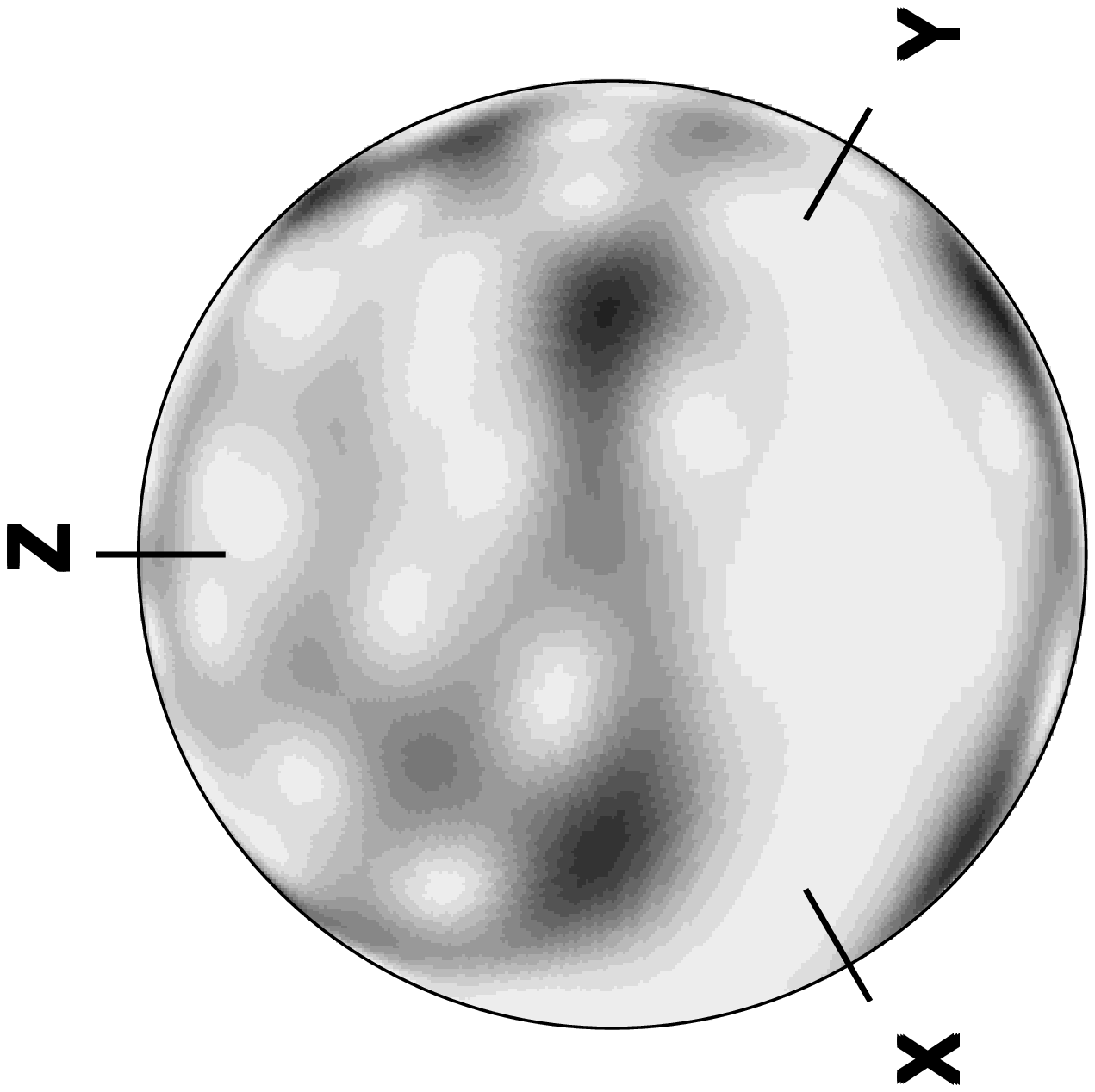}{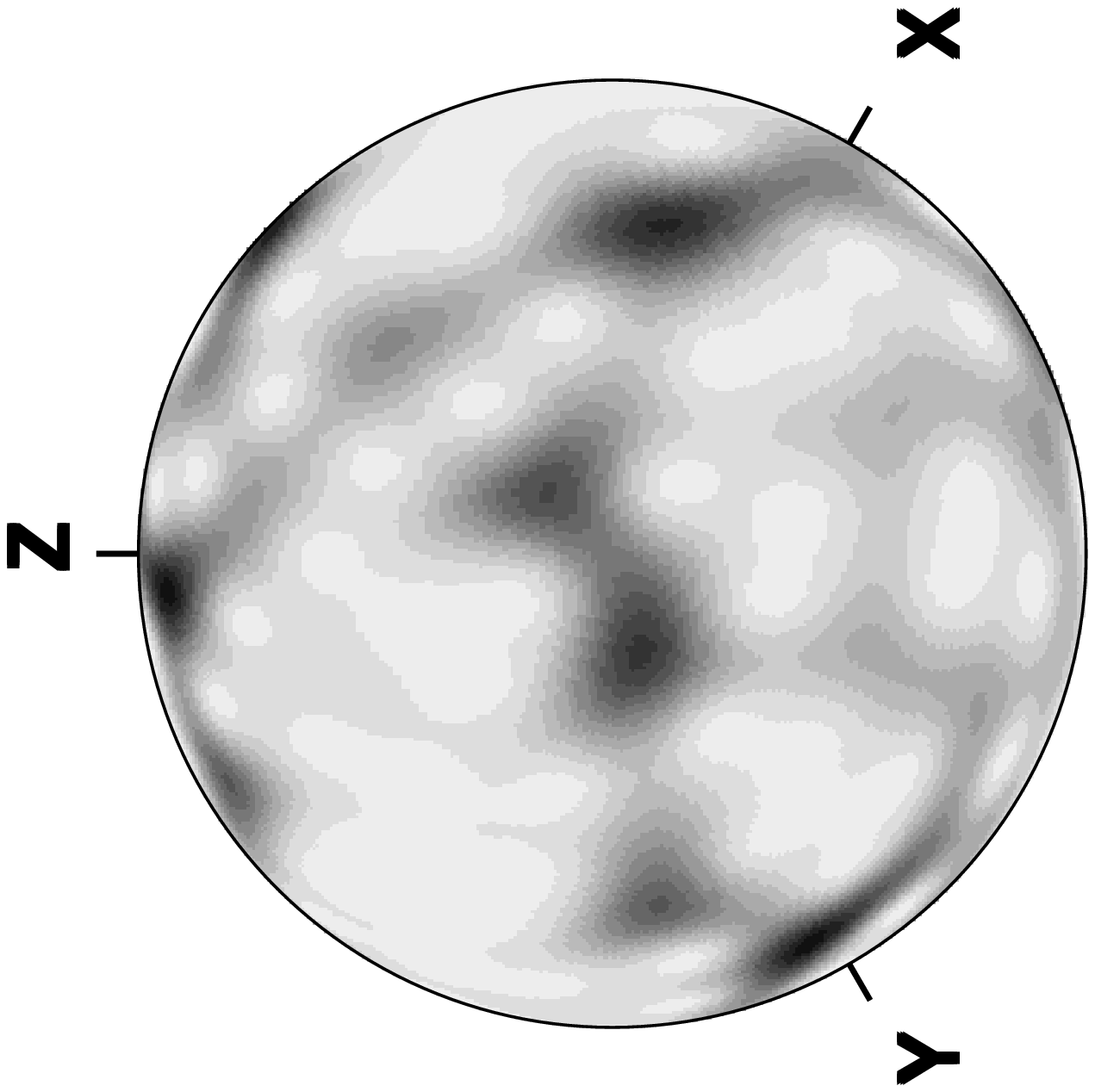}{a}{b}
\TwoByTwo{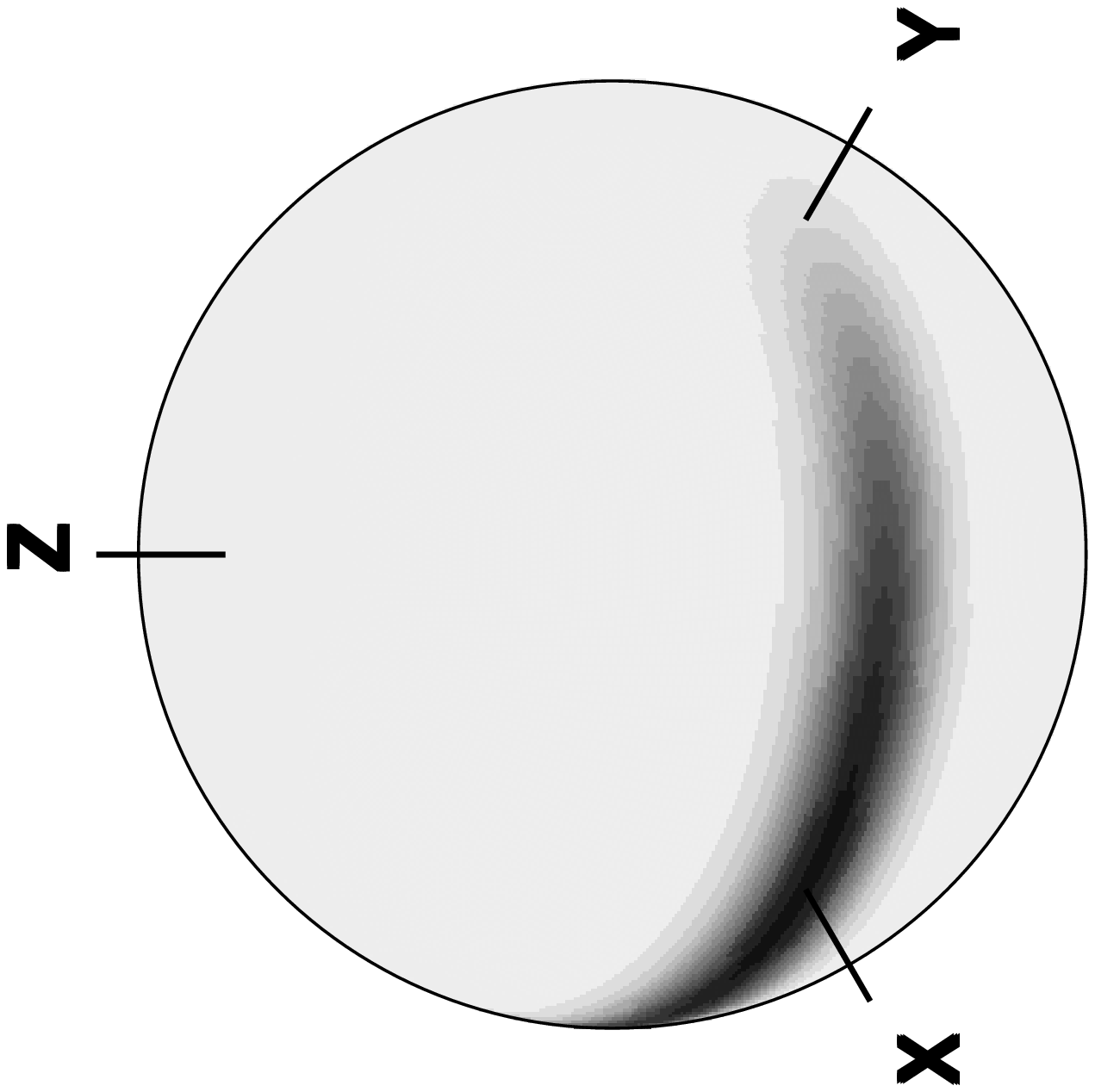}{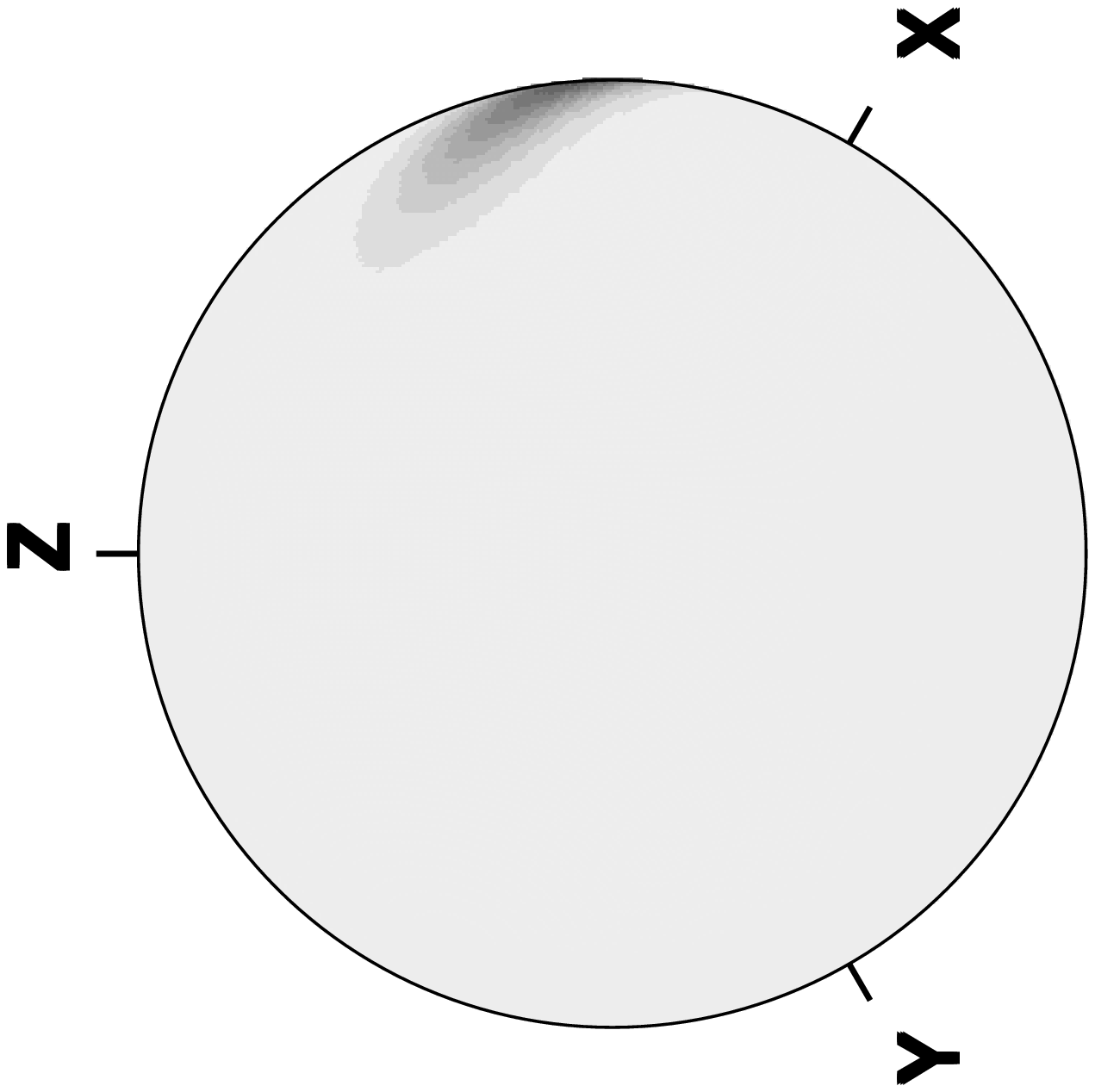}{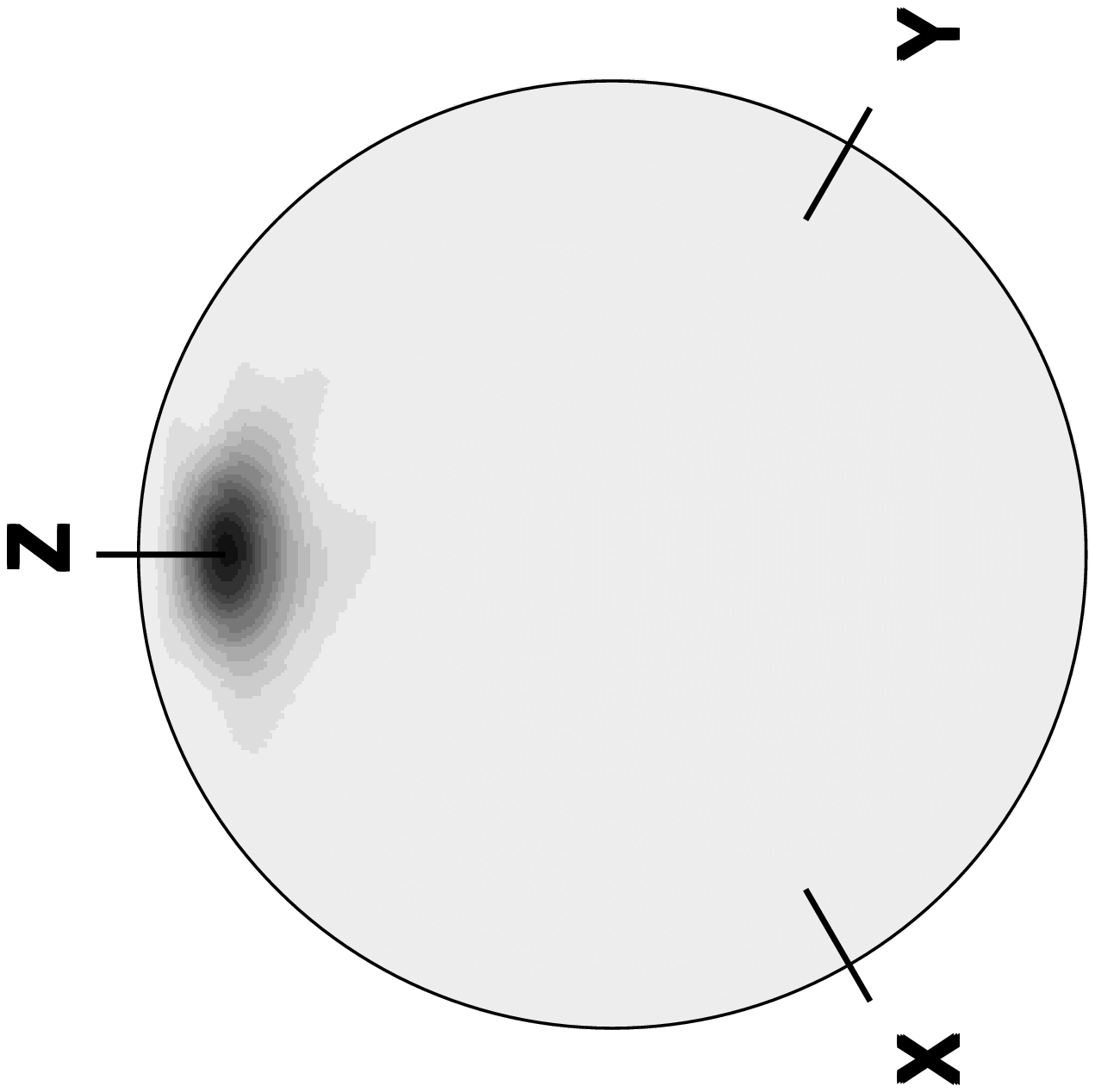}{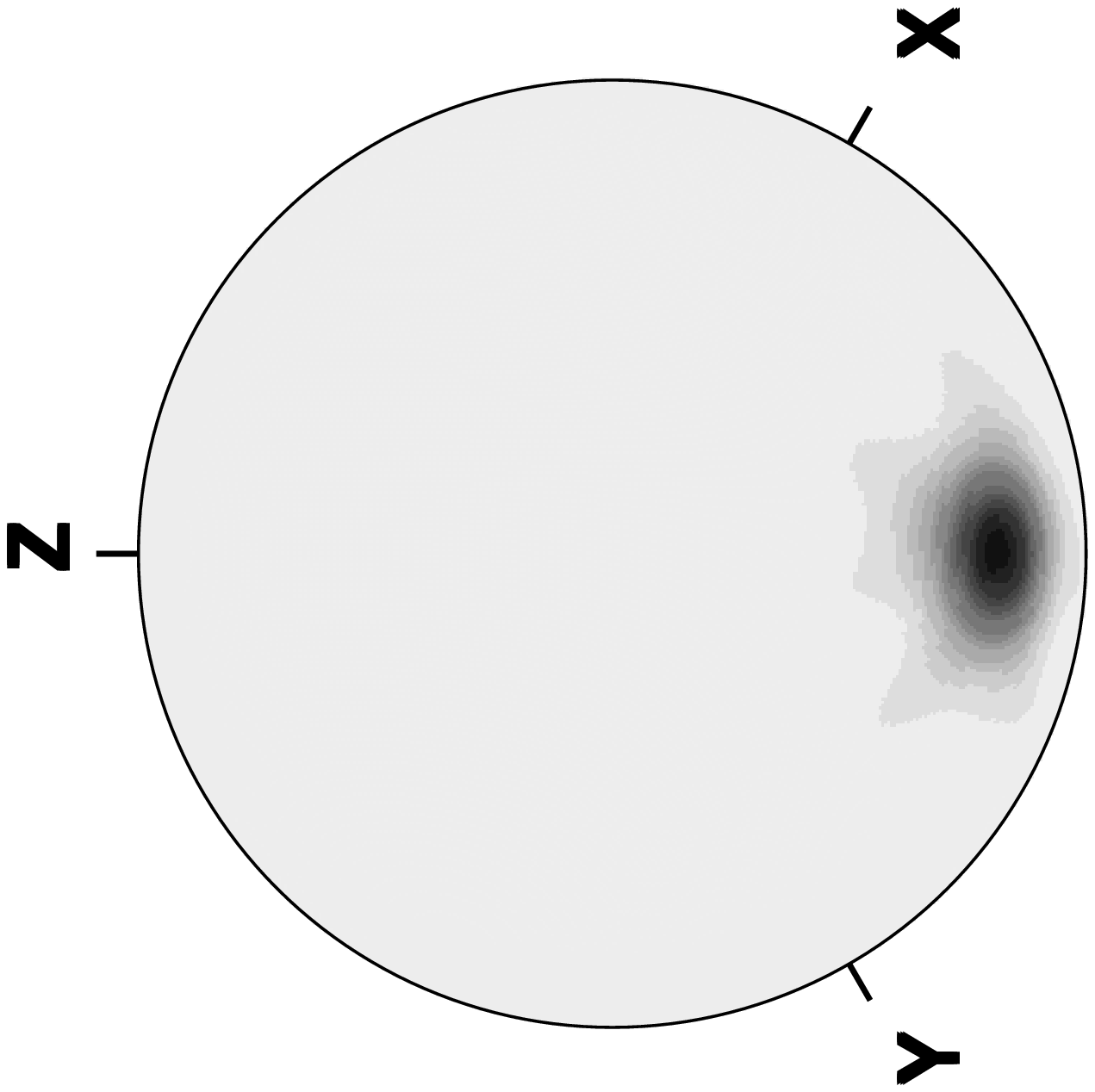}{c}{d}
\caption{\label{fig:JungMapStrong}
Combination of PM and JSM. Strong coupling. Left: top view, right: bottom view.
We do not use the stereographic projection as in
Figs.~\ref{fig:Poin1stOpen},~\ref{fig:PoinJungWeak} because the parabolic
manifold is badly seen with it.
(a) Classical map. Small dots: PM, larger dots: JSM. The open part is slightly
larger  than half a sphere. The PM has only a very small island of stability at
the back, too small to capture a quantum state with this value of $\hbar$. The
JSM has two regular islands around $\pm OZ$, and the parabolic manifold of
fixed points on the equatorial plane. All the remaining points lie in a single
chaotic sea obtained with only one starting point, iterated by the PM when
bound, and by the JSM when ionized.
(b) Husimi plot of a typical arbitrary state: it extends randomly over the
whole sphere
(c) Husimi plot of a quantum state quantized on the parabolic manifold.
(d) Husimi plot of a quantum state quantized on the $\pm OZ$ cap.}
\end{figure}

\begin{figure}[!htb]
\centering
\includegraphics*[angle=270,width=\figwidth]{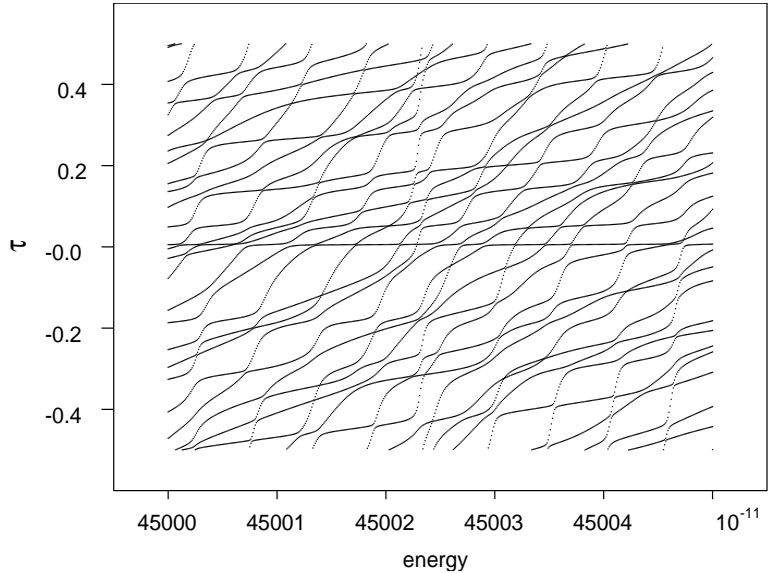}
\caption{\label{fig:Taustr}
Eigenphases as a function of energy. Strong coupling $\mu_{50}=9.81$}
\end{figure}

Choosing as coupling strength $\mu_{50}=9.81$ the open part slightly greater 
than
half a sphere. Fig.~\ref{fig:JungMapStrong}(a) shows the combined PM + JSM.
Except for the fixed points described below all points of this combined PM +
JSM pertain to a single trajectory. Due to the large value of the coupling this
trajectory jumps randomly in and out of the bounded part within a few
steps. The combined map is chaotic everywhere in this part.
Fig.~\ref{fig:Taustr} displays the phase shifts. One sees a lot of avoided
crossings between the phase curves, signaling a chaotic situation. A typical
Husimi plot of an arbitrary wave function is shown in
Fig.~\ref{fig:JungMapStrong}(b).

The important fixed points of this map (apart from isolated unstable fixed
points embedded in the chaotic sea) are
\begin{enumerate}
\renewcommand{\theenumi}{{\roman{enumi}}}
  \item the $-OX$ axis in the bound part (PM) surrounded by a very small island
of stability, too small with the present value of $\hbar$ to capture a single
quantum state.
  \item the $\pm OZ$ axis in the open part surrounded by a larger regular
island of the JSM. These are fixed points because the collision induces a
rotation around $OZ$, entirely embedded in the open part (JSM).
  \item all points in the equatorial plane in the open part (JSM). An electron
entering the collision region with $\theta_L=\pi/2$ and arbitrary $\varphi_L$,
up to the restriction that the electron energy should be positive, will emerge
from the collision region with the same angular momentum direction because the
increment $\delta\varphi_L = K\cos\theta_L$ equals zero. We thus have a
parabolic manifold of fixed points.
   \item if $\vert K \vert$ is chosen larger than $n\cdot 2\pi$, that is
$\mu_{L} > \frac{n L}{2}$ we would also find lines mapped onto themselves at
angles $\theta_L$ corresponding to $K\cos\theta_L=0, \pm 2\pi,\cdots,\pm n
\cdot 2\pi$
\end{enumerate}

In particular the parabolic manifold is embedded in the chaotic sea. A nearly
horizontal curve may be observed in the middle of Fig.~\ref{fig:Taustr}. 
It corresponds
to the fixed points of this parabolic manifold in the equatorial plane as shown
by Fig.~\ref{fig:JungMapStrong}(c)(see also ref.~\cite{Dietz:JPA96-95}). For
e.g. the stadium billiard, such manifolds of fixed points embedded in a chaotic
sea have significant quantum effects. There they give rise to the so called
bouncing ball states which in turn lead to deviations of the spectral
fluctuation properties in the eigenvalue spectrum from the random matrix
results \cite{Graf:PRL92-1296}.
We expect similar quantum effects of the parabolic manifolds on statistical
properties of the eigenphases of the scattering matrix. When considering
eigenphases of the scattering matrix as a function of the energy $E$ or the
coupling strength $\mu_L$, we find eigenphases which merely change.
The other fixed point near the $\pm OZ$ poles, surrounded by a significant
island of stability in the JSM, is less conspicuous in Fig.~\ref{fig:Taustr}.
It corresponds to a horizontal line near $\tau=-0.3$, barely visible due to
strong avoided crossings. It nevertheless traps a quantum state as seen in
Fig.~\ref{fig:JungMapStrong}(d).

\section{\label{sec:SemiClassics} Comparison between semi-classical and quantum
evolution on the surface of section}

\subsection{\label{sec:SemiClassics-principle} Principle}

We want to compare our results with the original work of Bogomolny
\cite{Bogomolny:CAMP90-67,Bogomolny:N92-805,Bogomolny:C92-5}, which was based
on the semi-classical formula
\begin{equation}\label{eq:Tphiphi'}
T(\varphi, \varphi^\prime) = \sum_\textrm{cl.tr.} \sqrt{\frac{\partial^2
S(\varphi,\varphi^\prime) / \partial\varphi \partial\varphi^\prime} {- 2 \pi
\rmi \hbar}} \exp \left( \frac{\rmi}{\hbar} S(\varphi,\varphi^\prime) \right),
\end{equation}
where, as in the remainder of this section, we use the notations of the review
paper of W.H. Miller \cite{Miller:ACP74-69}. In particular the $\rmi$ and the
signs in the argument of the square root function take care of the Maslov 
index.

We first notice that the simple minded way of using this equation, computing $S
= \int p \rmd q$ using the formulae obtained in the previous classical part of
this paper and the previous ones does not work, because our
``classical'' molecular reference frame is a moving frame, implying that
$L_{Z_C}$ is not conjugate to $\varphi_L^C$. Indeed consider first the
collision step. The change of orientation of $\vec{L}$ in this frame entails,
by conservation of angular momentum, a change of the orientation of $\vec{N}$,
and thus of the orientation of this frame, by an angle called
$\delta\varphi_L^\prime$ in section~\ref{sec:classical} and $\varphi_N^Q$ in
Appendix~\ref{sec:ClassQuantTransform}. This extra rotation of the frame
implies that $L_{Z_C}$ is no longer the generator of rotations of $\varphi_L^C$
around the $OZ_C$ axis. For the collision step this problem can be solved by
using the ``quantum'' molecular reference frame, which is fixed, since the
collision is taken as instantaneous, and that this frame is obtained from
the laboratory frame by a rotation with fixed Euler angles 
$\varphi_M, \theta_M,
0$ (see Appendix~\ref{sec:QCFrames}). That this analysis is correct is shown by
noting that the transformation produced by the collision in the ``quantum''
reference frame $\delta\varphi_L^Q = K \cos\theta_L^Q$ is canonical (the
exterior differential product $\rmd L_{Z_Q} \wedge \rmd \varphi_L^Q$ is
conserved), while the corresponding for the ``classical'' frame quantities 
is not.
Moreover the transformation from the ``quantum'' to the ``classical'' molecular
reference frame is not canonical as shown in
Appendix~\ref{sec:ClassQuantTransform}, where we have computed the momentum
conjugate to $\varphi_L^C$, such that this transformation becomes canonical
(eq.~\ref{eq:IC}). Nevertheless, since our goal is physical evidence we prefer
to stick to our previous practice of using $L_{Z_C}, \varphi_L^C$. The only
drawback of using non canonical coordinates is that the map is not area 
preserving on the sphere, a minor effect for $L/J=1/5$ or $1/2$ as chosen 
in this paper.

For the collision step we thus can solve this problem by using the ``quantum''
molecular frame, but the problem reappears for the Coulomb step, where this
frame is also moving. We could compute everything in the laboratory frame, but
it is more meaningful to use the ``quantum'' molecular frame for the collision
step, the laboratory frame for the Coulomb step, and two semiclassical formulae
for the change between these coordinate frames. This procedure follows closely 
the purely
quantum formulation, giving a semiclassical approximation for each of the steps
which enter in the discussion of section~\ref{sec:quantum}, as shown in
Table~\ref{table:PM}. It provides direct evidence that MQDT indeed is the
exact quantization of the Bogomolny formulation, in particular that
eqs.~(\ref{eq:det=1}, \ref{eq:1mEET}) are the exact counterpart of the semi
classical eq.~(1.20) of ref.~\cite{Bogomolny:N92-805}, remembering that $\matE$
describes half a turn, while $T$ describes a full turn.

As Poincar\'e section we select in coordinate space the intersection 
at radius $r_0$
with outgoing trajectories, i.e. an ``after collision'' Poincar\'e map. We thus
write:
\begin{equation}\label{eq:TfromUs}
T=U^{c} U^{QL} U^{f} U^{LQ}
\end{equation}
where from right to left the $U^i$'s indicate the unitary semiclassical 
operators for the transformation 
from the ``quantum'' molecular reference frame to the laboratory frame, free
motion in the laboratory frame, back to the ``quantum'' molecular frame and
collision in the ``quantum'' frame.
All these operators are built with $F_2$ / $F_3$ kind of generating functions,
i.e. they are of the form
\begin{eqnarray}\label{eq:Upsi}
 (U \psi)(\varphi) = \int\!\!\!\int && \sqrt{\frac{\partial^2
F_3(I^\prime,\varphi) / \partial I^\prime \partial\varphi} {2\pi\rmi\hbar}}
\exp\left(\frac{\rmi}{\hbar} \left(F_3(I^\prime,\varphi)\right)\right)
\nonumber\\
 & \times & \sqrt{\frac{-1}{2\pi\rmi\hbar}} \exp(-\frac{\rmi}{\hbar} I^\prime
\varphi^\prime) \nonumber\\
& \times & \psi(\varphi^\prime) \, \rmd I^\prime \, \rmd\varphi^\prime
\end{eqnarray}
where $I^\prime$ is the momentum conjugate to the coordinate $\varphi^\prime$
and the resulting integral is evaluated with the stationary phase
approximation. This generating function is of type $F_3$ as 
it is a function of the old momentum $I^\prime$ and the new coordinate 
$\varphi$. An
auxiliary Fourier transform is used to go from old coordinate $\varphi\prime$
to old momentum $I^\prime$. Evaluating the inner $\rmd I^\prime$
integral by the stationary phase approximation amounts to computing an 
$F_1$ type
generating function, and inverting the order of the  Fourier transform and
of the $F$ integral amounts to defining an $F_2$ generating function. Note 
that we have defined by this double integral a $U$ operator which goes from
coordinate $\varphi^\prime$ to coordinate $\varphi$ representation, while the
$F_3$ integral on its own gives an operator from the old momentum $I^\prime$ to
the new coordinate $\varphi$. Hence, we can compute eq.~(\ref{eq:TfromUs}) by
performing four integrations instead of eight by alternating between the 
coordinate
and the momentum representations. We can compute with the
same procedure a momentum coordinate $T_{\Lambda\Lambda^\prime}$ matrix 
by simply introducing two Fourier Transforms to the left and the right of the 
operator $T$. As momenta are
quantized $T_{\Lambda\Lambda^\prime}$ depends on discrete variables, while 
$T(\varphi,\varphi^\prime)$ is a function of continuous arguments
If there are no singularities all these possibilities give the
same result. We shall use them freely in order to obtain generating functions
which we
can compute analytically and we shall see in section~\ref{sec:singularities}
that this choice offers possibilities to circumvent singularity problems.

\subsection{\label{sec:computations} Computations}

First notice that the conjugate pair of coordinates in the ``quantum''
molecular frame is the known $\{L_{Z_Q},\varphi_L^Q\}$ pair. The momentum in
the laboratory frame is the modulus $N$ of $\vec N$.
Intuitively the conjugate angle is the angle $\varphi_N$ of rotation
of the core axis $\unitvec{M}$ around $\vec N$, referred to a reference
axis which only can be $\vec{J}$. Explicit formulae for computing the
transformation are given in Appendix~\ref{sec:QuantLaboTransform}, where we also
show that the transformation from $\{L_{Z_Q}, \varphi_L^Q\}$ to $\{N,
\varphi_N\}$ is indeed canonical, justifying our intuition concerning the 
definition
of $\varphi_N$.

The computation of the $F_3$ generating functions for the collision and the
free rotation steps follows the same scheme. In both cases the momentum is
conserved, while the angle is increased by an amount which is a function of the
corresponding momentum $I$ only $\delta\varphi = f(I)$, respectively
$\delta\varphi_L^Q=K L_{Z_Q}$ and $\delta\varphi_N = -\delta\beta = +2\pi
T_e/T_N$ (the free rotation is direct in the laboratory frame and seen with 
inverse orientation
in the ``classical'' molecular frame), which is a function of $N$ only for a
given total energy $E$. Using the equations defining $F_3(I^\prime, \varphi)$,
namely $\varphi^\prime=\partial F_3 / \partial I^\prime$ and $I = \partial F_3
/ \partial \varphi$ we obtain
\begin{equation}
F_3(I^\prime,\varphi)=\varphi I^\prime - \int f(I^\prime) \rmd I^\prime
\end{equation}
i.e. using \cite[eqs.~(A11,A12)]{Lombardi:JCP88-3479}:
\begin{eqnarray}
F_3(L_{Z_Q}^\prime,\varphi_L^Q) &=& L_{Z_Q}^\prime \varphi_L^Q - K \frac{
{L_{Z_Q}^\prime}^2} {2 L} = L_{Z_Q}^\prime \varphi_L^Q + 2 \pi
\mu_{L_{Z_Q}^\prime} \nonumber\\
F_3(N^\prime,\varphi_N) &=& N^\prime \varphi_N + 2 \pi \nu_{N^\prime}
\end{eqnarray}

The computation of the $F_2$ or $F_3(L_{Z_Q},\varphi_N)$ is more tricky. First,
whether it is a $F_2$ or $F_3$ generating function depends on which is the
original and the new frame, and we need both transformations. To go in the
opposite direction we use the unitarity of the operator $U$. We start with the
following formula which expresses $N$ as a function of the independent
variables:
\begin{equation}
N = \left\vert \sqrt{J^2-(L_{Z_Q}/\sin\varphi_N)^2} \pm
\sqrt{L^2-(L_{Z_Q}/\sin\varphi_N)^2} \right\vert
\end{equation}

This formula can be computed analytically by eliminating $\varphi_L^Q$ between
eqs.~(\ref{eq:phiN}) and (\ref{eq:phiLQ}) and using eq.~(\ref{eq:LwJ}). It is
more satisfying to notice that a little geometrical thinking on
Fig.~\ref{fig:RefFrames}(b) shows that it results from projecting the $J$ and
$L$ sides of the triangle given by the relation $\vec{J}=\vec{L}+\vec{N}$ onto
the $N$ side, because the common height is equal to $\vert L_{Z_Q} /
\sin\varphi_N \vert$.
By integration over $\varphi_N$ we obtain
\begin{eqnarray}\label{eq:F2LQ}
F_2(L_{Z_Q},\varphi_N) &=& \left( L_{Z_Q} \ArcXY\left( \sqrt{J^2-
(L_{Z_Q}/\sin\varphi_N)^2}, L_{Z_Q} \cot\varphi_N\right) \right.\\
&&\left. +J\ \ArcXY\left( J\cos\varphi_N, \sqrt{J^2- (L_{Z_Q}/\sin\varphi_N)^2}
\sin\varphi_N \right)\right) \nonumber\\
&\pm& \left( L_{Z_Q} \ArcXY\left( \sqrt{L^2-
(L_{Z_Q}/\sin\varphi_N)^2}, L_{Z_Q} \cot\varphi_N \right)\right.\nonumber\\
&&\left. +L\ \ArcXY\left( L\cos\varphi_N, \sqrt{L^2-
(L_{Z_Q}/\sin\varphi_N)^2} \sin\varphi_N\right)\right) \nonumber
\end{eqnarray}
where $\ArcXY$ is an unambiguous notation for the angle of a vector whose
projections on the $X$- and $Y$-axes are known (the order of parameters for the
analogous $\atand$ used in some programming languages is not fixed).

Notice that the derivative of $F_2$ with respect to $L_{Z_Q}$, which must give
$\varphi_L^Q$, equals the sum of the coefficients of $L_{Z_Q}$ in the first and
third terms of eq.~(\ref{eq:F2LQ}) (this also has a geometrical
interpretation). This shows that the sum of these terms is precisely equal to
$\varphi_L^Q L_{Z_Q}$. Using this result we can solve the problem of the 
integration
constant depending on $L_{Z_Q}$, which appears in the integration over
$\varphi_N$. We follow continuously the various angles over ranges greater than
$2\pi$ during the computations in order to compute the exact
values of the various $\ArcXY$.

Next we compute the actions $S(\varphi,\varphi^\prime)$ and
$S(\Lambda,\Lambda^\prime)$. We start from a pair of conjugate $\{
L_{Z_Q}^\prime, {\varphi_L^Q}^\prime \}$, i.e. a point on the Poincar\'e
sphere, and compute analytically the final $L_{Z_Q}$, $\varphi_L^Q$ as a
function of these. To compute, say $S(\varphi_L^Q,{\varphi_L^Q}^\prime)$, we
first invert numerically the relation between $\varphi_L^Q$ and
$L_{Z_Q}^\prime$ for fixed ${\varphi_L^Q}^\prime$. Usually, there are several
solutions. For each solution we then compute $S$ as a function of
${\varphi_L^Q}^\prime$ and $L_{Z_Q}^\prime$ by adding the
various $F_n$ generating functions which enter in the exponentials of the $U$
operators which are used to build $T$.

We do not need to evaluate all intermediate integrals by stationary phase to
obtain the amplitudes. Indeed, it was shown in ref.~\cite{Miller:ACP74-69} that
the formula for the amplitude given in eq.(\ref{eq:Tphiphi'}) is a direct
consequence of the action being a $F_1$ generating function of a canonical
transformation between the initial and the final states. We thus only need to 
compute the second derivative of $S$ with respect to $\varphi_L^Q$ and
${\varphi_L^Q}^\prime$. For this we use the derivatives of $S$ with respect to
initial variables $L_{Z_Q}^\prime$ and ${\varphi_L^Q}^\prime$, that we have
computed analytically by following the various steps, and general calculus
formulae for second derivatives of inverse functions. We have checked that
these amplitudes obey the relationship,
\begin{equation}\label{eq:AmpliDLz}
\left\vert \frac{\partial^2 S({\varphi_L^Q}^\prime,\varphi_L^Q)}
{\partial{\varphi_L^Q}^\prime \partial\varphi_L^Q} \right\vert = \left\vert
\left( \frac{\partial\varphi_L^Q} {\partial L_{Z_Q}^\prime}
\right)_{{\varphi_L^Q}^\prime} \right\vert^{-1}
\end{equation}
which was shown in ref.~\cite[eq.~(2.54b)]{Miller:ACP74-69} to be a simple
consequence of the conservation of probability. This is a very useful test
of our way to compute $S$ since the second term of this equation does not
depend on $S$.

\subsection{\label{sec:singularities} Singularities in the semi classical
computations}

\begin{figure}[!ht]
\centering
\TwoUnRelated{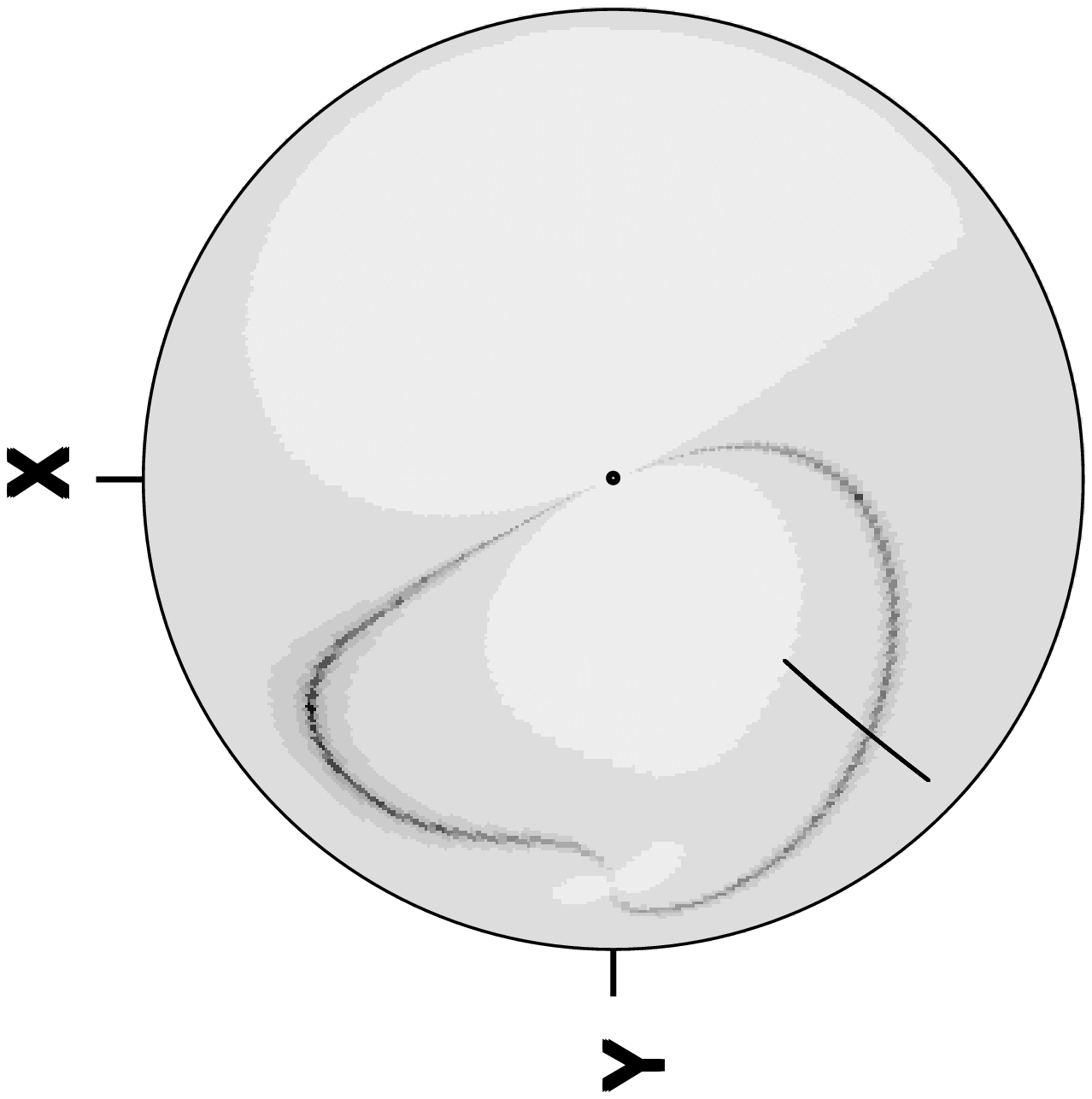}{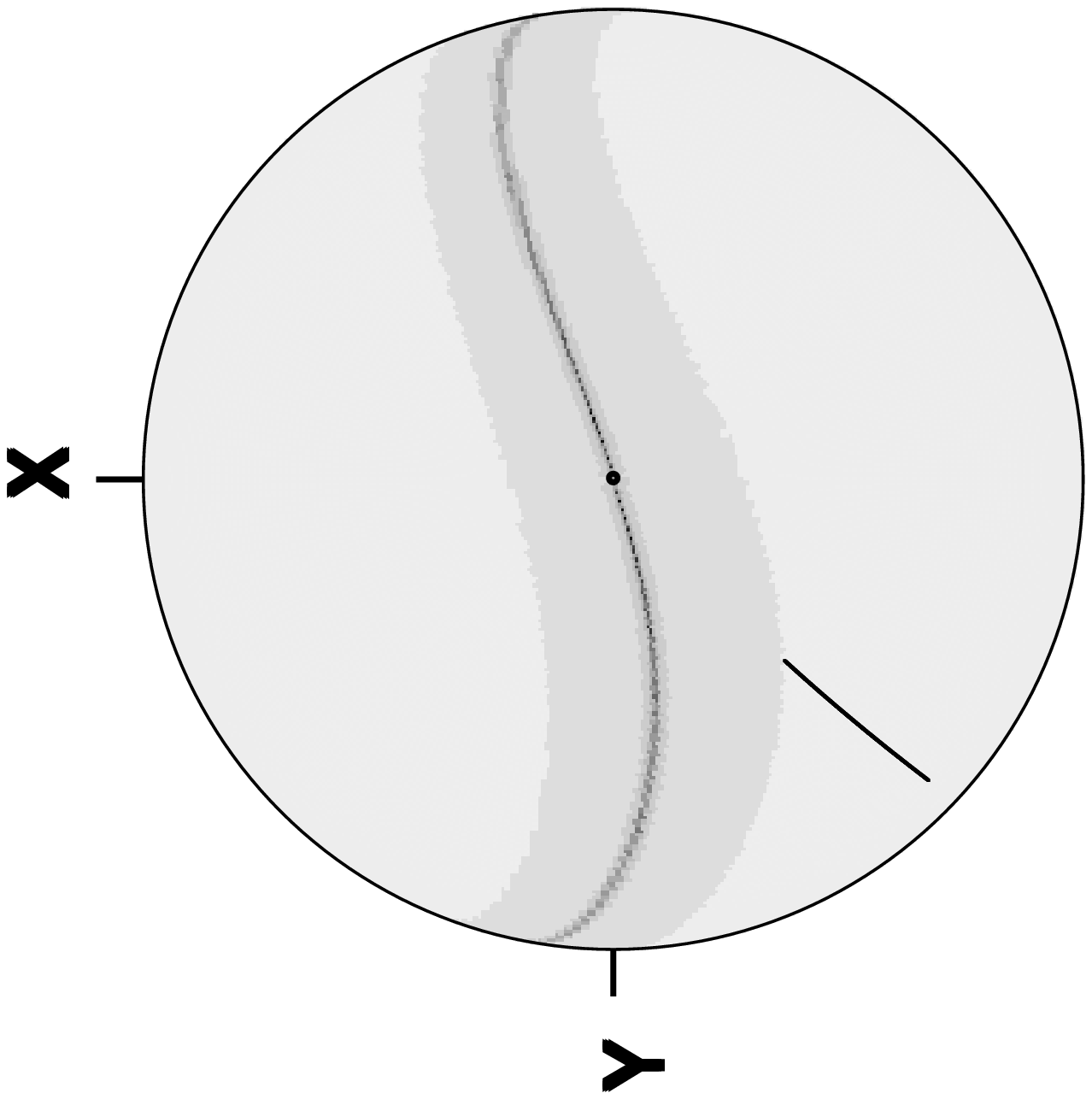}{a}{b}
\caption{\label{fig:Amplitudes} Divergence of the semiclassical amplitudes.
(a) $T(\varphi,\varphi^\prime)$. (b) $T(\Lambda,\Lambda^\prime)$. The part
of a straight line going through the $OZ$ pole in the lower left of the sphere
is the input corresponding to Fig.~\protect\ref{fig:FocusSing}(b): the fact
that it is slightly distorted is due to plotting in the ``classical'' reference
frame instead of the ``quantum'' reference frame.}
\end{figure}

We then use the resulting $T(\varphi,\varphi\prime)$ and 
$T(\Lambda,\Lambda^\prime)$ to compute the evolution of wave packets as 
was done in section~\ref{sec:Closed}. The first problem which appears is the
existence of divergences of the amplitudes. Fig.~\ref{fig:Amplitudes} plots
the amplitudes of $T$ as a function of the initial position on the Poincar\'e
sphere, which was used as a first step to compute them. The black lines of 
maxima
are actually divergences. They are at different positions for the two
representations.

\begin{figure}[!ht]
\centering
\TwoUnRelated{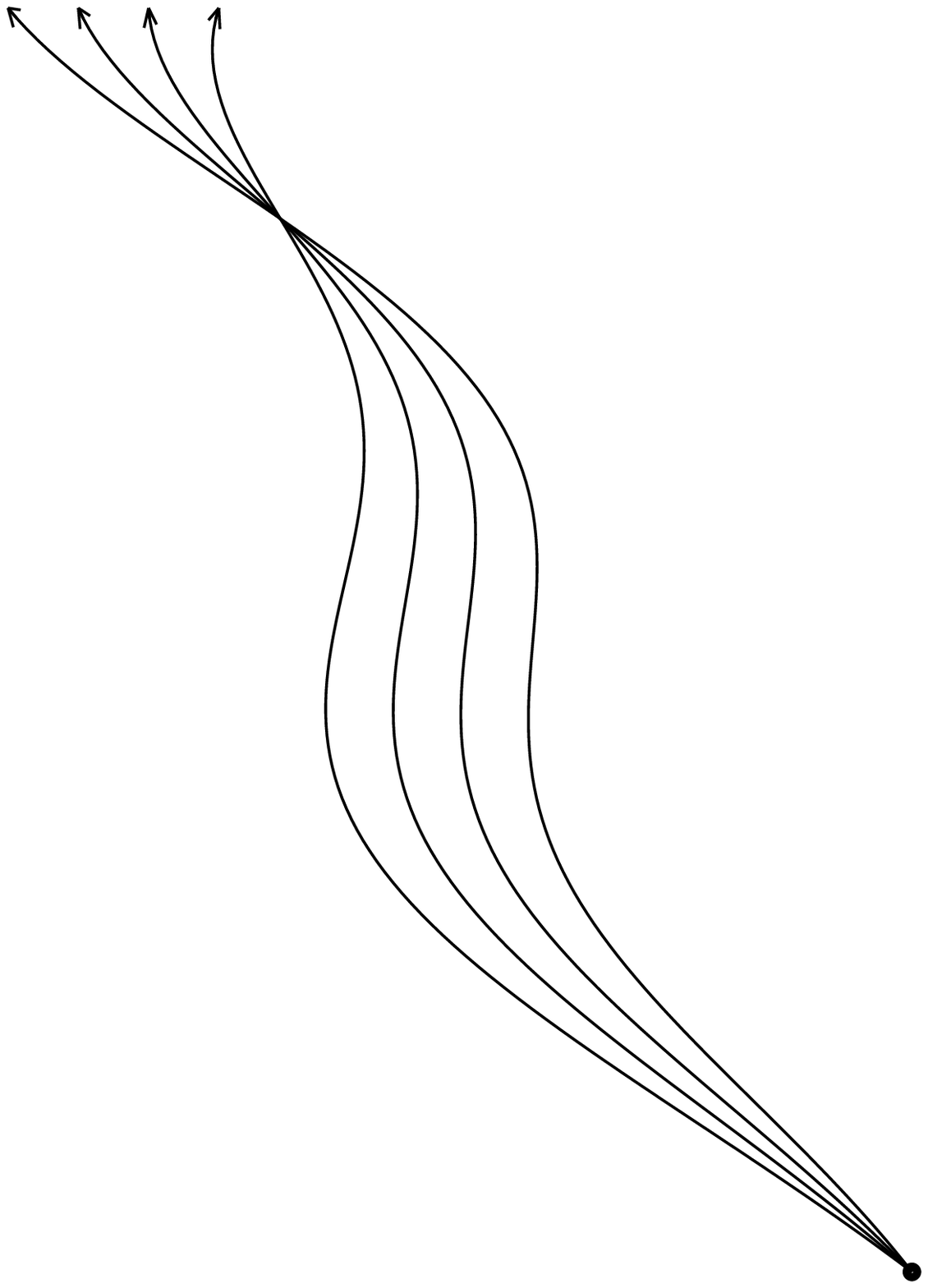}{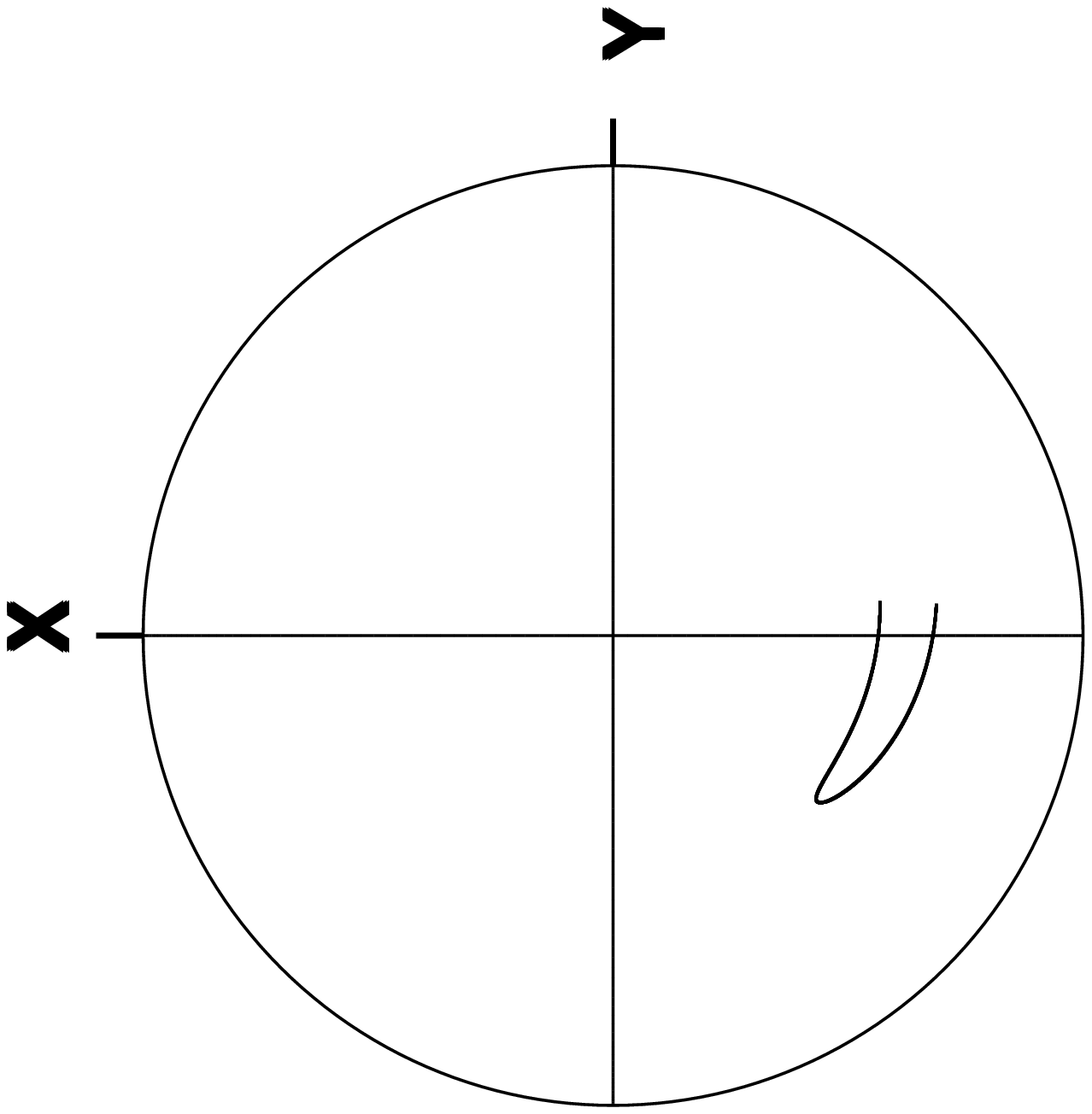}{a}{b}
\caption{\label{fig:FocusSing}Focus singularities. (a) represents such a
singularity in an ordinary (flat) coordinate space. (b) plots the image of the
straight oblique line in the lower left part of
Fig.~\protect\ref{fig:Amplitudes} (a) and (b) which cross the singularity line
in the $T(\varphi,\varphi^\prime)$ representation. All starting points on this
segment have the same coordinate $\varphi^\prime$, but different momenta
$L_Z^\prime$, i.e. they are equivalent to the starting points in (a). The back
folding of the image in (b), common especially to chaotic situations, is
the cause of the focus singularity: see text.}
\end{figure}

The physical meaning of these divergences is the existence of focus 
singularities.
We discuss the $\varphi$ representation first, by referring to
Fig.~\ref{fig:FocusSing}(a). Various trajectories starting from the same
coordinate point with different momenta refocus at the same point (on the map).
On the sphere the coordinate is $\varphi^\prime$ and the momentum is
$L_Z^\prime$. The points aligned on the straight line in the lower left part of
Fig.~\ref{fig:Amplitudes}(a) and (b), have the same value of $\varphi^\prime$ 
and a
varying $L_Z^\prime$. They thus constitute a fan of initial trajectories.
The image of this line is given in Fig.~\ref{fig:FocusSing}(b). The focus is
caused by the folding part of this image, precisely the point where it is
tangent to a radius from the $OZ$ pole: same position $\varphi$ independent to
first order of the initial momentum $L_Z^\prime$. The divergence of the
amplitude is a consequence of the presence of the term ${\partial\varphi} /
{\partial L_Z^\prime} \vert \varphi^\prime$ in the denominator of the
second part of eq.~(\ref{eq:AmpliDLz}).

The problem of singularities is common in any semi classical analysis. Their
solution, usually called ``regularization'' follow two main routes. The first,
common in physics textbooks, is to revert to a full quantum computation near
the singular point, and to match the results with the semi classical 
computation distant
from it. Notice that the MQDT analysis is such a fully quantum solution. The
second, common in some mathematical works, is to notice that we can circumvent
the problem by changing from coordinate to momentum representation through a
Fourier transform. The singularity is a result of a projection from phase
space onto coordinate space; there are no singularties in phase space
itself. The back bending in Fig.~\ref{fig:FocusSing}(b) which is tangent to a
meridian of the sphere is not tangent, at this point, to a parallel. The
starting points in Fig.~\ref{fig:Amplitudes} which project onto the same
$\varphi^\prime$ project to different $L_Z^\prime$. The lines of singularities
exist in both representations in Fig.~\ref{fig:Amplitudes}, but they appear at
different points. So one can compute the points
of the map by different routes, a possibility rooted on the possible choice at 
each $U$ step in the
computation of $T$ (eq.~(\ref{eq:TfromUs})) to select the $\varphi$ or $L_Z$
representation, both in the initial and the final state. The only remaining
problem is the fact that both lines of singularities share the $\pm OZ$ poles.
To solve it we would have to use a more general change of canonical
coordinates on the sphere, a rotation which can change the position of the
poles. We do not attempt to do this in the present paper.

\begin{figure}[!ht]
\setlength{\figwidth}{\textwidth}
\centering
\begin{tabular}{ccccc}
Start&Class.&MQDT&$T(\varphi,\varphi^\prime)$&$T(\Lambda,\Lambda^\prime)$\\
\\
\FiveNoLabel{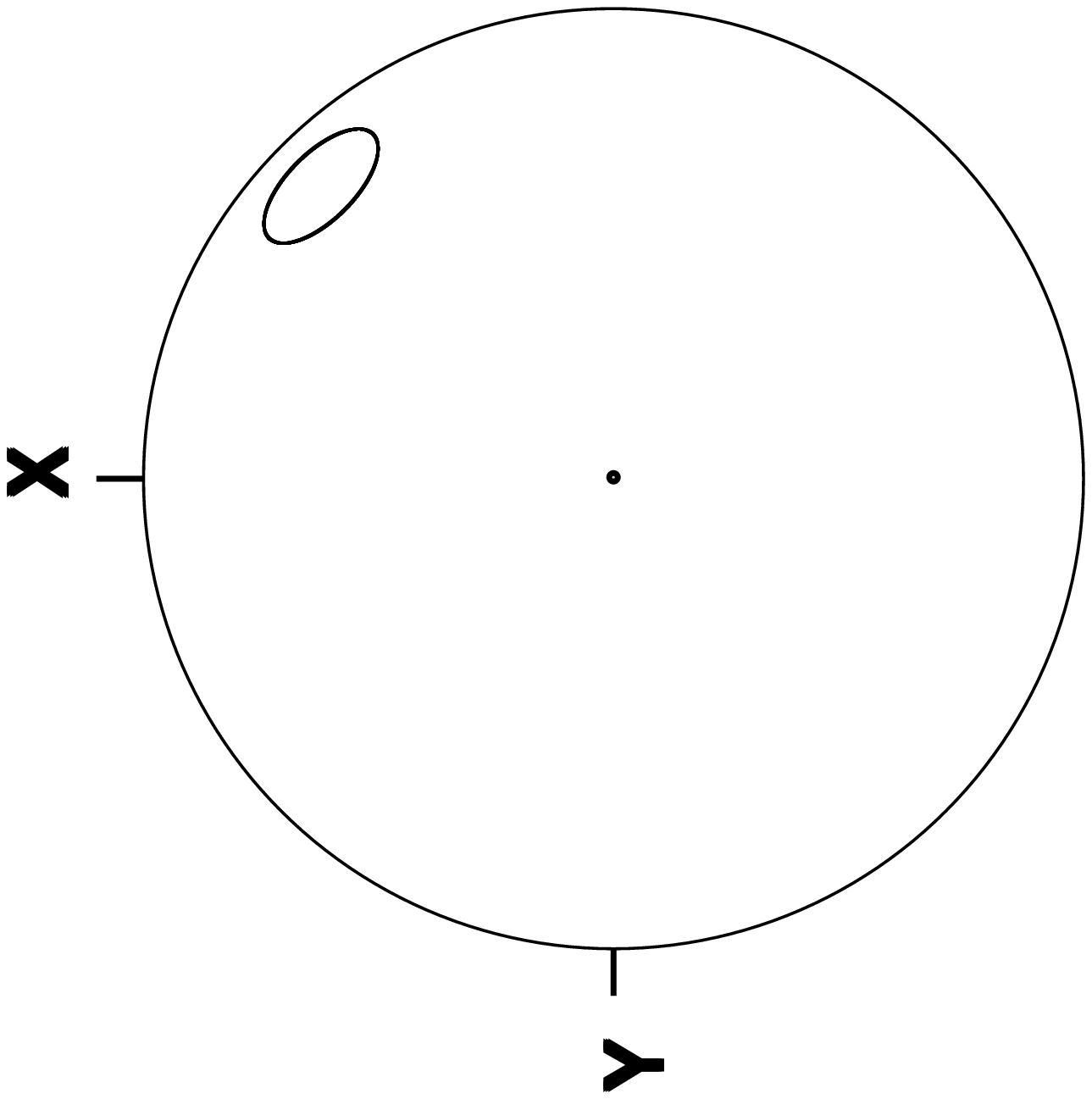}{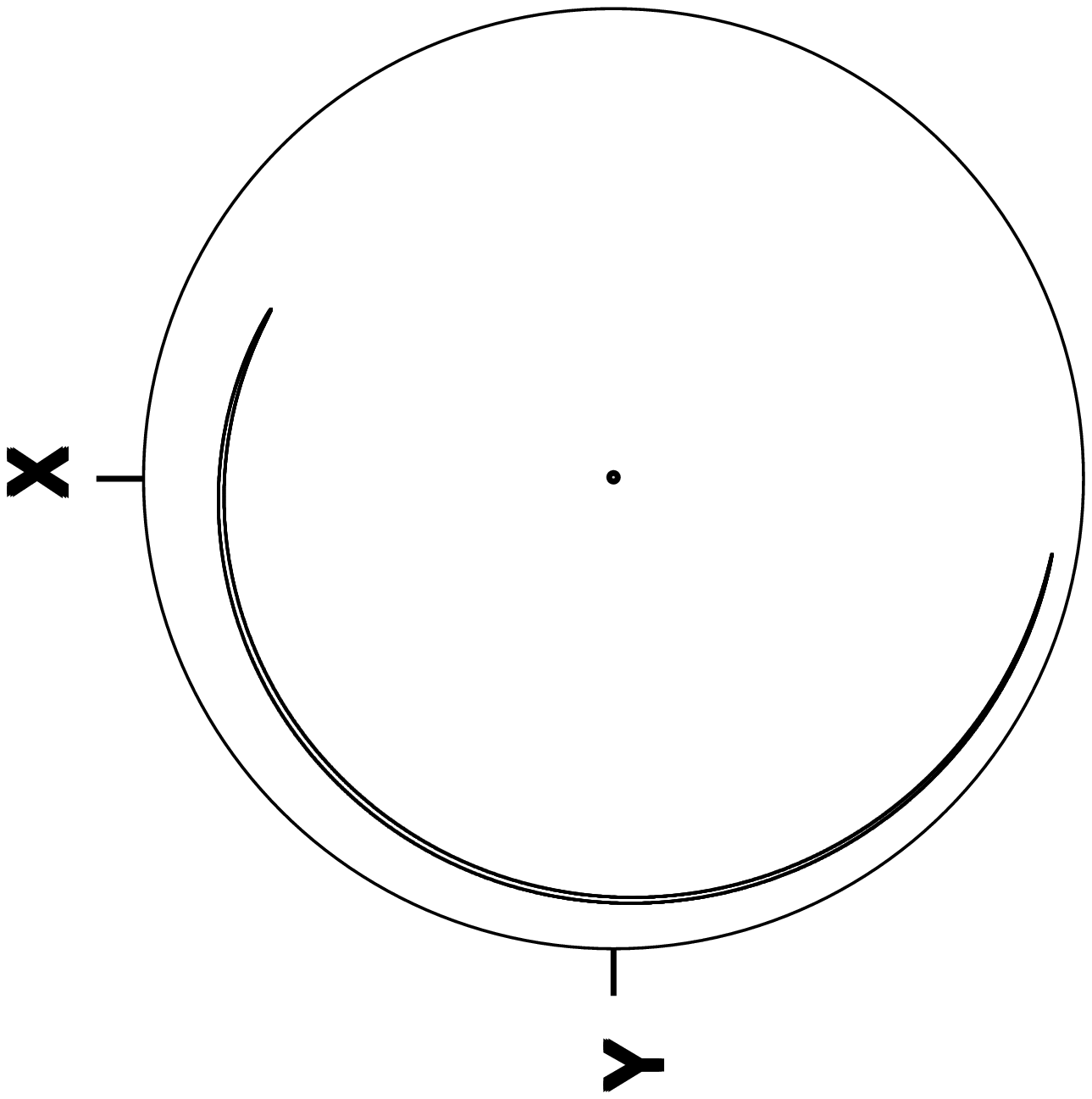}{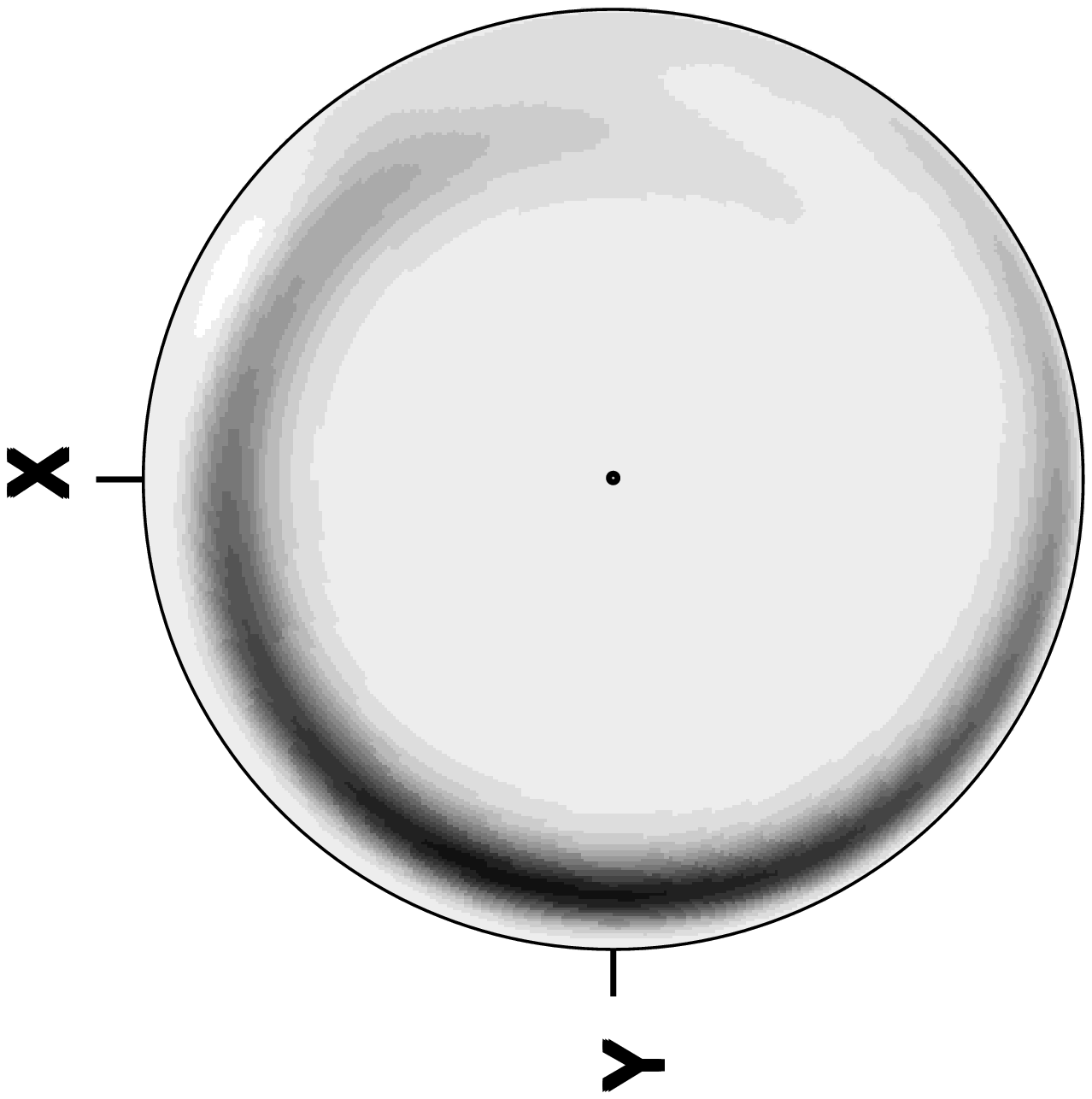}{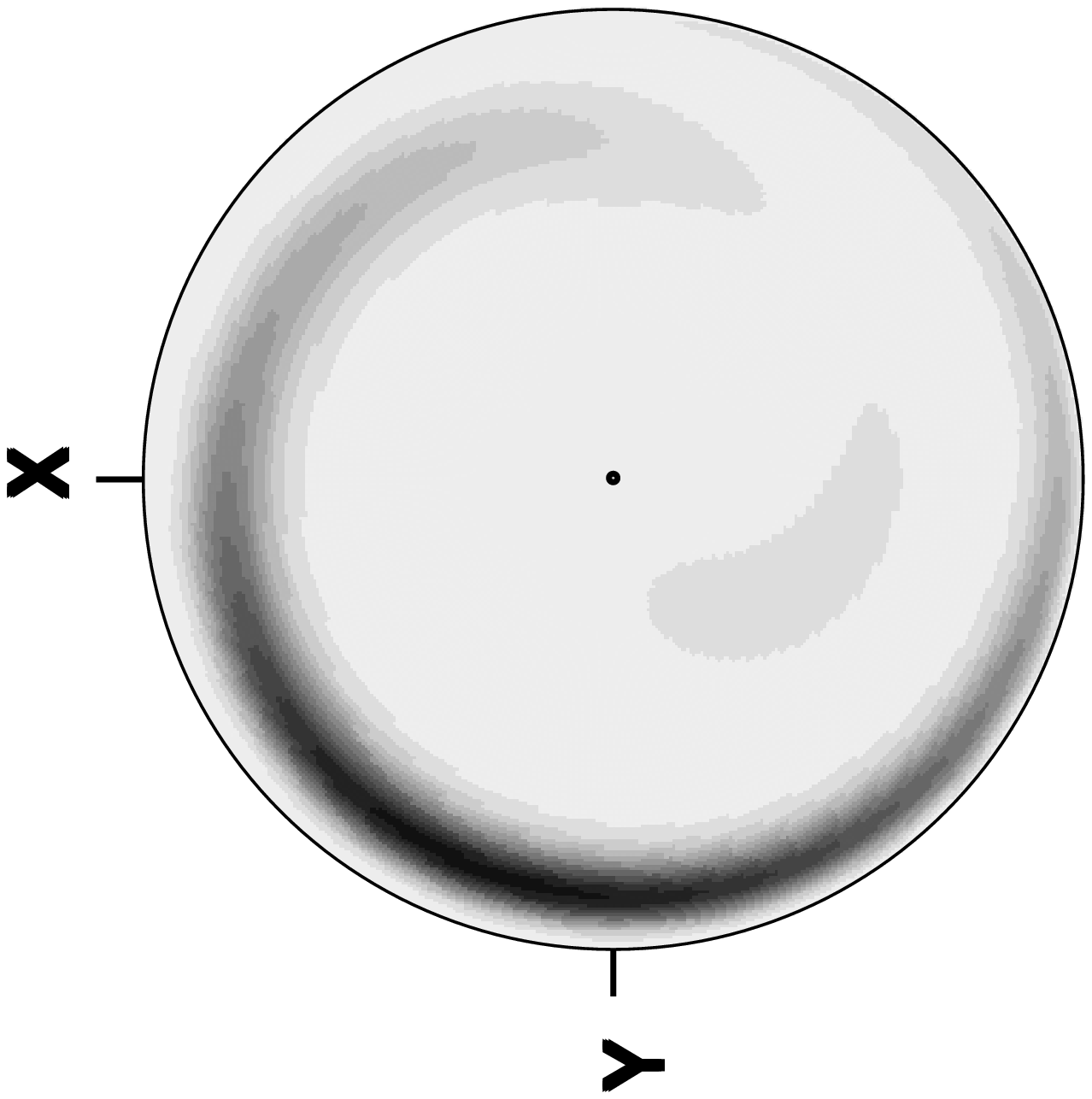}{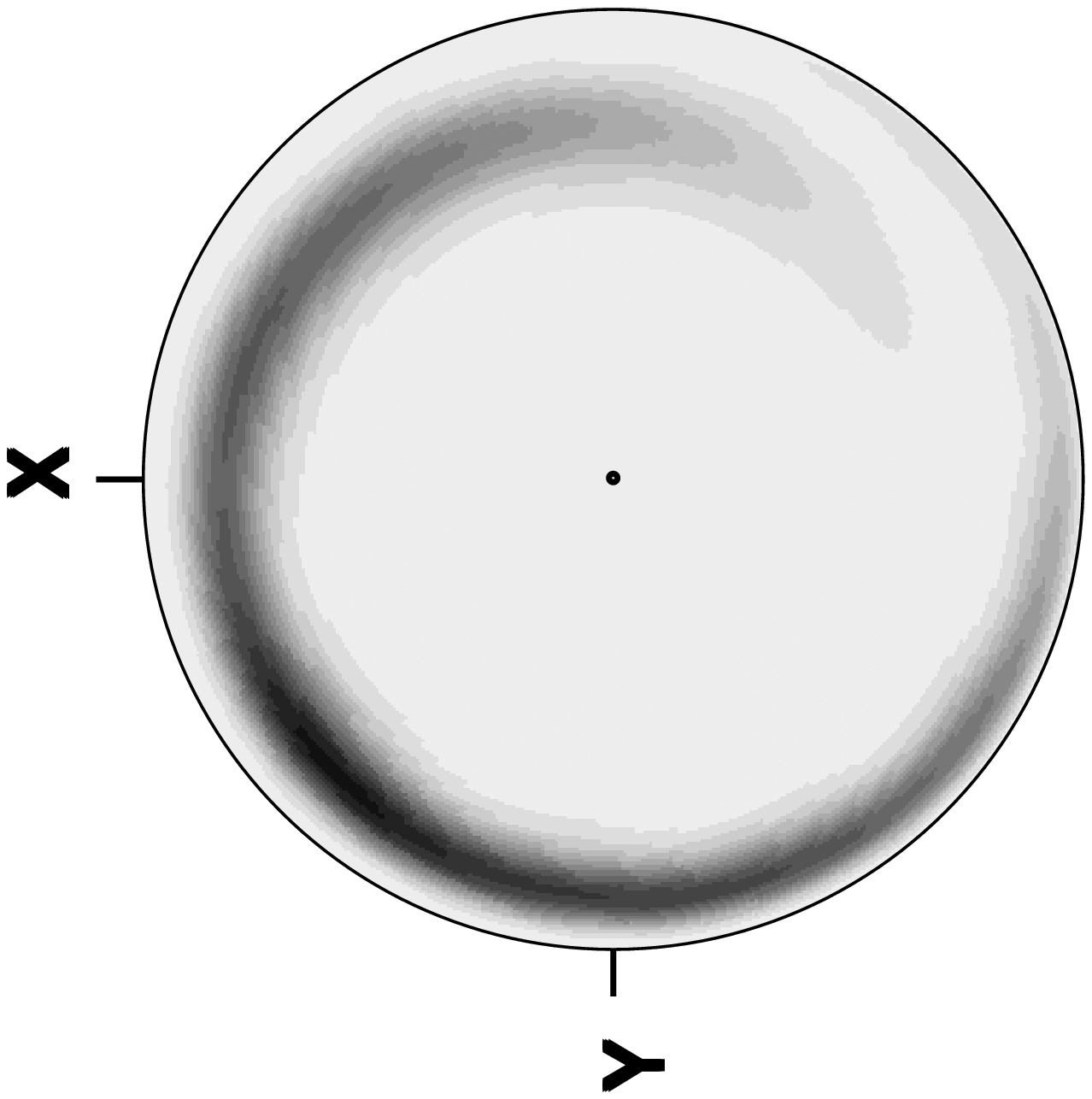}
\FiveNoLabel{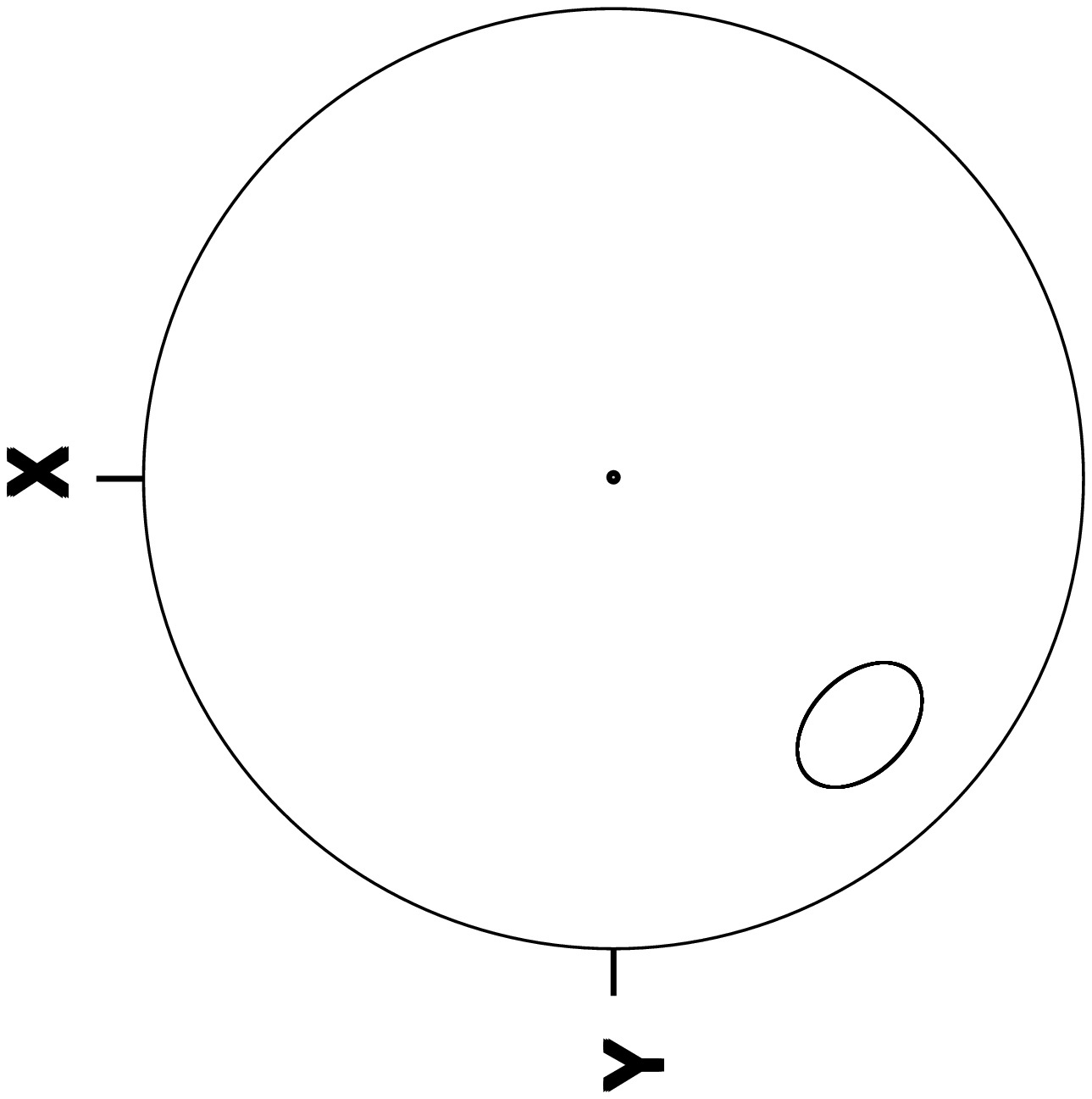}{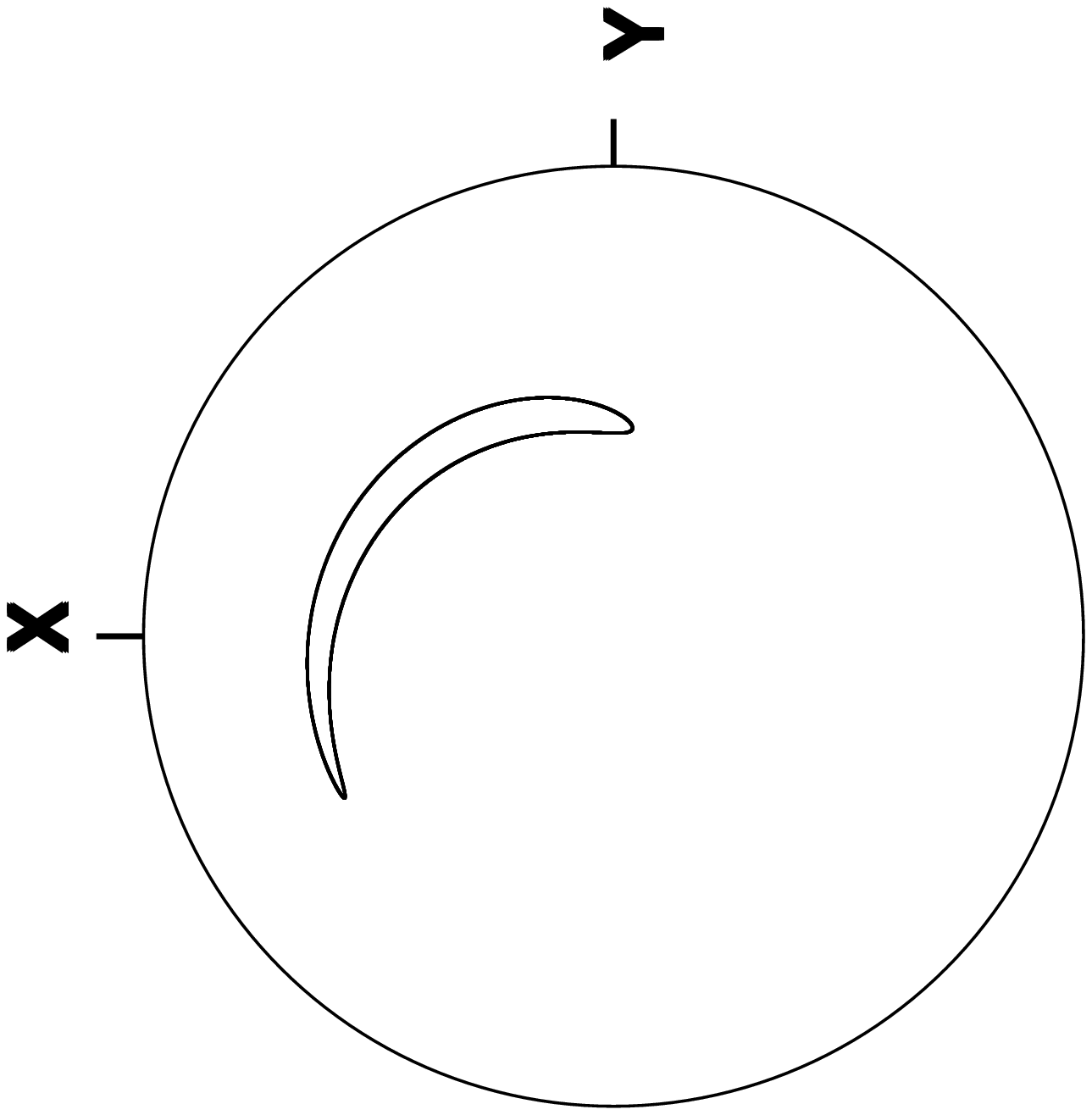}{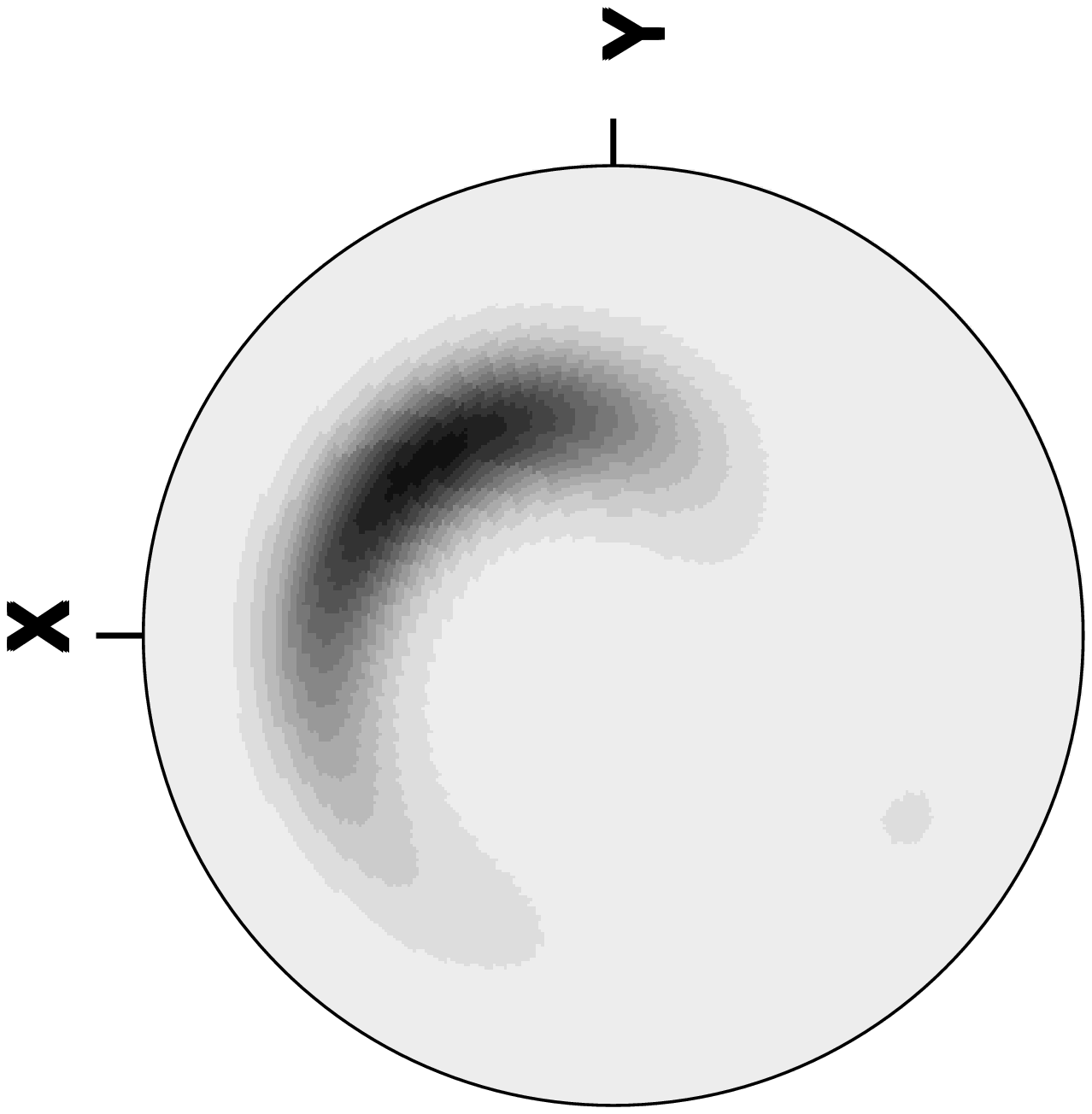}{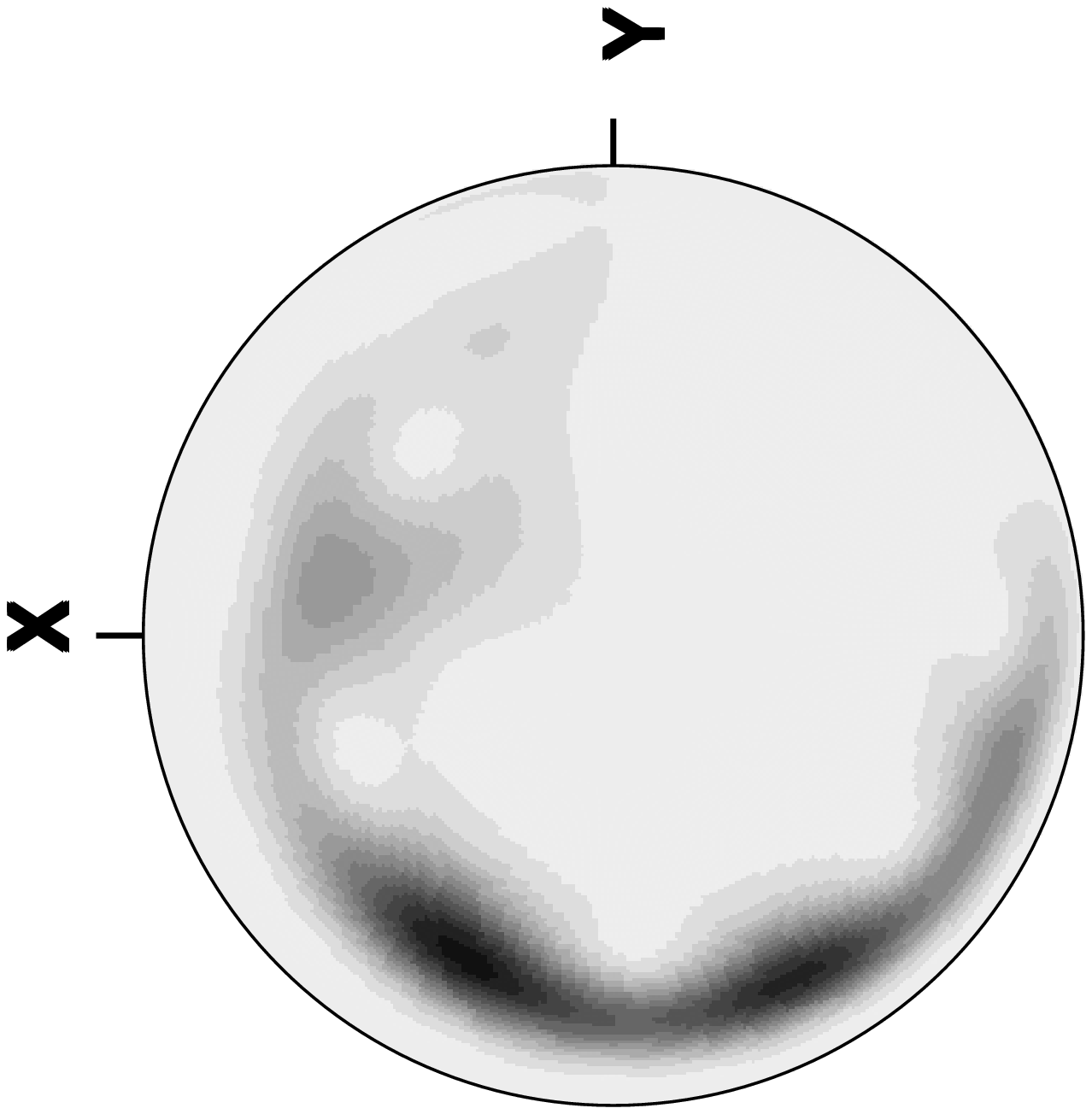}{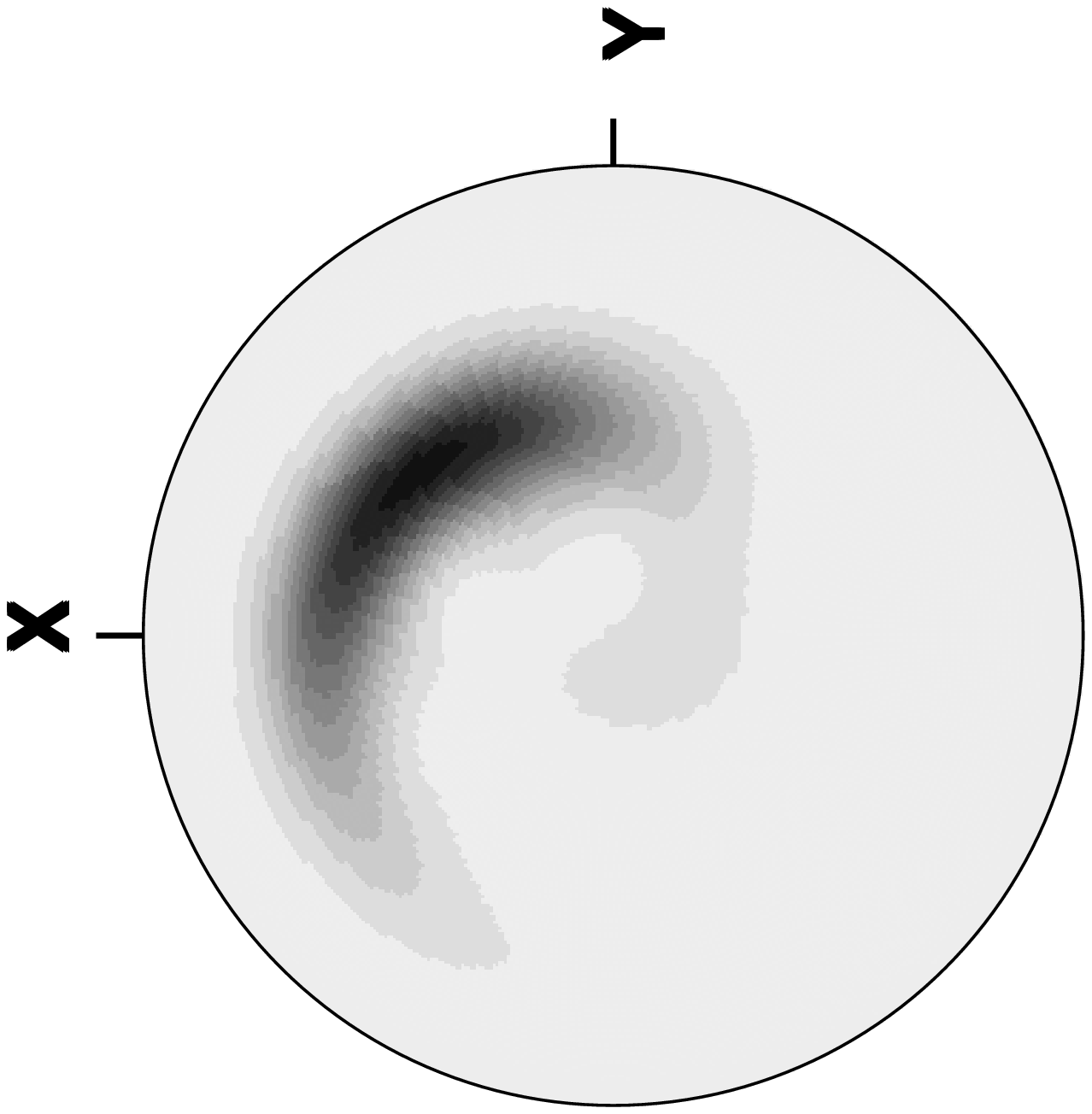}
\end{tabular}
\caption{\label{fig:NoWithSing} Effect of focus singularities. The starting
points lie on a circle. We show the first iteration by the Poincar\'e map,
and the corresponding Husimi
plots with different techniques. First row: starting in a point without
singularity. Second row: starting in a point with a $\varphi$ focus singularity,
at the intersection of the circle with the straight line in the lower left 
part of
Fig.~\protect\ref{fig:Amplitudes}(a) and (b) and the singularity line in the
$\varphi$ representation.}
\end{figure}

Illustration of this is given in Fig.~\ref{fig:NoWithSing}. The starting point
in the first row is selected to correspond to no singularity in any of the two
representations. All computations, classical, quantum and semi classical in
both representations agree nicely. In the second row the starting point lies on
the singular line in the $\varphi$ representation, but on a regular point in
the $\Lambda$ representation. The result of the $T(\varphi,\varphi\prime)$
computation is highly distorted, while that of the
$T(\Lambda,\Lambda^\prime)$ computation agree nicely with classical and MQDT
computations.

\section{Conclusion} We have formulated the multichannel quantum defect theory
for Rydberg molecules as the quantization of a classical symplectic map
composed of a standard surface of section for negative electron energies and a
Jung scattering map for positive ones. This formulation elucidates MQDT from an
entirely different point of view. Simultaneously we have shown it to provide
considerable computational advantages.

The method was applied in the context of a simplified model, where only
one rotational degree of freedom for the molecule was taken into account.
Simultaneously, the absolute value of the electron angular momentum was fixed.
This approximation allows for a two dimensional surface of section, and
simplified calculations, but is {\it not} essential to the representation of
MQDT as a quantum Poincar\'e map. These applications demonstrate the
simplifications achieved.
Finally we establish the relation to Bogomolny's
\cite{Bogomolny:CAMP90-67,Bogomolny:N92-805,Bogomolny:C92-5} semi classical
formulation of the quantum Poincar\'e map. Considering caustics we can clearly
see the basic difference between the exact and the semi-classical map.

\begin{appendix}

\section{\label{sec:QCFrames} Reference Frames}

We use three reference frames, the laboratory reference frame
$Oxyz$, and two molecular reference frames, called ``Classical'' $O X_C Y_C
Z_C$ and ``Quantum'' $O X_Q Y_Q Z_Q$ molecular reference frames.

In purely classical computations the ``Classical'' right handed molecular
reference frame is defined in such a way that $OZ_C$ lies along the inter 
nuclear axis
$\unitvec{M}$ of the molecular ionic core, and $OX_C$ lies along its angular
momentum $\vec{N}$. This is adequate because the two steps of the dynamics,
namely the collision and the free rotation, are rotations about theses itwo 
axes. 
We always use it
to display our results, as the physical content is most obvious within this
frame.

However, this is not the molecular reference frame selected in usual
definitions of quantum wave functions. There the molecular frame
$O{X_Q}{Y_Q}{Z_Q}$ is defined as the one deduced from the laboratory frame
$Oxyz$ by a rotation of three Euler angles (with the conventions of
\cite{MessiahII64}). The first two are well defined, namely $\varphi_M$,
$\theta_M$, the polar angles of $\unitvec{M}$. The third is arbitrary for a
diatomic molecule, since it corresponds to a rotation around the molecular
axis. It is given an arbitrary value $0$ or $\pi/2$, depending on authors; we
will select $0$ as in \cite{Fano:PRA70-253}.

Both reference frames share the same $OZ = \unitvec{M}$ axis, but have
different $OX$ axes. The classical $OX_C=\vec{N}$ axis is not well defined in
quantum mechanics, since $\vec{N}$ is a momentum conjugate to the angles
$\varphi_M$, and  $\theta_M$, which are the variables defining the rotational
molecular wave function in eqs.~(\ref{eq:CollisionWF}--\ref{eq:CoulombWF}).
Thus the classical reference frame defines the usual quantum wave functions
only in the semi classical limit. But conversely, the quantum reference
frame is also well defined in classical mechanics. So, if we want to compare
classical and quantum results, we do the quantum calculations in the usual
quantum reference frame, and then plot the resulting Wigner or Husimi
distributions in the classical reference frame, by applying the well defined
classical rotation which brings the quantum reference frame onto the classical 
one.

We shall now compute the formulae which connect these three frames. They are
based on the fact that we can compute the position of $\vec{J}$ and
$\vec{N}$ in $O{X_Q}{Y_Q}{Z_Q}$ once we know the position of $\vec{L}$ in this
frame, using simple geometrical relations: see Fig.~\ref{fig:RefFrames}(b).
We first need to define precisely the laboratory reference frame. Since
everything is invariant with respect to a rotation of the whole space, we may
choose the $Oz$ axis along $\vec{J}$ (Fig.~\ref{fig:RefFrames}(a)). Quantum
mechanically this amounts to studying the $\vert J,M_J \rangle = \vert J,J
\rangle$ state. Conversely, once we know the position of $\vec{J}$ in the
molecular frame, we know the position of the $Oz$ axis of the laboratory frame
in $O{X_Q}{Y_Q}{Z_Q}$, but there is no way to locate the $Ox$ and
the $Oy$ in this molecular frame. This is a consequence of the rotational
invariance of the problem. The corresponding angle $\varphi_M$ is conjugate to
$J_z$, and is one of the variables we have eliminated when reducing the
dimension of phase space from 10 to 4 (or 6 if $L$ is not kept constant).

The position of $\vec{J}$ in $O{X_Q}{Y_Q}{Z_Q}$ is a simple consequence of
our choice of $0$ as the third Euler angle of the rotation which brings the
laboratory frame $Oxyz$ onto the ``Quantum'' molecular reference frame
$O{X_Q}{Y_Q}{Z_Q}$. These three Euler angles are $\varphi_M$, $\theta_M$, and
$0$. The first Euler rotation of the angle $\varphi_M$ keeps $\vec{J} = Oz$
fixed and brings $Oy$ onto $O{Y_Q}$. The second Euler rotation around $OY_Q$ by
an angle $\theta_M$ brings $Oz$ onto $OZ_Q$. Conversely the position of
$\vec{J}$ in the molecular frame is obtained by a rotation of $-\theta_M$
around $OY_Q$. It is thus located in the $O{X_Q}{Z_Q}$ plane with coordinates
$J_{X_Q} = -J \sin \theta_M, J_{Y_Q}=0, J_{Z_Q} = J \cos \theta_M$, always
directed towards negative $OX_Q$ since $0 \leq \theta_M \le \pi$.

\subsection{\label{sec:ClassQuantTransform} Transformations between
``Classical'' and ``Quantum'' molecular frames}

Let us compute $\vec{N}$ in $O{X_Q}{Y_Q}{Z_Q}$. For a diatomic molecule, the
momentum $\vec{N}$ is perpendicular to the inter nuclear axis $\unitvec{M}$,
since the moment of inertia around the inter nuclear axis is zero. Thus
$\vec{N}$ is in the $O{X_Q}{Y_Q}$ plane. We then write
$\vec{N}=\vec{J}-\vec{L}$ in this frame

\begin{eqnarray}\label{eq:ConservQ}
 N_{X_Q} =  &-J \sin\theta_M &-L \sin\theta_L^Q \cos\varphi_L^Q \nonumber\\
 N_{Y_Q} =  & 0              &-L \sin\theta_L^Q \sin\varphi_L^Q\\
 N_{Z_Q} =  & J \cos\theta_M &-L \cos\theta_L^Q\nonumber
\end{eqnarray}

$N_{Z_Q}=0$ leads to

\begin{eqnarray}\label{eq:costheta}
  \cos\theta_M &=& \frac{L}{J} \cos\theta_L^Q\\
  \sin\theta_M &=& +\sqrt{1-\cos^2\theta_M}, \nonumber
\end{eqnarray}

because $ 0\leq\theta_M\le\pi$. This gives the position of $\vec{J}$ once
$\vec{L}$ is known in $O{X_Q}{Y_Q}{Z_Q}$.
Then $\vec{N}$ lies in the plane $O{X_Q}{Y_Q}$ at an angle

\begin{eqnarray}\label{eq:deltaphiQversC}
  \varphi_N^Q &=& \ArcXY{}\left(N_{X_Q},N_{Y_Q}\right)\\
  &=& \ArcXY{}\left(-\sqrt{J^2-L^2\cos^2\theta_L^Q}-L \sin\theta_L^Q
\cos\varphi_L^Q, -L\sin\theta_L^Q\sin\varphi_L^Q\right)\nonumber
\end{eqnarray}
where we have used eqs.~(\ref{eq:ConservQ}) and (\ref{eq:costheta}),
and where $\ArcXY$ is defined after eq.~(\ref{eq:F2LQ}).

Thus the ``quantum''  molecular reference frame angles for $L$ are related to
their corresponding ``classical'' reference frame value via:

\begin{eqnarray}
\theta_L^C &=& \theta_L^Q\\
\varphi_L^C &=& \varphi_L^Q-\varphi_N^Q\nonumber
\end{eqnarray}

The reverse is not obtained by simply inverting this relation. We
rather express the relation $\vec{J}=\vec{N}+\vec{L}$ in the ``classical''
reference frame, where we know that $\vec{N}$ is along $OX_C$.

\begin{eqnarray}\label{eq:ConservC}
J_{X_C} = & N & +L \sin\theta_L^C \cos\varphi_L^C \nonumber\\
J_{Y_C} = &   & +L \sin\theta_L^C \sin\varphi_L^C \nonumber\\
J_{Z_C} = &   & +L \cos\theta_L^C
\end{eqnarray}

Since the $OZ$ axis is common in the ``quantum'' and ``classical'' molecular
reference frames, relations \ref{eq:costheta} are conserved with $\theta_L^Q$
replaced by $\theta_L^C$. Then we recall that in the $O{X_Q}{Y_Q}{Z_Q}$
reference frame the projection of $\vec{J}$ into the $O{X_Q}{Y_Q}$ plane, which
is identical with the $OX_CY_C$ plane, is always directed along the negative
$OX_Q$ axis, which locates the $OX_Q$ axis in the ``classical'' frame. Thus

\begin{eqnarray}\label{eq:deltaphiCversQ}
\varphi_L^Q&=\varphi_L^C  - &
\ArcXY{}\left( -N -L\sin\theta_L^C \cos\varphi_L^C, -L \sin\theta_L^C
\sin\varphi_L^C\right)\\
&=\varphi_L^C  - & \ArcXY{} \left( -\sqrt{J^2-L^2(1-\sin^2\theta_L^C\,
\cos^2\varphi_L^C)}, -L \sin\theta_L^C\, \sin\varphi_L^C \right)\nonumber
\end{eqnarray}
where the second line is obtained by using eq.~(\ref{eq:IonisationCap}) which
was written in the ``Classical'' molecular frame.

Finally notice that the transformation between $\{~L_{Z_Q},\varphi_L^Q~\}$ and
$\{~L_{Z_C},\varphi_L^C~\}$ is not canonical, as can be checked by mere
differentiation from the non conservation of $\rmd L_Z \wedge \rmd\varphi_L$ 
(where
$\wedge$ denotes the exterior differential product). We have computed a
generating function, and it can be checked by differentiation that a momentum
conjugated to $\varphi_L^C$ would be
\begin{equation}\label{eq:IC}
\begin{split}
   & I(\varphi_L^C,\theta_L) = L\cos\theta_L + \\
   & \frac{
  (J^2-L^2) \mathbf{F} \left( \frac{\pi}{2}-\theta_L,
\frac{L^2\cos^2\varphi_L^C} {J^2 - L^2 \sin^2 \varphi_L^C} \right)
- (J^2-L^2\sin^2\varphi_L^C) \mathbf{E} \left( \frac{\pi}{2}-\theta_L,
\frac{L^2\cos^2\varphi_L^C} {J^2 - L^2 \sin^2 \varphi_L^C} \right)
    }{\cos\varphi_L^C \sqrt{J^2-L^2\sin^2\varphi_L^C}}
\end{split}
\end{equation}
where $\mathbf{E}$ and $\mathbf{F}$ are elliptic functions, instead of $L_Z^C =
L_Z^Q = L\cos\theta_L$. It can be checked that the extra term vanishes when
$L/J \ll 1$, as it should do since in this case the frame recoil is negligible.
The effect is thus small in this paper where we have selected $L/J=1/5$ or
$1/2$. But this formula is too unwieldy to be of any use, especially since our
main goal in using ``classical'' reference frame coordinates is physical
visualizing of the results.

\subsection{\label{sec:QuantLaboTransform}Transformation between ``Quantum''
molecular reference frame and laboratory reference frame}

The canonical momentum and angle in the ``quantum'' molecular reference frame
are $L_{Z_Q}$ and $\varphi_L^Q$. The canonical momentum and angle in the
laboratory frame (for the reduced phase space) are $N$ and its conjugate angle
$\varphi_N$, to be defined, and not to be confused with $\varphi_N^Q$
(eq.~\ref{eq:deltaphiQversC}), which is the angle of $\vec{N}$ in the plane
$O{X_Q}{Y_Q}$. We first express $N$ as a function of ``quantum'' molecular
coordinates (eq.~(\ref{eq:IonisationCap}) was with ``classical'' molecular
coordinates). Using eqs.~(\ref{eq:ConservQ}, \ref{eq:costheta}) we obtain
\begin{equation}\label{eq:N_Q}
N^2=J^2+L^2-2 L_{Z_Q}^2 + 2\, \sqrt{L^2-L_{Z_Q}^2} \sqrt{J^2-L_{Z_Q}^2}
\cos\varphi_L^Q
\end{equation}

Intuitively (we will justify it below) the angle conjugate to $N$ is the angle
$\varphi_N$ of rotation of the molecular axis $\unitvec{M}$ around $\vec{N}$.
The reference for this angle can only be the position of $\vec{J}$. Notice that
the relative positions of $\vec{J}$ and $\vec{N}$ have nothing special. We
choose as reference axis around $\vec{N}$ the unit normal $\vec{n}$ to the
plane \{$\vec{N}$, $\vec{J}$\}:
\begin{equation}
   \vec{n} = \frac{\vec{N}\times\vec{J}} {\vert\vec{N}\times\vec{J}\vert} =
\frac{\vec{L}\times\vec{J}} {\vert\vec{L}\times\vec{J}\vert}
\end{equation}
(using $\vec{N}=\vec{J}-\vec{L}$, $\times$ denotes the skew vector product) and
the unit binormal
\begin{equation}
\vec{m}=\frac{\vec{N}}{N} \times \vec{n}
\end{equation}
so that, after a little algebra
\begin{eqnarray}\label{eq:phiN}
 \varphi_N &=& \ArcXY(\vec{n}\cdot\unitvec{M}, \vec{m}\cdot\unitvec{M})\\
  &=& \ArcXY\left( \frac{(\vec{L}\times\vec{J})_{Z_Q}}
{\vert\vec{L}\times\vec{J}\vert}, \frac{-N J_{Z_Q}}
{\vert\vec{L}\times\vec{J}\vert}\right) \nonumber\\
 &=& \ArcXY\left( \frac{- \sqrt{L^2-L_{Z_Q}^2} \sqrt{J^2-L_{Z_Q}^2}
\sin\varphi_L^Q} {\vert\vec{L}\times\vec{J}\vert},\frac{-N L_{Z_Q}}
{\vert\vec{L}\times\vec{J}\vert} \right)
\end{eqnarray}
with
\begin{equation}\label{eq:LwJ}
  {\vert\vec{L}\times\vec{J}\vert}^2 = L^2 J^2 - \left( \frac{J^2+L^2-N^2}{2}
\right)^2
\end{equation}

That this is the correct definition of $\varphi_N$ is checked by showing by mere
derivation that $\rmd L_{Z_Q} \wedge \rmd\varphi_L^Q = \rmd N \wedge
\rmd\varphi_N$.

Reciprocal relations are obtained from the right part of eq.~(\ref{eq:phiN})
\begin{equation}\label{eq:LzQ}
L_{Z_Q}= -\frac{\sin\varphi_N}{N} {\vert\vec{L}\times\vec{J}\vert}
\end{equation}
and by extracting $\cos\varphi_L^Q$ from eq.~(\ref{eq:N_Q}) and
$\sin\varphi_L^Q$ from the left part of eq.~(\ref{eq:phiN}), and replacing
$L_{Z_Q}$ by the previous expression, which leads to:
\begin{equation}\label{eq:phiLQ}
  \varphi_L^Q=\ArcXY\left(\frac{N^2-J^2-L^2}{2}+\frac{\sin^2\varphi_N} {N^2}
\vert\vec{L}\times\vec{J}\vert^2, -\cos\varphi_N \vert\vec{L}\times\vec{J}\vert
\right)
\end{equation}

\end{appendix}

%GATHER{paper.bib}
%GATHER{paper.bbl}
\bibliography{paper}

\end{document}